# Assessing the impact of social activity permissiveness on the COVID-19 infection curve of several countries

**Gerardo L. Febres**, gerardofebres@usb.ve

Departamento de Procesos y Sistemas, Universidad Simón Bolívar, Sartenejas, Baruta, Miranda 1080, Venezuela

**Abstract:** This document aims to estimate and describe the effects of the social distancing measures implemented in several countries with the expectancy of controlling the spread of COVID-19. The procedure relies on the classic Susceptible-Infected-Removed (SIR) model, which is modified to incorporate a permissiveness index, representing the isolation achieved by the social distancing and the future development of vaccination campaigns, allowing the math model to reproduce more than one infection wave. Simulating the contagious process attaining various stages with constant permissiveness connected with short transitions, allowed estimating the shape of the social permissiveness curve for 38 countries. The adjusted SIR models are used to study the compromise between the economy's reactivation and the resulting infection spreading increase. The document presents a graphical abacus that describes the convenience of progressively relax social distancing measures while a feasible vaccination campaign develops.

## 1. Introduction

There is a long list of pandemics in World history. Some pandemics have charged relevant fractions of the human population at the time they occurred. While COVID-19 is currently in the middle of its development, it will not likely compete with the deadliest pandemics the world has seen. COVID-19 is about the ninth place in the deadliest pandemics of history [1]. However, the SARS-Cov-2 has proven to be an uncommonly contagious virus. Some authors conceive a perspective locating SARS-CoV-2 among the most dangerous in history[2]. Arguably, the COVID-19's reproduction number is high as compared to other viruses [2,3]. Despite its medium-range death rate, it has become one of the most universal and challenging contemporary times. Under similar circumstances, a less deadly but more contagious disease tends to produce a more profound impact than a less contagious disease. The contagiousness is the main parameter causing an exponential growth of the infection spread. The characteristic death rate, on the contrary, reflects severe effects with an impact that grows linearly with the characteristic index.

Ever since the set of papers by Kermack, W.O. and McKendrick [4–6], A.G., published between 1927 and 1933, which first formulated the infectious phenomenon as the SIR model, the disease reproduction number, which at the time the authors associated with the population density, has been a dominant factor in modeling the spread of infectious diseases. Kermack and McKendrick even analyzed their formulation around a term they called "critical density." The Kermack and McKendrick's formulation has been notoriously successful. Almost a century after their publications, the Susceptible-Infected-Retired (SIR) model, with its refined versions, is still the cornerstone of all infectious disease mathematical modeling. In his books [7,8], J.D. Murray presents a good selection of the SIR models and their derivatives. Advances in applying the SIR model incorporate physical barriers and geographical discontinuities that affect the viruses' dissemination. Detailed consideration of these spatial effects is the subject of many studies. However, given the worldwide reaction to COVID-19, which has limited the inhabitants' national and international mobility, the present study relies on the basic SIR model, thus disregarding diffusion and spatial effects.

The fear of immersing into an uncontrollable deathly spread of the infection has led to an unprecedented world's slowdown of social and economic activities. The initially infectious disease problem COVID-19 has unfolded into a multi-aspect problem combining extended medical and economic fight fronts. An overview of the disease's development worldwide is justified to configure solutions for the global situation COVID-19 currently represents.

The high contagiousness of COVID-19 and the distinctive worldwide impact compared to other previously experienced contagious situations makes the study of the compromise between social distancing measures, the limited speed for vaccinations, and the pace of the economic pace reactivation a relevant need for any country.

One objective of this paper is to develop a method to describe and determine the effectiveness of a sequence of social distancing plans. By simulating the SIR models adjusted to each studied country, we determine approximate sequences of social distancing plans which must have been implemented to produce the results represented in the historical data available. Finally, we use the adjusted models to describe the compromise between the reactivation of economic activities versus the permissiveness of social distancing measures and the feasible vaccination plan.

## 2. Adjustment to the classical SIR Model

The following equations correspond to the classical SIR model presented by Kermack-McKendrick in 1927 [4] that describes the dynamics of contagious disease in a population. The Susceptible individuals (S) may get Infected (I) and eventually become part of the group of Removed (R) individuals.

$$\frac{dS}{dt} = -r\ S\ I\ , \tag{1a}$$

$$\frac{dI}{dt} = r\ S\ I - a\ I\ , \tag{1b}$$

$$\frac{dR}{dt} = a\ I, \tag{1c}$$

$$N = S + I + R. \tag{1d}$$

The parameter $a$ is the inverse of the characteristic time it takes for an infected individual to recover or die. Thus, $a$ is named the removed rate and describes the flow of individuals being removed from the infected group and entering the removed group. The parameter $r$, named the infection rate, establishes a pressure for the susceptible population $S$ to pass to the infected group $I$. This pressure is proportional to the product $S \cdot I$.

The basic reproduction rate $Ro$ of the SIR model represented in Equations (1abcd) is the number of people infected before the first infected individual gets into the retired group. Since at the start of the process, all individuals are susceptible, the value of $Ro$ is determined as

$$Ro = \frac{r\ N}{a}.$$

### 2.1. The assumption of an homogeneus density.

The classical SIR Model is the backbone behind the dynamics of a contagious disease propagation process. *It is an elegant set of equations that replicates any infectious process developing in a space homogeneously populated*. This model does not consider contagiousness change or external intervention, thus implying the system's parameters as $r$ and $a$. The population involved also is considered constant, as well as its distribution on the geographical area.

### 2.2. The SIR Model adjusted for social distancing meassures and vaccination

The classical SIR Model (1abc) is the unarguable reference model for all infectious diseases. However, it lacks some degrees of freedom needed to explicitly model different conditions in the unfolding of COVID-19 in each country. The SIR Model (1abc) cannot produce more than one wave. The derivative of the infected people $dI/dt$ changes its sign only. It occurs at the peak at the peak in the infectious curve when the removal rate $a$ becomes larger than *the product* $r\ S$. But it never changes again. To properly Model COVID-19's behavior, the SIR Model (1abcd) must be adjusted. We see the potential for enhancing the adaptability of the SIR Model by incorporating the following parameters:

*Confinement Permissiveness*, $e(t)$. The confinement permissiveness $e(t)$ represents reducing of the propagation speed achieved by the social distancing and lockdown programs. The measures and policies to fight the propagation of the epidemic develop on the way and change over time. We use a time-series $e(t)$ of values between zero and one to represent how these policies vary over time. A value of $e(t)$ close to zero represents nearly complete isolation, whereas values close to one represent no isolation or a pre-COVID-19 normal social behavior.

*Vaccination plan*, $v(t)$. By December 2020, several vaccines have been developed. A COVID-19 vaccine is expected to immunize individuals, therefore reducing the susceptible population. In this study, a sequence of stages of vaccination activity is represented by $v(t)$.

*Variable Removal Rate*, $a(t)$. The removal rate, represented as the constant $a$ in the Model (1abcd), is related to the speed of removal of infected individuals. The removal occurs when an individual goes to an isolated environment —a hospital or a place for treatment— or when he or she dies. Ironically, a high removal rate may indicate a fast transit through sickness and early death. Nevertheless, it may also be due to the authorities' high capacity to test and quickly identify infected people and, therefore, properly isolate and medically treat them. The removal rate may change along the stages of the pandemic in any country. Thus, we include the variable removal rate $a(t)$ in Model (2abcd) for the sake of completeness.

*Removal delay*. For COVID-19 it takes about five days for the symptoms to manifest, thus, we introduce a time delay $T$ on the removal rate and present function I with its time argument as $I(t - T)$ where necessary.

After incorporating these parameters, the adjusted SIR Model becomes

$$\frac{dS}{dt} = -r \ S \ I \ e(t) - S \ v(t) \, , \tag{2a}$$

$$\frac{dI}{dt} = r \ \text{e(t)} \ S \ I - a(t) \ I \, , \tag{2b}$$

$$\frac{dR}{dt} = a(t) \ I(t - T) + S \ v(t), \tag{2c}$$

$$N = S + I + R. \tag{2d}$$

Adding the time delay in the number of infected, as indicated in Equation (2c), will introduce inner, smaller scaled, oscillations in the process. We have observed these oscillations in all countries where we have obtained reasonably reliable data. These neatly defined high-frequency oscillations are a remarkable process characteristic which has received, so far, little attention from the scientific community. However, we do not want to mix-up this detailed aspect of the modeling with the more general model-adaptation issue about this study's subject. We may treat this oscillatory aspect of the pandemic in another study in the future.

We include the function $e(t)$ in Equations (2a) and (2b) to represent the effects of a non-constant population exposed to the virus. There are three effects in the resulting model. First, the oscillatory behavior and multiple waves are now possible in Model (2abcd). Second, the flattering of the infection curve by representing the social-distancing measures in the permissiveness function $e(t)$. Third, the non-constant character of an imaginary transmission surface separating the infected and the susceptible groups.

Vaccination works by increasing the retirement of people from the susceptible condition. Thus, in the math model, the derivative of the retired $dR/dt$ adds up the term $S \cdot v(t)$ while the same term must be subtracted from the derivative of the susceptible $dS/dt$. The function $v(t)$ is the fraction of people vaccinated over the basis of the susceptible people.

Finally, we may think of possibilities for some variations of $a(t)$, but assuming there are little variations in the geographical conditions and the intrinsic social behaviors of any country, we model the process considering the removal rate $a(t)$ as constant, thus $a(t) = a$, and the Model is expressed as:

$$\frac{dS}{dt} = -r \ S \ I \ e(t) - S \ v(t) \, , \tag{3a}$$

$$\frac{dI}{dt} = r \ \text{e(t)} \ S \ I - a \ I \, , \tag{3b}$$

$$\frac{dR}{dt} = a\,I + S\,v(t), \tag{3c}$$

$$N = S + I + R. \tag{3d}$$

A continuous assessment of the reproduction rate is useful because it signals whether or not the epidemic is decreasing at some time. By numerically integrating the number of actually susceptible people after some social distancing program and a vaccination program, we obtain an expression for the reproduction rate $Rp$ that is applicable all along the epidemic process.

$$Rp(t) = \frac{r\,S(t)}{a}\ . \tag{4}$$

### 2.3. Estimating the death rate.

The whole data registering process in the disease propagation event may be inaccurate. The disease itself has long incubation times that foster imprecisions. For highly contagious diseases, as COVID -19, the testing programs usually cannot cope with the infection process's impetus, resulting in many unacknowledged infected individuals. On the other hand, to catch and isolate every individual contaminated, the comprehensive testing programs may regard as infected, people showing some symptoms but not infected with COVID-19. Nevertheless, because of the legal implications in most countries, the death people count is expected to be accurate in number and time. The cumulative recovered individuals should also be a reliable number concerning the contaminated people detected. Unfortunately, some countries report numbers about recovered patients in a time-shifted fashion or do not report them at all. Thus, in most cases, an approximation of the death rate concerning the infected is possible by comparing the cumulative deaths with the sum of the deaths and recovered while considering the delay between the dead and infected statistics. The curve of the death rate gives a clear idea about medical progress in treating the disease. Figure 1 illustrates the death rate curve for Italy built with data from the Center for Humanitarian Data [9].

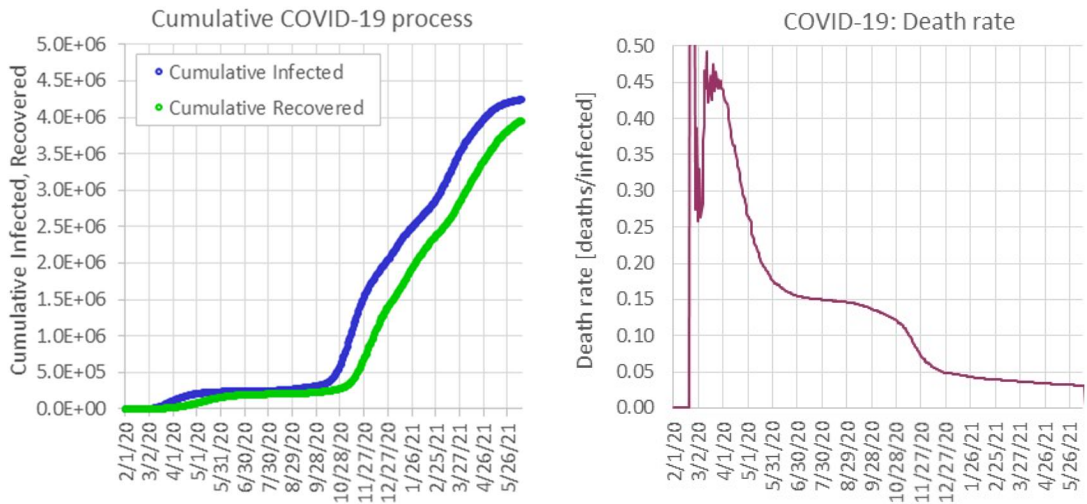

**Figure 1.** Left: Cumulative infected, cumulative recovered caused in Italy by Covid-19. Right: Death rate (computed as Deaths / (Infected + Recovered) ) caused in Italy by Covid-19, Data from [9] on the 26th of May, 2021)

Proper estimation of the death rate is of critical relevance to obtain a representative evaluation of the medical treatment effectiveness given to the patients. Figure 1 Right shows the time series of death rate relative to the infected people. These figures come from values referred to in the time series shown in Figure 1 Left. The graph in Figure 1 Right exhibits an initial stage where unstable numeric values suddenly oscillate until the process settles down to values that reflect a proper estimation of the death rate. Additional signals about the medical learning process come from comparing the time delay between the infected and the recovered time series for some countries. In Italy, for example, at the beginning of the pandemic, the recovery time was about one and a half months. The delay shortens to 25 days by October of 2020, thus suggesting relevant improvements in treating patients toward their recovery.

## 3. Simulating the COVID-19

To build and adjust the Model to replicate the case-history of several specific countries, we carefully combine the values of four parameters: the contagious rate $r$, the removal rate $a$, the actual number of infected individuals by the time the first death was registered, and the isolation effectivity time-series $e(t)$. Changing each parameter alters the shape of the simulated infected individuals in different directions. Therefore, the method requires the iterative test of parameter-value sets until we observe a satisfactory fit against the daily infected curve.

The contagious rate determines the steepness of the epidemic curve during the exponential growth epoch. The height of the curve is also affected by the contagious rate $r$. The removal rate $a$ mainly affects the slope of the descending contagious epoch. Increasing the number of individuals detected as infected by the time of the first death serves to move the whole infectious curve forward in time, thus setting the rise of the infection earlier in time. The social distancing measures are represented by setting a time series containing the social isolation. The isolation effectivity time-series must be carefully set in order to adjust the shape of the daily infection curve.

The fitting process starts from early times toward late times. Orienting the fitting in this direction makes sense not only because changes at a time should not change any behavior in the past but also because it reduces the number of parameters being simultaneously adjusted, thus easing the fitting process.

### 3.1. Simulation platform

We implemented the Model (3abcd) in a spreadsheet. This environment offers advantages in its user interface allowing to overview the vast scenario of simultaneously simulating COVID-19 for dozens of countries. On the other hand, the spreadsheet environment is a limited flexibility system environment. Encapsulating parts of the code, as routines do in classical coding languages, is difficult in spreadsheets. As the model grows in size and complexity, the spreadsheet environment becomes difficult to control, losing the advantage of its manageable user interface. Nevertheless, we could build the COVID-19 simulator in a spreadsheet environment, containing the models of several countries, each one treated as an independent system, and COVID-16 data obtained from the Center for Humanitarian Data [9]. The simulator applies Euler's method to numerically integrate the equations of Model (3abcd) contained in the cells.

A group of cells is arranged to work as data-input panels. Figure 2 shows how these cells look. A first panel serves to input the country population $N$ (which is equal to the initial susceptible value $So$), the infection rate $r$, and the removal time $1/a$. The next panel serves to try dates, values, and transition times that approximately solve Equations (3abcd) for the permissiveness function $e(t)$. A third panel allows investigating projected behaviors of Model (3abcd) into the near future by setting a parameter of a vaccination plan $v(t)$.

| Permissivenes Pattern | | | |
|---|---|---|---|
| Stage Number | Stage Start Date | Permissiveness | Transition time [days] |
| 0 | - | 1 | - |
| 1 | 28-Jan-20 | 0.7 | 3 |
| 2 | 19-Mar-20 | 0.02 | 7 |
| 3 | 25-Apr-20 | 0.3 | 5 |
| 4 | 21-May-20 | 0.15 | 6 |
| 5 | 7-Sep-20 | 0.24 | 7 |
| 6 | 18-Nov-20 | 0.04 | 21 |
| 7 | 18-Dec-20 | 0.18 | 21 |
| 8 | 27-Feb-21 | 0.25 | 21 |
| 9 | 8-Apr-21 | 0.07 | 21 |

| General parameters | | |
|---|---|---|
| Population | N | 83,992,949 |
| Infection rate | r | 3.61E-09 |
| Removal Time | 1/a [days] | 21 |
| Removal rate | a [1/days] | 0.047619 |
| Basic Reprod. Rate | Ro | 6.368 |

| Vaccination plan | | |
|---|---|---|
| | Vaccination Start | Vaccination [indiv/Day] | Vaccination [%/Day] |
| 0 | - | 0 | 0 |
| 1 | 1-Apr-21 | 180000 | 0.0021 |
| 2 | 1-May-24 | 155000 | 0.0018 |
| 3 | 1-Sep-24 | 190000 | 0.0023 |
| 4 | 1-Sep-24 | 26000 | 0.0003 |
| 5 | 1-Nov-24 | 26000 | 0.0003 |
| 6 | 1-Nov-24 | 0 | 0.0000 |
| 7 | 1-Nov-24 | 0 | 0.0000 |

**Figure 2**. Input data cells of the COVID-19 simulation of the adjusted SIR Model (3abcd). Cells with green numbers and dates work as data-input cells. Cells with data too far into the future are grey to indicate irrelevance for the simulation

### 3.2. Model Fit

The difficulty to adjust the model to fit the historical data for a country resides in the changing conditions for the transmission of COVID-19. The alternating conditions between relaxed social behavior and rigorous lock-down social measures impact the shape of the infectious curve. Thus, unlike traditional implementations of the SIR model, we investigate the equation-solutions for the non-constant contagious-permissiveness rate expressed in equations (3abcd). At this stage the problem is, therefore, to determine the oscillating permissiveness pattern that leads to the minimal model error for the registered data for each country.

The first step is to estimate the process removal rate $a$. Since the inverse of the removal rate is the average infectious period, the time phase between the infected and recovered curves provides an estimate of the value. Typically, when there is not enough registered data, the removal rate is estimated using the incubation time. Recent assessments of the COVID-19's incubation time indicate a range between 2 and 14 days [10]. For COVID-19, however, there is plenty of data available for most countries and we prefer to inspect the phases between confirmed cases and their corresponding recovery date. In Figure 1.Left, for example, the elapsed time between the infectious curve and the recovery curve is near one month, while the graphs for Germany and Austria, included in the Appendix, show an elapsed time between the two curves of about a half of a month (2 weeks). Estimating the removal rate by this graphical method provides larger values if compared to the incubation period. This may be explained by considering the incubation time method does not include many delaying events of the recovering process. When reading the graphs to select a removal time $1/a$, we select from three different values: 14, 21, and 28 days. In countries where the data about recovered is not clear enough, we use $1/a = 21$.

To set a contagious rate value $r$ we focus on the initial stage of the contagious process for each country: the first-case date for each country. By January, February, and the first weeks of March, no rigorous social measures were yet in effect. Excluding some specific Asian countries, worldwide traveling was allowed, and except for the use of masks, life went mostly normal. At that time started, however, an undeniable sense of cautiousness and a generalized spread of media messages asking for modifications of our social behavior. We use the term permissiveness to reference the overall impact of our social behavior changes and the social distancing measures later imposed. The permissiveness may vary upon time and appears as $e(t)$ in Equations (3ab). Assuming a uniform incidence of the new social behavior, we choose the value of the isolation permissiveness at 0.7 for all modeled countries' initial COVID-19 stages. Following, we adjust the infection rate $r$ to make the Model (3abcd) replicate the initial growth, before the first peak, of the infection process. This adjustment of the initial stages also requires trials on the date the first actual infection occurred. Most times, it must have occurred few days before the reported date of the first detection.

Once the infection rate $r$ is established, the next stages of the process are fitted by adjusting the remaining freedom degrees: the value of $e(t_i)$ for each stage $i$ and the stage starting date $t_i$. We adjust the isolation permissiveness transition value for successive stages by specifying the number of days it takes to go from one value to the next. The value transition is assumed linear. By introducing variations to the isolation permissiveness, the date each stage starts, and the transition number of days, the numbers of infected individuals at all stages are adjusted to obtain a reasonably good fit to the registered data. The results obtained after applying this model fit procedure to performed 38 countries of all continents are included in the Appendix. In general, no structural differences are observed in the behavior of these countries; in all cases, a repetition of infection waves caused by the oscillatory nature of the permissiveness dominates the overall shape of the infection curves. However, focusing on a smaller scale, higher frequency oscillations can be seen in every country. Studying these seven-day-period oscillations is beyond the scope of this paper. Nevertheless, Figure 3 shows this behavior along with the adjusted model of infected individuals for Iran, Mexico, and Nigeria. The blue dots represent the registered data [9], and the continuous blue line represents the model. The orange line shows the estimated permissiveness resulting from the social distancing measures in the three countries.

### 3.3. Assessment of the confinement-permissiveness-factor $e(t)$

The problems the COVID-19 pandemic has created are not limited to the health system's overwhelm and the resulting mortality increase. The economic consequences of the social distancing and lockdown measures compete with the direct effects of the disease. An endless discussion about the convenience of adopting extremely rigorous lockdown measures has begun in most countries. The dilemma between relenting a country's economy or assuming a high cost of lives while prosecuting a decent economic activity level has proven to be a compromise challenging to resolve.

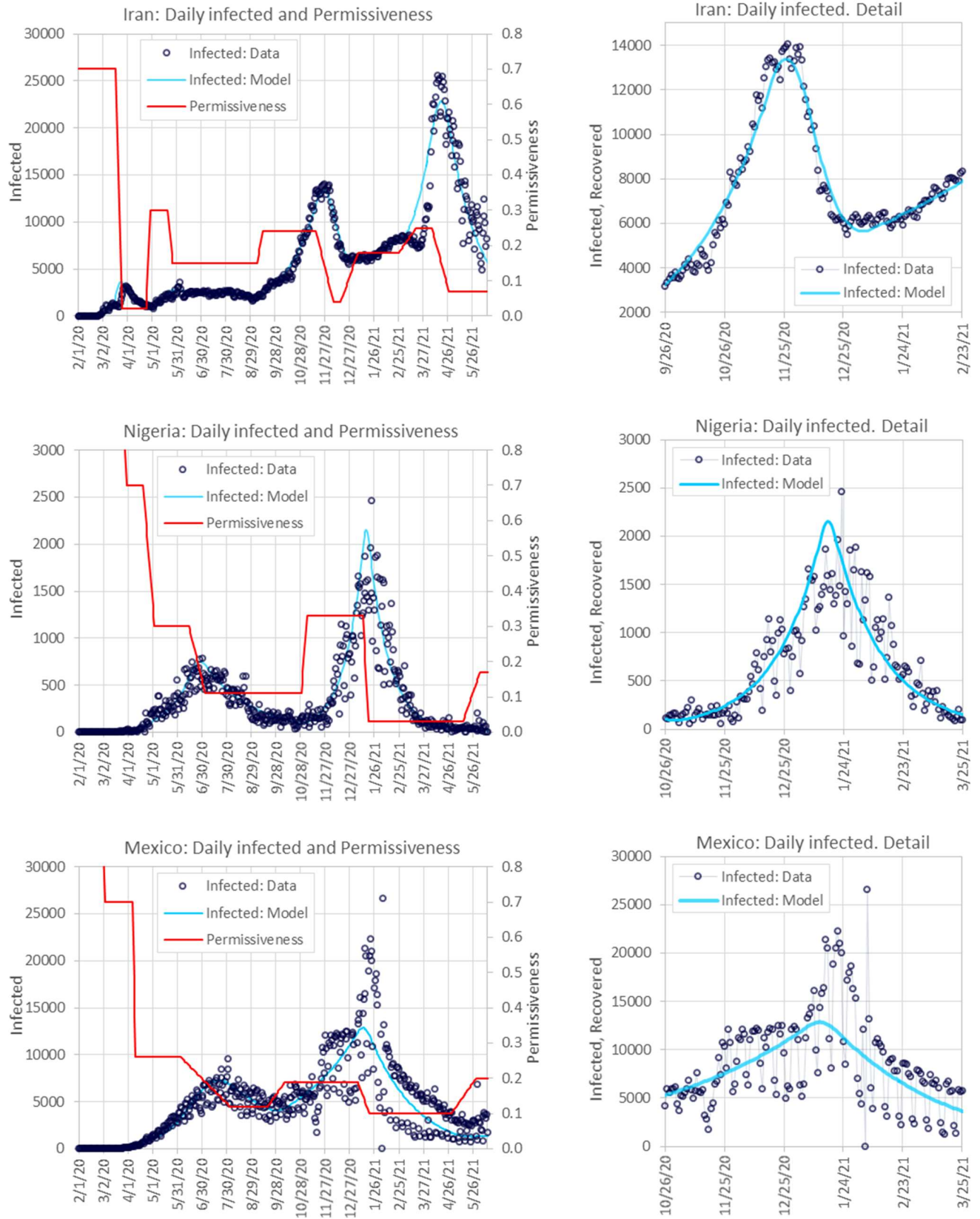

**Figure 3**. COVID-19. Daily infected, model fit, and computed permissiveness for Iran, Nigeria and Mexico during year 2020 and first months of year 2021. Graphs on the right show detailed infection lapses for the corresponding countries to illustrate different minor scale oscillation patterns. Data from the Center for Humanitarian Data [9]

The reproduction rate is the number of newly infected people by each previously infected individual. When there is only one individual infected, the reproduction rate value is also the critical value defining whether or not the disease propagates. When there are many infected individuals, this numerical connection is not so evident. However, by the time there are many individuals infected, there should be an approximation of the basic reproduction rate. Therefore, reasoning the contagiousness is reduced as the permissiveness lowers and any vaccination program implies, should ease the computing the reproduction rate for any time. Expression (4) previously presented shows this effect. Using Model (3abcd) fitted for a specific country, we can graphically see the social distancing measures' impact in reducing the contagious process. Moreover, looking at the reproduction rate, we can estimate the condition concerning the level where the number of infected tends to increase or decrease, making it a valuable tool to assess the process trend and adopt wise decisions regarding the timing and depth of social distancing measures. Figure 2 illustrates the permissiveness pattern e(t) leading to the fit of the infectious process in Iran that is shown in Figure 3. The appendix includes permissiveness patterns for all countries studied.

### 3.4. Death rate

The Appendix includes graphs showing the evolution of the death rate for most countries studied. The death rate curves are built accounting for the daily infected individuals, the daily deaths, and the reported daily recovered. When there is missing data is due to counting pauses, for example during weekends, the death rate curve is built assuming an equally distributed number of deaths within the missing-data period. For some countries, like the UK and Sweden, the data does not include the recovered people. For other countries, like Switzerland and the USA, the registering of recovered people stopped at some time. When using the model to project future COVID-19's behaviors, the death rate considers the most recent days of recovered-people reliable data. When using the model to project future COVID-19's behaviors, the death rate considers the most recent days of recovered-people reliable data.

### 3.5. The infection rate r and the country conditions

The COVID-19 infection rate's model $r(t)$ pivots on the base reproduction rate $Ro$. We obtain the $Ro$ value by testing parameters on Model (3abcd) that replicate, on an outer scale, the contagious process of 38 countries during 2020. An attempt to assess the $Ro$ value needs to account for the country's population and indexes representing the average distance between two susceptible individuals. These indexes may be a function of the population and its density. However, for a fixed population density, the inhabitants' mobility might also be a dominant parameter. This mobility mostly depends on the younger mature population $Y$, who are the most intimately connected to the country's productive activity, and the Gross National Product $GNP$. Then, a goal is to obtain an expression $Ro \approx f(D, Y, GNP)$ which links these presumably dominating parameters to $Ro$. After visualizing several sets of values in 2D graphs, we did not find a clear connection between the populations' density and the resulting value $Ro$. Perhaps the very intuitive effect of the population density and the contagiousness of COVID-19 is shadowed or distorted by other more determinant factors. Considering the density of cities is hard to determine because the geographical limits of cities is seldom stablished, we decided to model $Ro$ as a function of only two parameters: the younger mature population $Y$ and the Gross National Product $GNP$, as Expression (5) indicates.

$$Ro \approx f(Y, GNP) \ . \tag{5}$$

To input population and economic facts, we used information available in [11]. The younger mature population $Y$ is an inference that does not directly appear in [11]. To obtain $Y$, we decided the splitting age between younger and older people is 65. The age of 65 years is close to the retiring age in most countries. It is also an age that some data repository sites, as [11], refer to as a transition age with an eight percent of the world population. Thus the world's percentage of younger mature people is approximately 92%. The value $Y_i$ for each country is adjusted by considering a linear displacement of the percentage of people in the country and accounting for the difference between the country's median age and the overall world's median age. To combine the two arguments in Equation (5) into a single index, we introduce the Population Activity Index $PAI$ that is determined as follows:

$$PAI = \frac{Y \cdot \text{GNP}}{1000} \ ,$$

where $Y$ is the younger mature population in millions and $GNP$ is the Gross National Product per capita. Table 1 shows the data relative to the population, median age, and $GNP$ for the counties studied. Following the objective of comparing the reproduction rates for several countries, Table 1 also includes the base reproduction rate $Ro$, and the computed Population Activity Index $PAI$.

**Table 1**. Model fit parameters for the assessed countries

| Country | Population [millions] | Median Age [years] | Percentage of people under 65 years | Population under 65 years [millions] | 2018 GNP / Capita [$/inhabitant] | Base Reproduction rate $Ro$ | $PAI$ |
|---|---|---|---|---|---|---|---|
| Austria | 9.01 | 43 | 89.1 | 8.0 | 51525 | 6.431 | 414 |
| France | 65.27 | 42 | 89.3 | 58.3 | 41631 | 6.895 | 2428 |
| Germany | 83.8 | 46 | 88.4 | 74.1 | 47639 | 7.724 | 3529 |
| Greece | 10.4 | 46 | 88.4 | 9.2 | 20324 | 5.122 | 187 |
| Italy | 60.5 | 47 | 88.2 | 53.3 | 34520 | 7.732 | 1841 |
| Norway | 5.4 | 40 | 89.8 | 4.9 | 81734 | 5.749 | 398 |
| Russia | 145.9 | 40 | 89.8 | 131.1 | 11371 | 8.244 | 1490 |
| Spain | 46.8 | 45 | 88.7 | 41.4 | 30338 | 7.020 | 1257 |
| Sweden | 10.1 | 41 | 89.6 | 9.0 | 54589 | 6.363 | 494 |
| Switzerland | 8.7 | 43 | 89.1 | 7.7 | 82818 | 6.670 | 639 |
| UK | 66.7 | 40 | 89.8 | 61.0 | 43043 | 7.278 | 2624 |
| Argentina | 45.2 | 32 | 91.7 | 41.4 | 9912 | 4.091 | 411 |
| Brazil | 212.6 | 33 | 91.4 | 194.4 | 9001 | 7.633 | 1750 |
| Canada | 37.7 | 41 | 89.6 | 33.8 | 46313 | 7.133 | 1566 |
| Chile | 19.1 | 35 | 91.0 | 17.4 | 15925 | 4.135 | 277 |
| Colombia | 50.9 | 31 | 91.9 | 46.8 | 6719 | 4.199 | 314 |
| Ecuador | 17.6 | 28 | 92.6 | 16.3 | 6296 | 4.742 | 103 |
| Guatemala | 17.9 | 23 | 93.8 | 16.8 | 4473 | 4.966 | 75 |
| Mexico | 128.9 | 29 | 92.4 | 119.1 | 9673 | 7.334 | 1152 |
| Panama | 4.3 | 44 | 88.9 | 4.0 | 15593 | 4.685 | 62 |
| Paraguay | 7.1 | 26 | 93.1 | 6.6 | 5806 | 4.119 | 39 |
| Peru | 33.0 | 31 | 91.9 | 30.3 | 6941 | 5.055 | 210 |
| USA | 331.0 | 38 | 90.3 | 298.8 | 62997 | 8.758 | 18825 |
| India | 1380.0 | 28 | 92.6 | 1,278.0 | 2006 | 7.854 | 2564 |
| Iran | 84.0 | 32 | 91.7 | 77.0 | 5520 | 6.368 | 425 |
| Japan | 126.5 | 48 | 88.0 | 111.2 | 39159 | 7.782 | 4356 |
| Malaysia | 32.4 | 41 | 89.6 | 33.8 | 11373 | 5.784 | 385 |
| S. Korea | 51.3 | 44 | 88.9 | 45.6 | 33340 | 7.364 | 1519 |
| Taiwan | 32.4 | 40 | 89.3 | 62.7 | 7295 | 5.878 | 457 |
| Thailand | 69.8 | 42 | 89.8 | 21.3 | 25893 | 7.137 | 551 |
| Australia | 25.5 | 38 | 90.3 | 23.0 | 57396 | 5.641 | 1321 |
| New Zealand | 4.8 | 38 | 90.3 | 4.4 | 41793 | 6.056 | 182 |
| PNGuinea | 9.9 | 22 | 94.5 | 9.4 | 2720 | 4.118 | 26 |
| Algeria | 43.9 | 29 | 92.4 | 40.5 | 4115 | 4.404 | 167 |
| Egypt | 102.3 | 25 | 93.3 | 95.5 | 2549 | 5.136 | 243 |
| Kenya | 53.8 | 20 | 94.4 | 50.8 | 1708 | 4.302 | 87 |
| Nigeria | 206.1 | 18 | 94.9 | 195.7 | 2033 | 5.844 | 398 |
| S. Africa | 59.3 | 28 | 92.6 | 54.9 | 6374 | 5.542 | 350 |

## 4. Results

### 4.1. Approximating values for the base reproduction rate $Ro$ and the infection rate $r$

The obtained base reproduction rate $Ro$ values are used to build an analytic expression to estimate the infection-rate $r$ for a known country and its conditions. We believe the transmission process is sensitive to the population activity level and the number of active people moving around within a particular geographical context. On the other hand, by attributing most of the economic activity to pre-retired people, we associate this sector with the disease's propagation. Thus, we link this effect by introducing the Population Activity Index $PAI$ retrieved as the product $Y \cdot GNP$. The term $Y$ represents millions of pre-retired inhabitants and the $GNP$ in thousands of dollars per capita. The dimensions of the $PAI$ is [MM inhabitants/M$ GNP x Capita]. Since the disease propagates according to the number of infected people, this process's tendency is likely to be well represented by a power function. A good fit of the values of $Ro$ resulting from the simulations, and the data included in Table 1, were obtained for the approximation proposed in Expression (6).

$$Ro \approx 2.93 \cdot PAI^{0.12} \quad , \tag{6}$$

Expression (6) is useful to estimate the disease's infectiousness in countries densely populated or economically very active. For countries with a low $PAI$, other factors dominate the value $Ro$, and as a result, within this range, the Expression (6) does not offer useful results. Figure 4 compares $Ro$ values obtained after simulating the COVID-19 infection process for several countries and the $Ro$ values obtained with Expression (6).

To complete this analysis, the infectious rate $r$ is estimated in terms of the active population $Y$, the Gross National Product $GNP$ and the removal rate $a$, by recalling Expression (5). Then,

$$r(t) = \frac{a \cdot Ro}{S} \approx \frac{a \cdot 2.93 \cdot PAI^{0.12}}{S} = \frac{a \cdot 2.93 \cdot (Y \cdot GNP)^{0.12}}{S}. \tag{7}$$

Expression (7) suggests a significant relationship between population, social activity, and virus contagiousness used in the next section dealing with vaccination plans and economic reactivation. Writing Expression (7) in terms of the pre-pandemic National Gross Product $GNP_o$, we obtain

$$r(t) \approx \frac{a \cdot 2.93 \, (Y \cdot GNP_o \cdot e(t))^{0.12}}{S}.$$

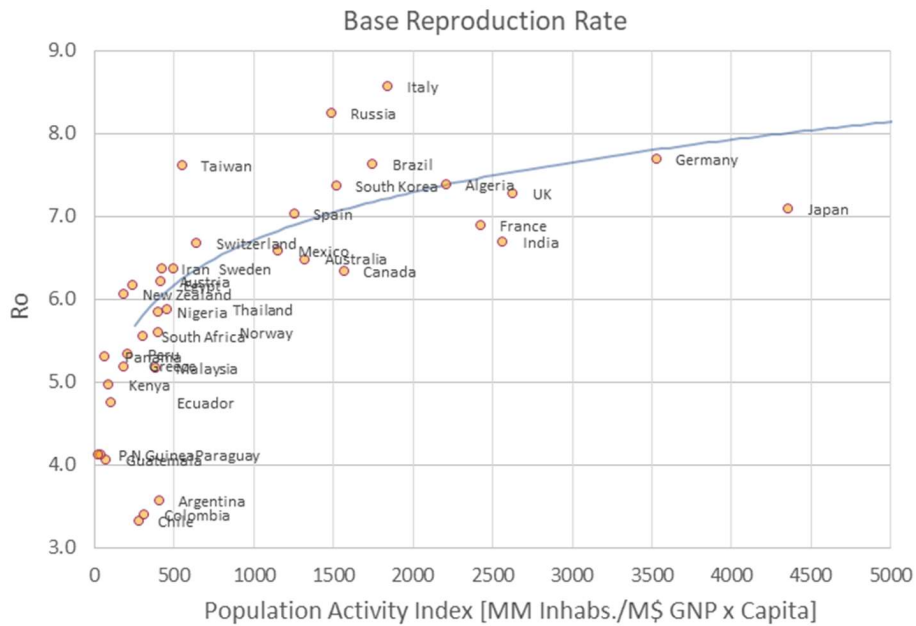

**Figure 4**. COVID-19 Base Reproduction Rate versus the Population Activity Index (PAI)

## 4.2. Death rate and medical treatment advances

Though being just two time-series, the collected data offer possibilities for making some inferences about the progress of the medical treatment provided. The death rate is the ratio between the individuals infected and the individuals who never recover from the infection. There is a distortion effect when computing this ratio because of the lag existing between the two statistics. Nevertheless, comparing the curves of cumulative infected and cumulative recovered, included in the Appendix, show a delay of less than a month, suggesting a similar length for the treatment and recovery time. Therefore, the death rate, computed after ten months of the pandemic, is a valid index, especially after the numerical instability that is typical of the days following the infection's take off in each country.

Registers about the infected individuals are generally available daily from all countries. The register of recovered patients, however, is not available or not reliable for some countries. When data about infected or recovered individuals are not reliable, we tag the death rates as NA (Not Available).

Table 2 presents the initial death rate and the death rate registered in December 2020 for the countries studied. Countries are grouped by continent and the time sequence when the pandemic started in every region. The groups are Asia and Europe, Oceania, America, and Africa. Table 2 includes a quantification of the improvement of the death rate. The percentage of the initial death rate that would be become a "recovered rate" by the end of the period studied expresses the improvement. The Appendix includes graphs showing the death rate's evolution for each studied country. When there is no reliable data, the graph's line is colored grey.

**Table 2**. COVID-19: death rate change for the assessed countries. Grouped by continent and ordered by death rate at the end of the studied period

| Death rate $[\frac{Deaths}{Infected}]$ | | | | | | | |
|---|---|---|---|---|---|---|---|
| Country | Start [%] | Dec. 2020 [%] | Improve [%] | Country | Start [%] | Dec. 2020 [%] | Improve [%] |
| Malaysia | 1.3 | 0.6 | 52.2 | Papua N. G. | 5.4 | 1.2 | 78.4 |
| Taiwan | 2.9 | 1.2 | 57.6 | N. Zealand | 1.2 | 1.3 | -3.3 |
| Thailand | 2.3 | 1.5 | 32.7 | Australia | 1.5 | 3.4 | -122.8 |
| Japan | 19.2 | 1.7 | 91.4 | | | | |
| South Korea | 2.8 | 1.8 | 34.9 | Panama | 23.6 | 2.1 | 91.2 |
| Iran | 8.1 | 6.7 | 17.2 | Chile | 2.5 | 2.8 | -12.3 |
| India | NA | NA | NA | Colombia | 16.8 | 2.9 | 82.5 |
| | | | | Paraguay | 8.0 | 2.9 | 63.5 |
| Austria | 4.2 | 1.4 | 66.0 | Argentina | 14.8 | 3.0 | 79.8 |
| Norway | 3.0 | 1.8 | 38.3 | Brazil | 14.4 | 3.0 | 79.5 |
| Switzerland | 7.1 | 1.9 | 72.9 | Guatemala | 18.2 | 3.6 | 80.1 |
| Russia | 10.5 | 2.2 | 79.2 | Canada | 16.1 | 3.8 | 76.3 |
| Germany | 5.0 | 2.2 | 56.7 | Peru | 9.2 | 3.9 | 57.8 |
| Italy | 33.0 | 6.7 | 79.7 | USA | 29.3 | 4.9 | 83.2 |
| France | 35.47 | NA | NA | Ecuador | 35.7 | 7.4 | 79.4 |
| Spain | 20.56 | NA | NA | Mexico | 13.7 | 11.4 | 17.1 |
| Sweden | NA | NA | NA | | | | |
| UK | NA | NA | NA | Nigeria | 8.74 | 1.82 | 79.2 |
| | | | | Kenya | 12.52 | 2.58 | 79.4 |
| | | | | S. Africa | 3.92 | 2.87 | 26.7 |
| | | | | Algeria | 10.08 | 4.30 | 57.3 |
| | | | | Egypt | 13.49 | 6.09 | 54.9 |

*4.3. Combining social distancing measures and vaccination plans*

Assessing the relationship among reproduction rate, economic activity, and social distancing programs is the main objective of many studies. This pandemic's aspect is fundamental to the decision making toward the unpostponable economy reactivation and recovery process. Some authors announce there will hardly be a full recovery of the economic output in some countries. Developed countries seem to be the most affected in this regard. Pichler et al. explore this relationship in detail for the UK. According to Pichler et al. [12], the GNP reduction due to the UK's lockdown was 26%. Pichler et al. also estimated the recovery of the economy would produce a recovery of 84% of the GNP lost during the pandemic lockdown. As a counterpart effect, the social distancing measures needed to bind the speed of the infection's spread, slow down the economic activity. In the present study, the results indicate that in most countries, a confinement permissiveness as low as 5% and even lower were set to break the infection's growth at some times. These values are somewhat coincident with the results of a study by Jarvis et al. [13], which computed a reduction of 74% in the reproduction rate by implementing social distancing measures in the UK.

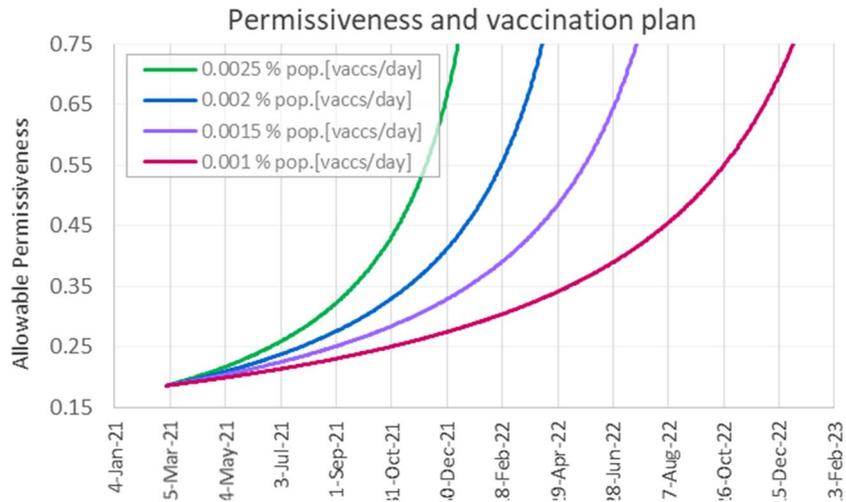

**Figure 4**. Maximum permissiveness allowed for limited daily infection in Mexico. All curves represent the permissiveness allowed for the continuous and constant vaccination rates specified in the Graph's legend

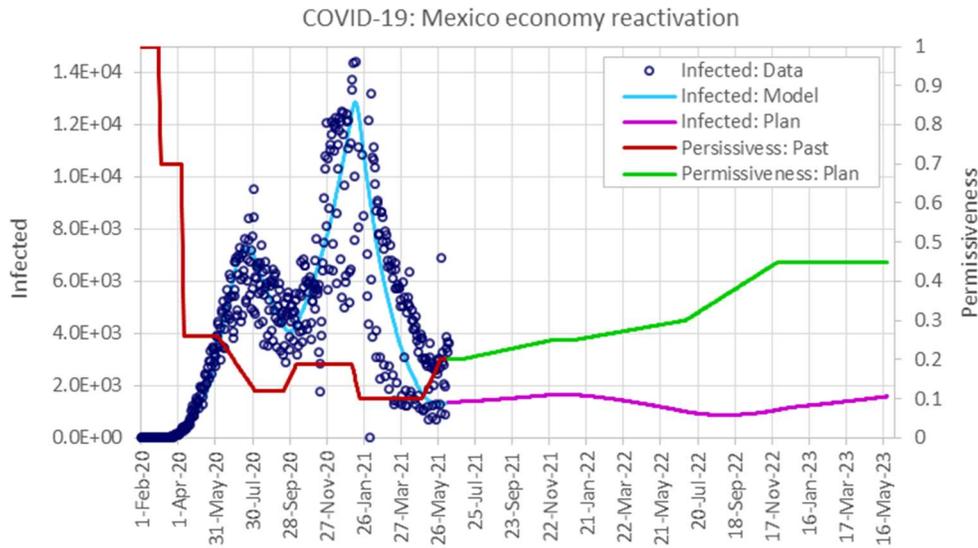

| **Permissiveness Pattern** | | | |
|---|---|---|---|
| Stage Number | Stage Start Date | Permissiveness | Transition time [days] |
| 0 | | 1 | |
| 1 | 1-Mar-20 | 0.7 | 4 |
| 2 | 7-Apr-20 | 0.26 | 4 |
| 3 | 5-Jun-20 | 0.12 | 60 |
| 4 | 19-Sep-20 | 0.19 | 21 |
| 5 | 7-Jan-21 | 0.1 | 14 |
| 6 | 1-May-21 | 0.20 | 30 |
| 7 | 1-Jul-21 | 0.25 | 150 |
| 8 | 1-Jan-22 | 0.30 | 180 |
| 9 | 1-Jul-22 | 0.45 | 150 |

| **Vaccination Pattern** | | | |
|---|---|---|---|
| | Vaccination Start | Vaccination [Indiv/Day] | Vaccination [%/Day] | Vaccine Effectivity |
| | | 1 | 1 | |
| 1 | 1-Jan-21 | 194000 | 0.001505 | 0.95 |
| 2 | 1-Aug-21 | 194000 | 0.001505 | 0.95 |
| 3 | 1-May-22 | 194000 | 0.001505 | 0.95 |
| 4 | 1-Aug-22 | 194000 | 0.001505 | 0.95 |
| 5 | 1-Jan-23 | 0 | 0.000000 | 0.95 |
| 6 | 1-Feb-23 | 0 | 0.000000 | 0.95 |
| 7 | 1-Mar-23 | 0 | 0.000000 | 0.95 |

**Figure 5**. Mexico: Estimated infection process for fastest economic reactivation with a constant vaccination campaign of 0.0015% (194000 individuals/day) of the population vaccinated each day. Notice other vaccination speeds are deactivated by setting the starting date deep into the future

Some criteria are set to solve the compromise between economic reactivation and potential infection growth. The first criterion is to adjust any reactivation plan to be implemented to avoid any increase in the rate of daily infections. Another assumption is that the vaccination programs are continuous and at a constant rate. Then the derivative of the infection rate in Expression (3b) is zero, and thus, under these circumstances

$$e(t) = \frac{a}{r\,S} \ .$$

The susceptible is represented by subtracting the total number of vaccinated people from the population. The quotient $a/r$ is the reproduction rate at any time. Thus, we write

$$e(t) = \frac{a}{r\,(N - V(t))} = \frac{N}{R\left(N - \int_{t_s}^{t} v(t)\,dt\right)} \ ,$$

where the sub-index s refers to the integral limit indicating the time when the vaccination campaign starts. Using the relationship between permissiveness and social activity suggested by Equation (7), we can include into the expression the estimation of the economy's reactivation, thus obtaining the maximum permissiveness allowed during the process of re-establishing normality.

$$e(t) = \frac{1}{PAI^{0.12}\left(N - \int_{t_s}^{t} v(t)\,dt\right)} \ . \tag{8}$$

We use Equation (8) to graph the abacus shown in Figure 4, which includes the fastest plans for a country's economic reactivation for several vaccination campaigns that could be afforded

## 5. Discussion

Adjusting the parameters to fit the models to these registered numbers is by no means an easy task. However, fitting the curve to the daily infected individual's data was satisfactorily accomplished after some trials, and obtaining an intuitive sense of the change each parameter exerts over the shape of the curve. During this process, we noticed substantial independence of the parameters in molding the model towards the shape of the data on a 2D graph, suggesting a difficulty to fit the objective curve by changing the value of the wrong parameter. We think this remarks the adjusted model is a faithful representation of the real situation for each country. Simulating models for separated countries implies no migration between any two countries, a condition that typically separates simulations from actual processes. However, the circumstantial reduction of most international traveling during the first stages of the pandemic in 2020 enhances our approach's applicability.

### 5.1. COVID-19's Reproduction rates

The SIR model fit was adjusted to reproduce the recent history of the contagious curve in 38 countries. In addition to the basic reproduction rate, a step-wise sequence of contact permissiveness was considered until a good fit was reached for each country. To represent the generalized fear caused before the virus infected most countries, a permissiveness of 0.7, instead of 1, representing total normality, was set as the initial value. The reproduction rate values obtained with this method confirms the fact SARS CoV-2 is highly contagious. The reproduction rate values obtained with this method confirms the fact SARS CoV-2 is highly contagious. Our results occupy the upper segment when compared with other studies' findings [12,14–17]. A fact that helps explain these differences is considering the constrained permissiveness of 0.7 by the time the contagious started in every country studied. We think this assumption is better to represent reality than considering, as other studies do, total normality when the infection started in every country. The estimation here presented includes the effect of the population's cautiousness when computing the reproduction rate. Thus, reducing the permissiveness leads to correctly computing the reproduction rate values replicating the historically registered infected numbers. On the other hand, studies not considering this effect are underestimating the reproduction rate values.

*5.2. COVID-19's general infection patterns during the year 2020*

*Asian countries*

Most Far East Asian countries, South Korea, Malaysia, Taiwan, and Thailand, increased the number of infected people during February and March. Japan managed to delay the infection rise until May. At the beginning of their first wave, Taiwan and Malaysia promptly initiated effective social distancing measures and kept a low first infection peak. Afterward, they have maintained the reproduction number around the corresponding critical values. Currently, a second peak is menacing to develop. In the Middle East, the Iranian infection process started in February. So far, Iran has kept the infection number around critical values but has not lowered it down close to zero, and thus, there is no clear separation of waves. India shows a late start and a slow growth in the daily infections.

*Oceanic countries*

Oceanic countries are among the counties that exhibited the best performances in controlling the pandemic. New Zealand, Vietnam and Malaysia are regarded as the most successful countries dealing with the pandemic [18]. Papua New Guinea, which is geographically closer to the first disease's epicenter, started its first wave with considerable delay concerning other region countries. This contrast is perhaps due to the low Papua New Guinea's population activity index.

*European countries*

Except for Russia, all European countries studied show a similar pattern of infection waves consisting of a first peak, initiated during March and April, and a second larger infection peak during November and December of 2020. After the first peak, these countries drove the lowered the daily infections to levels that in the graphs look close to zero. The intense concern caused by the uncertainty of reaching a peak before the pandemic overwhelms the medical infrastructure and drives these countries to apply strict social distancing measures and lockdowns. The social distancing permissiveness is below 5% just before and after the first peak. However, this strict social distancing period did not last enough, and a second peak appeared in November.

After the first infection peak in Russia, the infection curve never was driven to a low level. Therefore, the increase of a slope toward the second peak started earlier than in other European counties.

*American countries*

A notorious epidemic upraises started in the USA by the end of March. In the USA, after the first infection peak that occurred by April, the country has never kept a low reproduction number. Therefore, in the USA, the possibility of reducing the COVID-19 epidemic was practically null. Canada encountered its first peak during May. During the following months, the epidemic was controlled by an effective social distancing plan and a lower than one reproduction number. Later, these measures were somewhat relaxed, and thus, we expect the second peak to occur during the first months of 2021. Latin American countries experienced the beginning of COVID-19 in months and reaching their corresponding first infection peak from May to October. None of the studied Latino American countries controlled the epidemic after the first peak. Therefore, there is a low expectancy to overcome the pandemic before the vaccination program works out a solution. The graphs show clear progress in the understanding of COVId-19 from the medical point of view.

*African countries*

In African countries, the first wave started two months later than the USA and Canada, and about the same time as Latin-American countries. None of the African and Latin-American countries have managed to sustain rigorous distancing measures for three months or longer. Thus, furthering the possibility of winning the battle against COVID-19 by the enforcing of social distancing programs.

A qualitative view of infection patterns at large geographical levels suggests the long distance travel effect's impact. There have been lockdowns in countries. However, we do not see any structurally different infection rate patterns when comparing countries. On the contrary, focusing on large regions and continents, differences emerge in the shape of the infection curves. The infection growth in Latin American countries is not exponential. The growth starts and increases its slope until a certain degree, and then it becomes essentially linear until the first peak appears. This infection growth type is especially noticeable in Argentina, Ecuador, Guatemala, Mexico, and Panama. Performing a similar inspection in European countries, we observe an illustrative representation of exponential growth. See

Austria, France, Greece, Italy, Norway, Spain, and the UK. All exhibit an infection curve shaped as the increasing-slope due to the exponential growth.

The similarity between the infection curve's shape of the United States and Russia is remarkable. Both countries show a sequence of peaks with modest reductions in the infection rate between any two peaks. We think this effect results from moving the infection epicenter from one large city to another in a vast country. Thus, in these cases, the country's scale and the superposition of several pandemic outbreaks do not allow us to appreciate the typical infection curve. Brazil is another country that may fit within this pattern behavior, but not long enough time has elapsed to recognize the shape of Brazil's pattern. In the United States, the hot spots shifted from the Northeast to Midwest and then to Florida, California, and Texas [19]. In the United States, the lockdown measures have considerably diminished urban social activity. However, the reduction of long-distance trips by airplanes and land transit services has been much smaller. The long distances involved in this transit signal the diffusion mechanisms' relevant presence, which is not included in the classical SIR model [20]. Thus, applying the classical SIR model for very large and populated countries as the United States and Russia may be beyond the classical SIR model's limits.

5.3 Medical treatment progress

The cumulative infected and recovered graphs included in the Appendix are not reliable for all countries. In some cases, like the UK, the data reported recovered people only for some periods. For example, in other cases, in Sweden, the number of recovered was not reported to the data source we used. Nevertheless, for most countries, we do have enough data to build the cumulative graphs for infected, recovered, and dead people with time. Therefore, for most countries, we were able to compute the death rate over time.

The current death rate has reached a steady-state for all countries here reviewed, allowing us to state the overall death rate is near 3 %. Nevertheless, at the beginning of the pandemic, the death rate was considerably higher for most countries. Arguably a learning process, which includes medical treatment and hospital logistics improvements, has contributed to this sensible death rate reduction [21,22]. Several factors explain the remarkable decrease in the death rate experienced this year: (1) the health centers' decongesting. (2) The learning curve about treating the illness. (3) An evolution in the criteria to handle and classify data. (4) Changes in our immunology system towards COVID-19.

There are several facts Table 2's numbers indicate. Firstly, the high number of infected individuals by the time of the first death was produced announces that Italy was surprised by this pandemic's contagiousness. The model simulations suggest that once the contagious widespread started, it is practically impossible to stop the infection's rapid propagation. A second remark is the high removal rate in Germany. The high removal rate, combined with the timely opportune warning, explains the relatively low impact of Germany's infection.

Qualitative analysis indicates the COVID-19 is exceptionally contagious. A radically worse situation did not occur thanks to the barrier built with the unprecedented social distancing and quarantine programs implemented. The simulation of the model evidences the sensitivity of the resulting number of deaths to the isolation effectivity. According to these models, increasing the permissiveness, even slightly, will most likely produce an upturn in the number of deaths.

*5.4. Perspectives for social and economic normalization*

The contagiousness of COVID-19 is an outstanding characteristic of the SARS-CoV-2 virus. During the year 2020, we all witnessed how continental, international, and regional isolation efforts have been only marginally useful to control the virus' spread [19]. Isolation at large scales, as the postponing or canceling of the international people interchange, delayed the first wave in some countries. However, it did not stop the virus from spreading in any country that we know. Additionally, the COVID-19's proved not to be as sensitive to seasonal temperature changes as it was once widely thought. The resulting situation leaves vaccination and social distancing measures at the scale of individuals or compact families as the only identified mechanisms we can administer to pursue the renormalization of economy and life before 2020.

The virus' spread and the economic sectors' shutdowns are "the sword and the wall" of today's world situation. The obvious fact about the effects of economic activity and virus spread imposes a compromise around which the programmed normalization of activities should rotate. The graph shown in Figure 5 is a valuable tool to discuss the possible locations for every country to adopt to restart the activities toward the normalization of social and economic life. Figure 5 represents the Mexican case. Nevertheless, to facilitate the process comparison among countries, the vaccination campaign and permissiveness of the social measures are included in their normalized version. The graphic abacus represented in Figure 5 highlights the hardness of the vaccine availability problem we are facing now. As suggested in [12], the social distancing measures only delay the impact of the COVID-19. Therefore, the vaccination campaign must cover just about the whole population. A social and economic normalization may, and we think it should, start right now, but the completion of this process before at least two years is not foreseeable. Especially for countries of limited vaccines availability.

## 6. Conclusions

An assessment of the overall learning curve regarding death fraction and medical treatment of the infection was possible by inspecting the cumulative data of infected, recovered, and deaths for each country. With the introduction of several vaccines, we may expect lower infection rates. However, the death rate and the treatment time show a stable condition where no much further advance should be expected.

This study provides strong evidence of the enormous sensibility of the newly infected to the relaxation of the social distancing measures. The contagiousness of the COVID -19 is exceptionally high. Besides keeping the social contact permissiveness at low levels and implementing an aggressive vaccination campaign, there is no practical way to contain the infection. Additionally, vaccination has its logistic barriers, and thus, we are skeptical about the possibility of vaccinating the required percentage of people shortly. Therefore, allowing for the permissiveness to increase will almost certainly produce the infections' steep growth again.

Most graphs related to COVID-16 show the effects of minor-scale changes, which may affect the resulting phenomena at major scales manifestations in a non-dissipative-space, as the time-endless wave an infectious disease creates. This behavioral aspect of COVID-19 is likely to be so clearly visible due to this virus's remarkable contagiousness. The detailed study of this specific behavior in a pandemic is justified.

We have proven a good fit for the SIR model is feasible for the enduring and recurrently varying permissiveness of the COVID-19 process experienced in several countries. The generic adjusted SIR model is a useful tool to determine the sequence of social distancing measures followed by a country —or geographical connected space. An immediate advantage of having this assessment method is the possibility of comparing results among different countries and forecasting likely results for properly described future scenarios. The simulations performed offer three important conclusions regarding the impact of the permissiveness over the infections curves: (i) There is a delay of about one month between a permissiveness change and the corresponding answer in infection curves. This is in important parameter to be aware of and to avoid being over confidence about economy reactivation plans. (ii) Social distancing measures are effective only when the permissiveness drops below 0.45. (iii) When the permissiveness is in the range of 0.5 to 0.8, it does not break nor slows down the contagious disease. However, these permissiveness values have an undesirable cost in the economy while no benefits justify them. The recommendation is, therefore, to design recovery plans maintaining a low-value permissiveness until a large fraction of the population is vaccinated. The relationship between the virus contagiousness and the economic activity, so far just a statistical connection with Gross National Product, paves the way to include economic reactivation within the model and assess different paths to reactivate the economy in any country we wish to study in further detail. Yet, the present model is useful to draw guidelines to conveniently synchronize the feasible vaccination plans and the permissiveness' increase associated with the economic reactivation.

Future works may include developing tools with algorithms to automatically find the best fit of a k-stage plan of social distancing measures at a vaccination program regime. The output would be an adjusted graphic-abacus, like the one shown in Figure 5, for the desired country or system. Acknowledging the vaccines' limited production and availability, combined with the difficulties of imposing social behavior changes, having such a tool to forecast near-future scenarios would be of great value to making intelligent government decisions.

**Funding:** No external founding was received for this research.

**Conflicts of Interest:** The author declares no conflicts of interest.

## References

1. LePan N. Visualizing the History of Pandemics. Available: https://www.visualcapitalist.com/history-of-pandemics-deadliest/

2. Orient JM. A perspective on SARS-CoV-2, the most dangerous virus in history. J Am Physicians Surg. 2020;25: 38–46. Available: https://www.jpands.org/vol25no2/orient.pdf

3. Yuan J, Li M, Lv G, Lu ZK. Monitoring transmissibility and mortality of COVID-19 in Europe. Int J Infect Dis. 2020;95: 311–315. doi:10.1016/j.ijid.2020.03.050

4. KERMACK W, MCKENDRICK A. Contributions to the mathematical theory of epidemics—I. Bull Math Biol. 1991;53: 33–55. doi:10.1016/s0092-8240(05)80040-0

5. Kermack WO, Mckendrick a G. CONTRIBUTIONS TO THE MATHEMATICAL THEORY OF EPIDEMICS--II . THE PROBLEM OF ENDEMICITY * Laboratory of the Royal College of Physicians , Edinburgh , U . K . 1 . Introduction . In a previous communication ( Kermack and McKendrick , 1927 ) an attempt was mad. Bull Math Biol. 1991;53: 57–87.

6. Kermack WO, Mckendrick a G. Contributions to the mathematical theory of epidemics – III. Further studies of the problem of endemicity. Bull Math Biol. 1991;53: 89–118.

7. Murray JD. Mathematical Biology I. An Introduction. Third Edit. Marsden SS, J.E. A, Sirovich L, Wiggins S, editors. Springer-Verlag; 2002.

8. Murray JD. Mathematical Biology. II: Spatial Models and Biomedical Applications. Third Edit. Springer-Verlag; 2003. doi:10.1007/b98869

9. Center for Humanitarian Data. [cited 16 Nov 2020]. Available: https://data.humdata.org/dataset/novel-coronavirus-2019-ncov-cases

10. Elias C, Sekri A, Leblanc P, Cucherat M, Vanhems P. The incubation period of COVID-19: A meta-analysis. Int J Infect Dis. 2021;104: 708–710. doi:10.1016/j.ijid.2021.01.069

11. Macrotrends.net. Available: https://www.macrotrends.net/countries/ranking/gnp-gross-national-product

12. Pichler A, Pangallo M, Maria del Rio-Chanona R, Lafond F, Doyne Farmer J. Production networks and epidemic spreading: How to restart the UK economy? arXiv. 2020. doi:10.2139/ssrn.3606984

13. Jarvis CI, van Zandvoort K, Gimma A, Prem K, Klepac P, Rubin GJ, et al. Quantifying the impact of physical distance measures on the transmission of COVID-19 in the UK. medRxiv. 2020; 1–10. doi:10.1101/2020.03.31.20049023

14. Hilton J, Keeling MJ. Estimation of country-level basic reproductive ratios for novel coronavirus (sars-cov-2/ covid-19) using synthetic contact matrices. PLoS Comput Biol. 2020;16: 1–10. doi:10.1371/journal.pcbi.1008031

15. McCulloh I, Kiernan K, Kent T. Inferring True COVID19 Infection Rates From Deaths. Front Big Data. 2020;3: 1–10. doi:10.3389/fdata.2020.565589

16. Zhao PJ. A Social Network Model of the COVID-19 Pandemic. medRxiv. 2020; 2020.03.23.20041798. doi:10.1101/2020.03.23.20041798

17. Liu Y, Gayle AA, Wilder-Smith A, Rocklöv J. The reproductive number of COVID-19 is higher compared to SARS coronavirus. J Travel Med. 2020;27: 1–4. doi:10.1093/jtm/taaa021

18. Hoh G, Ling T, Amiera N, Leng PC, Yeo LB, Cheng CT. Factors Influencing Asia-Pacific Countries ' Success Level in Curbing COVID-19⊛: A Review Using a Social – Ecological System ( SES ) Framework. 2021; 1–29.

19. Mike Stucka, Janie Haseman, Ramon Padilla JZ and CP. How the South and Southwest became the global epicenter of the COVID-19 pandemic. Usa Today. 2020. Available: https://www.usatoday.com/in-depth/graphics/2020/07/10/maps-show-covid-19-hot-spot-surge-south/5397551002/

20. Paeng SH, Lee J. Continuous and discrete SIR-models with spatial distributions. J Math Biol. 2017;74: 1709–1727. doi:10.1007/s00285-016-1071-8

21. Horwitz LI, Jones SA, Francois F, Greco J. Trends in Covid-19 risk-adjusted mortality rates in a single health system. medRxiv. 2020. doi:10.1101/2020.08.11.20172775

22. Annila A, Kuismanen E. Natural hierarchy emerges from energy dispersal. BioSystems. 2009;95: 227–233. doi:10.1016/j.biosystems.2008.10.008

Appendix

# Using adjusted SIR models to assess the compromise between COVID-19's spread and economic reactivation

# Country

## Country. Permissiveness pattern

| | Permissivenes Pattern | | |
|---|---|---|---|
| Stage Number | Stage Start Date | Permissiveness | Transition time [days] |
| 0 | | 1 | 0 |
| 1 | 1-Mar-20 | 0.70 | 4 |
| 2 | 7-Apr-20 | 0.26 | 4 |
| 3 | 5-Jun-20 | 0.12 | 60 |
| 4 | 15-Sep-20 | 0.19 | 40 |
| 5 | 15-Jan-21 | 0.110 | 15 |
| 6 | 1-Mar-21 | 0.190 | 120 |
| 7 | 1-Jul-21 | 0.24 | 120 |
| 8 | 1-Nov-21 | 0.350 | 120 |
| 9 | 1-Mar-22 | 0.57 | 120 |

## Country. COVID-19 general parameters

| | General parameters | |
|---|---|---|
| Population | N | 128,932,753 |
| Infection rate | r | 2.43E-09 |
| Removal Time | 1/a | 21 |
| Removal rate | a | 0.04761905 |
| Basic Reprod. Rate | Ro | 6.579 |

## Country. Social distancing and infected model

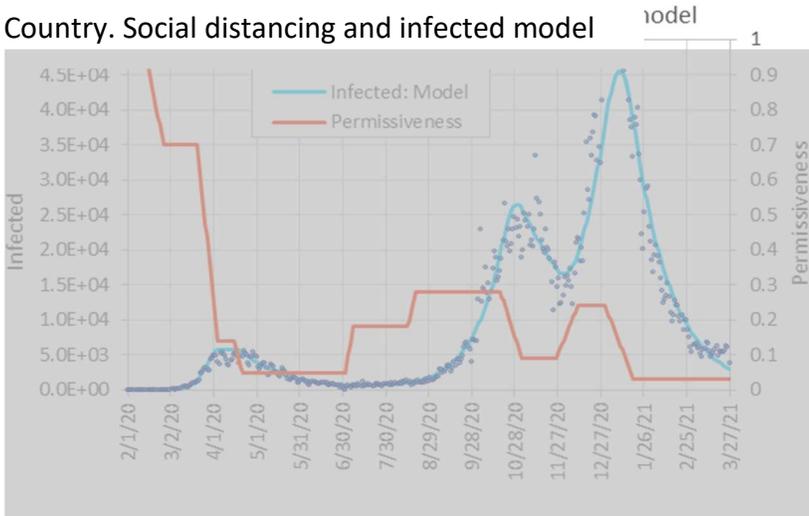

## Country. Cumulative process

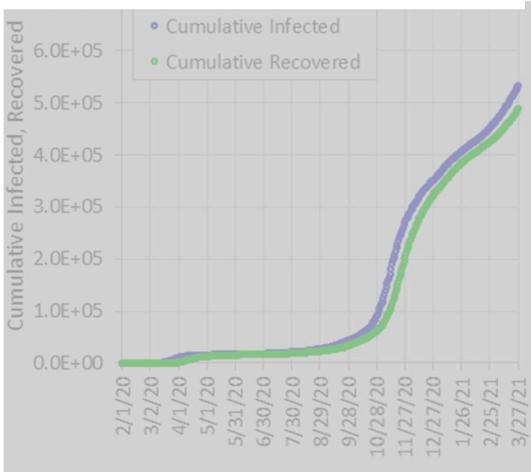

## Country. Death rate

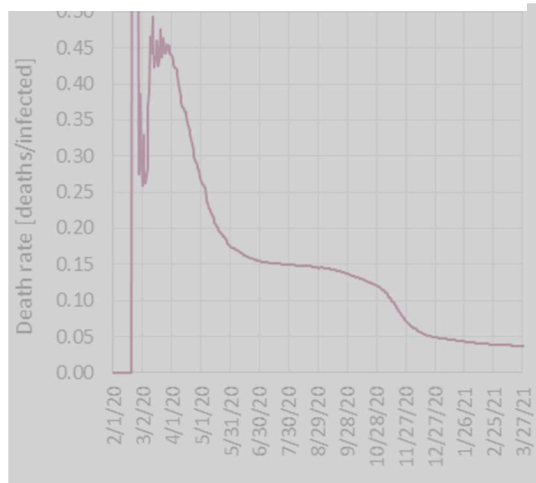

# Austria

## General parameters

| | | |
|---|---|---|
| Population | N | 9,006,398 |
| Infection rate | r | 4.93E-08 |
| Removal Time | 1/a [days] | 14 |
| Removal rate | a [1/days] | 0.071429 |
| Basic Reprod. Rate | Ro | 6.216 |

## Permissiveness Pattern

| Stage Number | Stage Start Date | Permissiveness | Transition time [days] |
|---|---|---|---|
| 0 | | 1 | |
| 1 | 22-Feb-20 | 0.7 | 7 |
| 2 | 22-Mar-20 | 0.01 | 4 |
| 3 | 5-Jul-20 | 0.34 | 4 |
| 4 | 9-Nov-20 | 0.09 | 4 |
| 5 | 5-Jan-21 | 0.2 | 21 |
| 6 | 15-Feb-21 | 0.2 | 21 |
| 7 | 25-Mar-21 | 0.1 | 7 |
| 8 | 18-Nov-22 | 0.04 | 20 |
| 9 | 21-Dec-22 | 0.16 | 7 |

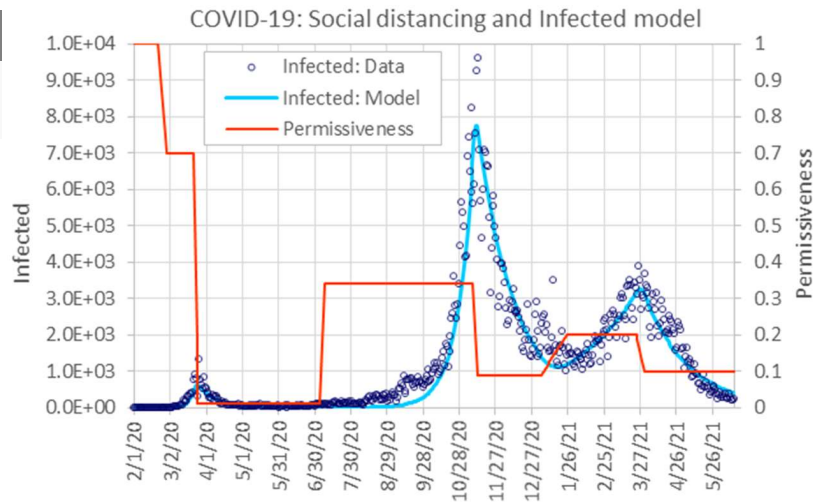

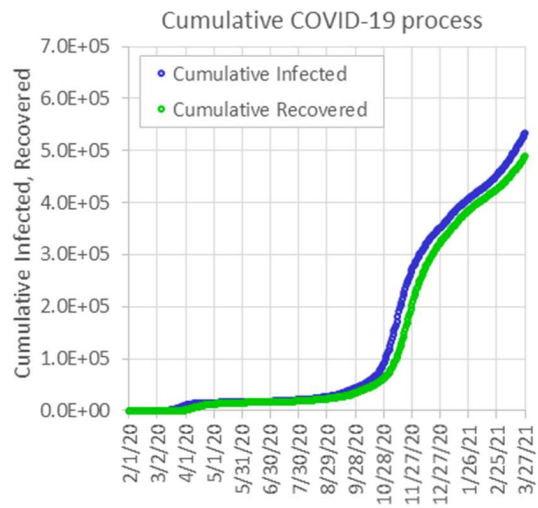

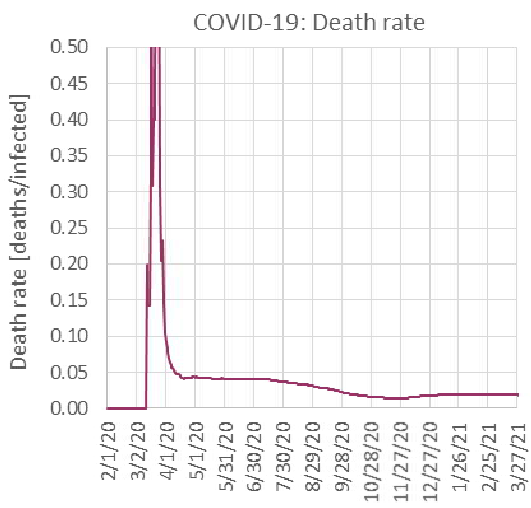

# France

## General parameters

| | | |
|---|---|---|
| Population | N | 65,273,511 |
| Infection rate | r | 5.03E-09 |
| Removal Time | 1/a [days] | 21 |
| Removal rate | a [1/days] | 0.047619 |
| Basic Reprod. Rate | Ro | 6.895 |

## Permissiveness Pattern

| Stage Number | Stage Start Date | Permissiveness | Transition time [days] |
|---|---|---|---|
| 0 | | 1 | |
| 1 | 7-Feb-20 | 0.7 | 7 |
| 2 | 20-Mar-20 | 0.02 | 15 |
| 3 | 14-Jun-20 | 0.28 | 7 |
| 4 | 22-Oct-20 | 0.05 | 14 |
| 5 | 13-Dec-20 | 0.18 | 14 |
| 6 | 2-Apr-21 | 0.05 | 14 |
| 7 | 7-Sep-22 | 0.24 | 7 |
| 8 | 18-Nov-22 | 0.04 | 20 |
| 9 | 21-Dec-22 | 0.16 | 7 |

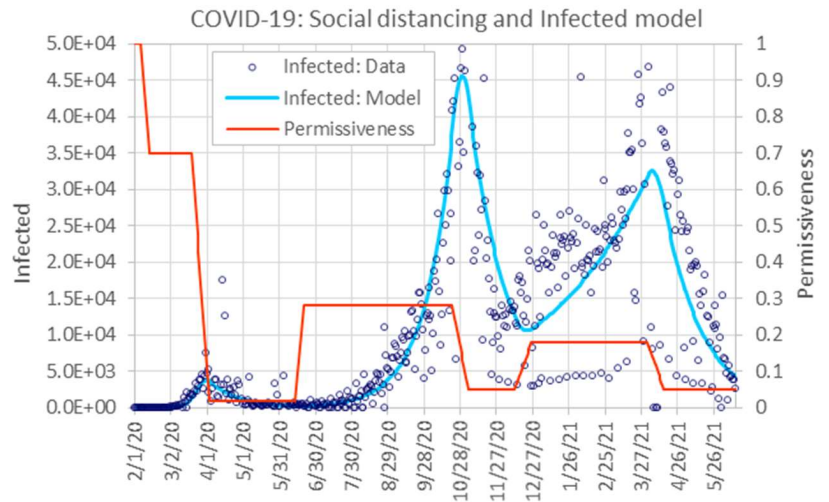

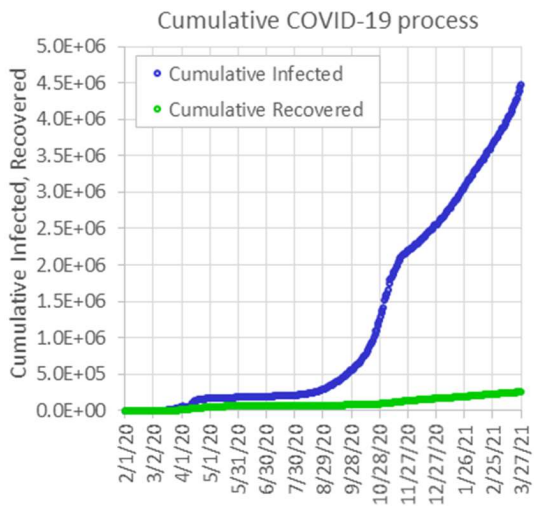

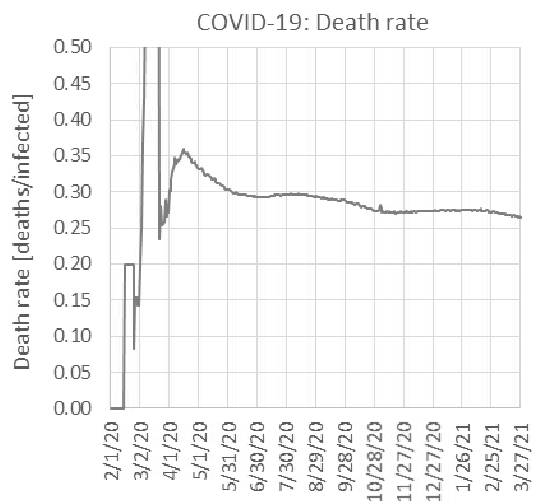

# Germany

| General parameters | | |
|---|---|---|
| Population | N | 83,783,942 |
| Infection rate | r | 6.55E-09 |
| Removal Time | 1/a [days] | 14 |
| Removal rate | a [1/days] | 0.071429 |
| Basic Reprod. Rate | Ro | 7.683 |

| Permissivenes Pattern | | | |
|---|---|---|---|
| Stage Number | Stage Start Date | Permissiveness | Transition time [days] |
| 0 | | 1 | |
| 1 | 20-Feb-20 | 0.7 | 7 |
| 2 | 19-Mar-20 | 0.04 | 7 |
| 3 | 20-Jul-20 | 0.26 | 7 |
| 4 | 31-Oct-20 | 0.15 | 7 |
| 5 | 10-Dec-20 | 0.1 | 15 |
| 6 | 17-Feb-21 | 0.17 | 21 |
| 7 | 5-Apr-21 | 0.07 | 21 |
| 8 | 18-Nov-22 | 0.04 | 20 |
| 9 | 21-Dec-22 | 0.16 | 7 |

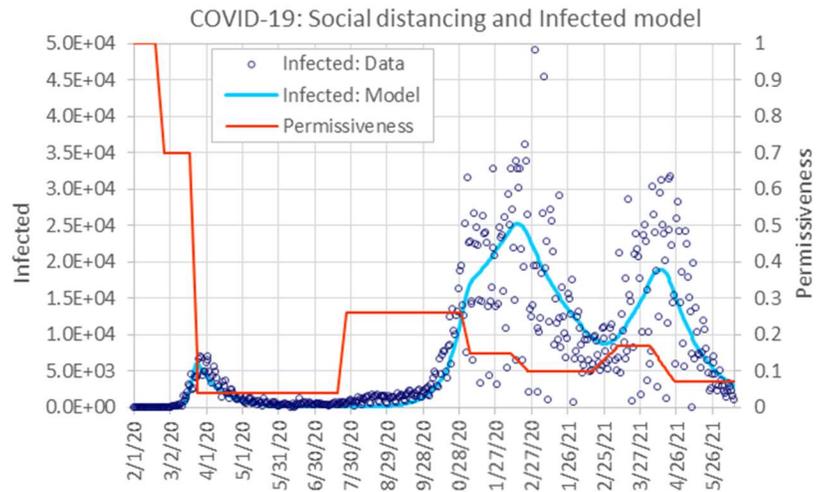

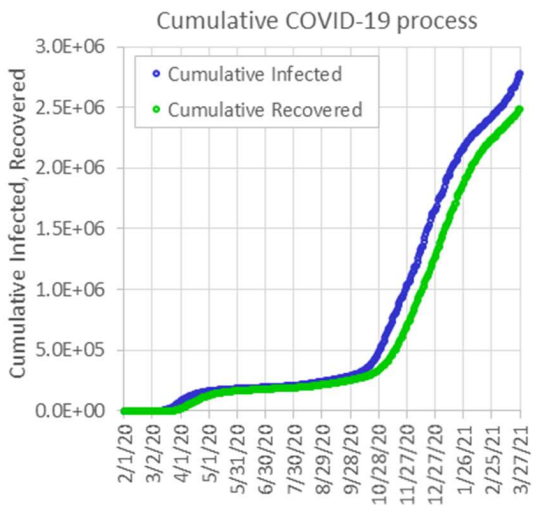

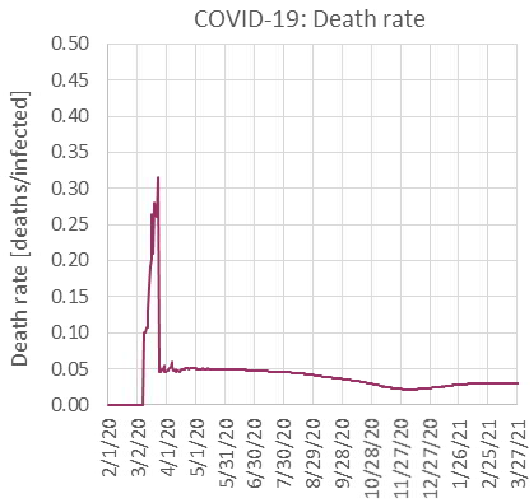

# Greece

| General parameters | | |
|---|---|---|
| Population | N | 10,423,054 |
| Infection rate | r | 2.37E-08 |
| Removal Time | 1/a [days] | 21 |
| Removal rate | a [1/days] | 0.047619 |
| Basic Reprod. Rate | Ro | 5.188 |

| Permissiveness Pattern | | | |
|---|---|---|---|
| Stage Number | Stage Start Date | Permissiveness | Transition time [days] |
| 0 | | 1 | |
| 1 | 8-Feb-20 | 0.7 | 7 |
| 2 | 20-Mar-20 | 0.07 | 4 |
| 3 | 22-Jun-20 | 0.37 | 14 |
| 4 | 12-Nov-20 | 0.06 | 4 |
| 5 | 5-Jan-21 | 0.32 | 15 |
| 6 | 15-Mar-21 | 0.15 | 21 |
| 7 | 7-Sep-22 | 0.24 | 7 |
| 8 | 18-Nov-22 | 0.04 | 20 |
| 9 | 21-Dec-22 | 0.16 | 7 |

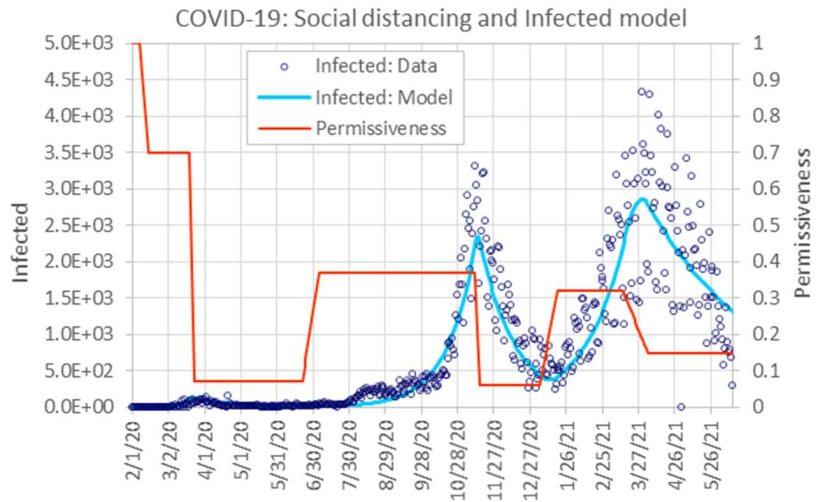

COVID-19: Social distancing and Infected model

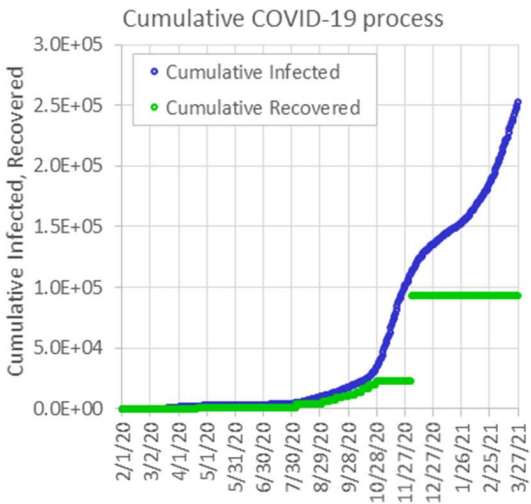

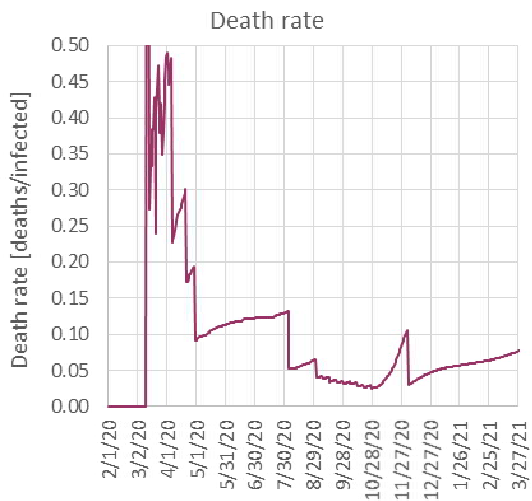

# Italy

## General parameters

| | | |
|---|---|---|
| Population | N | 60,461,826 |
| Infection rate | r | 5.06E-09 |
| Removal Time | 1/a [days] | 28 |
| Removal rate | a [1/days] | 0.035714 |
| Basic Reprod. Rate | Ro | 8.566 |

## Permissiveness Pattern

| Stage Number | Stage Start Date | Permissiveness | Transition time [days] |
|---|---|---|---|
| 0 | | 1 | |
| 1 | 25-Jan-20 | 0.7 | 7 |
| 2 | 14-Mar-20 | 0.04 | 7 |
| 3 | 1-Aug-20 | 0.3 | 10 |
| 4 | 3-Nov-20 | 0.07 | 7 |
| 5 | 20-Dec-20 | 0.1 | 15 |
| 6 | 10-Feb-21 | 0.19 | 21 |
| 7 | 15-Mar-21 | 0.04 | 14 |
| 8 | 18-Nov-22 | 0.04 | 20 |
| 9 | 21-Dec-22 | 0.16 | 7 |

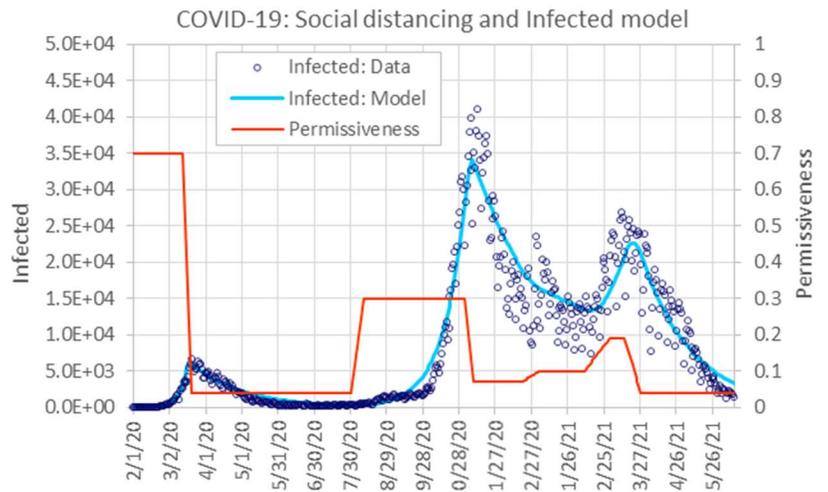

COVID-19: Social distancing and Infected model

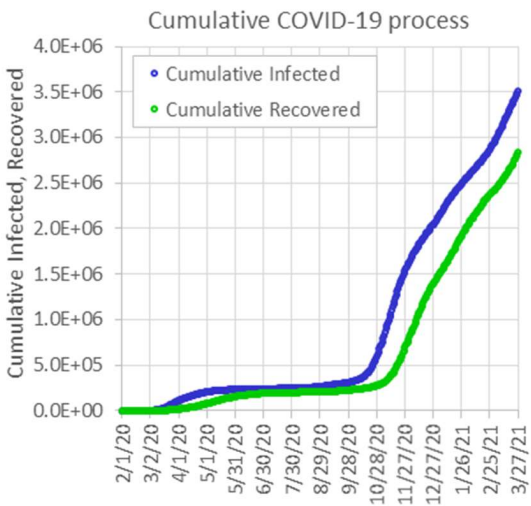

Cumulative COVID-19 process

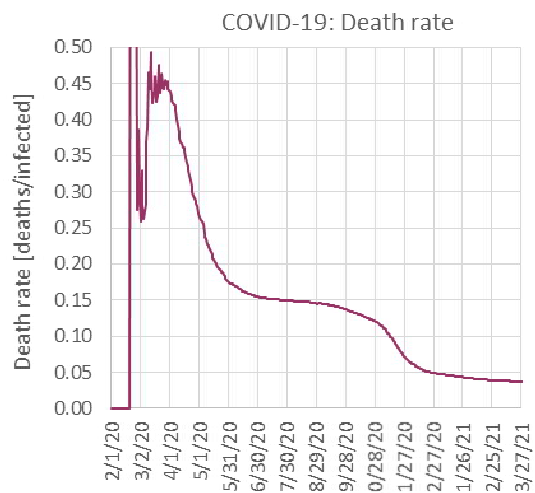

COVID-19: Death rate

# Norway

| General parameters | | |
|---|---|---|
| Population | N | 5,421,241 |
| Infection rate | r | 4.92E-08 |
| Removal Time | 1/a [days] | 21 |
| Removal rate | a [1/days] | 0.047619 |
| Basic Reprod. Rate | Ro | 5.601 |

| Permissiveness Pattern | | | |
|---|---|---|---|
| Stage Number | Stage Start Date | Permissiveness | Transition time [days] |
| 0 | | 1 | |
| 1 | 12-Feb-20 | 0.7 | 6 |
| 2 | 24-Mar-20 | 0.03 | 3 |
| 3 | 4-Jul-20 | 0.45 | 60 |
| 4 | 25-Sep-20 | 0.39 | 15 |
| 5 | 8-Nov-20 | 0.1 | 7 |
| 6 | 5-Dec-20 | 0.27 | 7 |
| 7 | 2-Jan-21 | 0.04 | 7 |
| 8 | 5-Feb-21 | 0.33 | 7 |
| 9 | 17-Mar-21 | 0.1 | 7 |

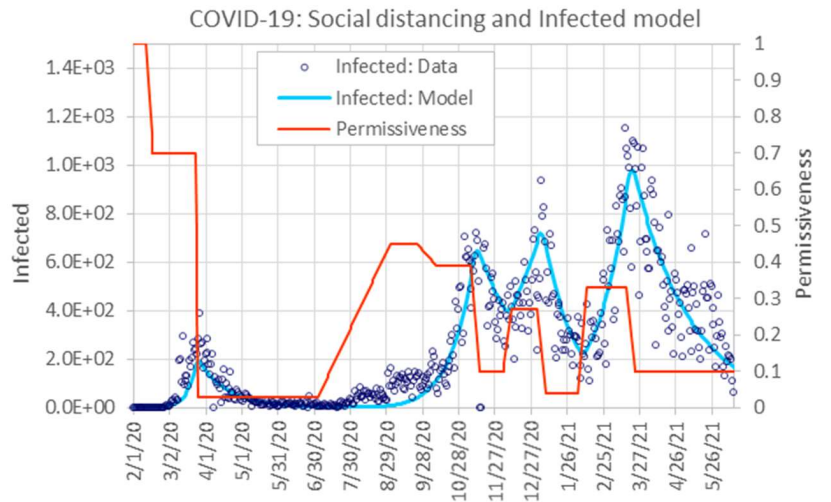

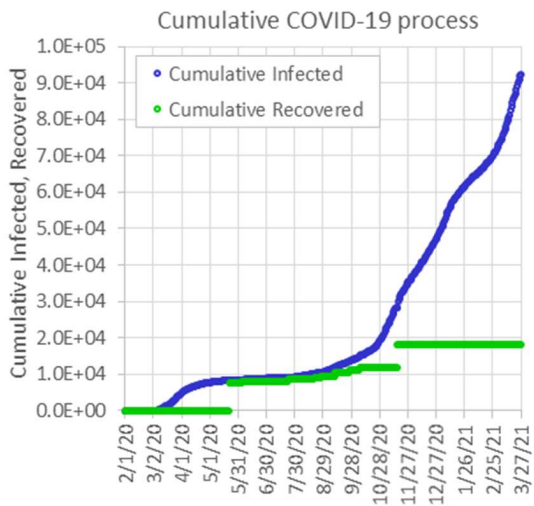

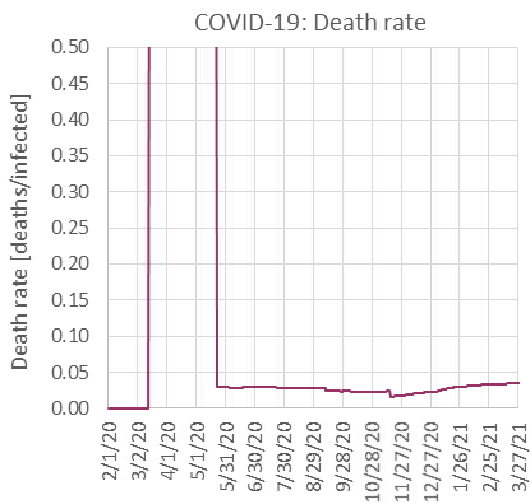

# Russia

| General parameters | | |
|---|---|---|
| Population | N | 145,934,462 |
| Infection rate | r | 2.69E-09 |
| Removal Time | 1/a [days] | 21 |
| Removal rate | a [1/days] | 0.047619 |
| Basic Reprod. Rate | Ro | 8.244 |

| Permissiveness Pattern | | | |
|---|---|---|---|
| Stage Number | Stage Start Date | Permissiveness | Transition time [days] |
| 0 | | 1 | |
| 1 | 28-Feb-20 | 0.7 | 4 |
| 2 | 18-Mar-20 | 0.45 | 14 |
| 3 | 20-Apr-20 | 0.1 | 14 |
| 4 | 20-Aug-20 | 0.2 | 30 |
| 5 | 15-Oct-20 | 0.15 | 6 |
| 6 | 1-Dec-20 | 0.13 | 7 |
| 7 | 20-Dec-20 | 0.09 | 7 |
| 8 | 5-Apr-21 | 0.15 | 30 |
| 9 | 21-Dec-22 | 0.16 | 7 |

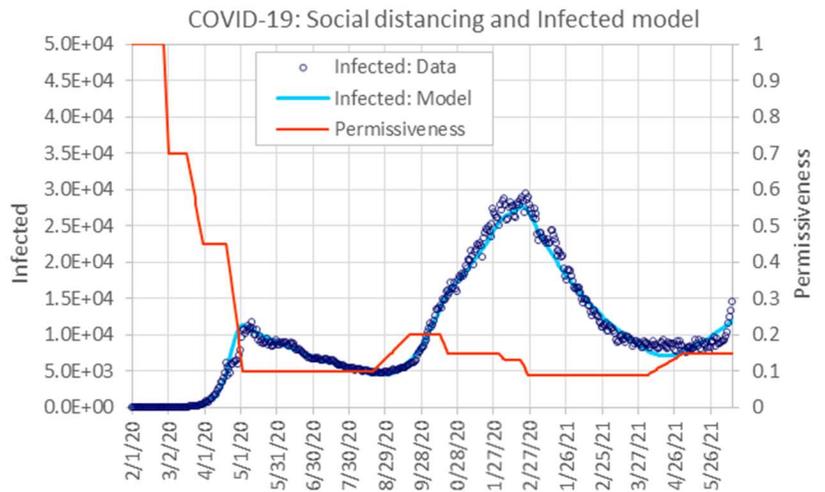

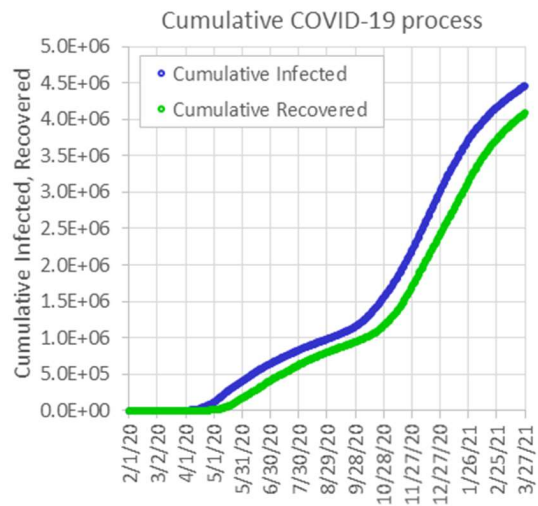

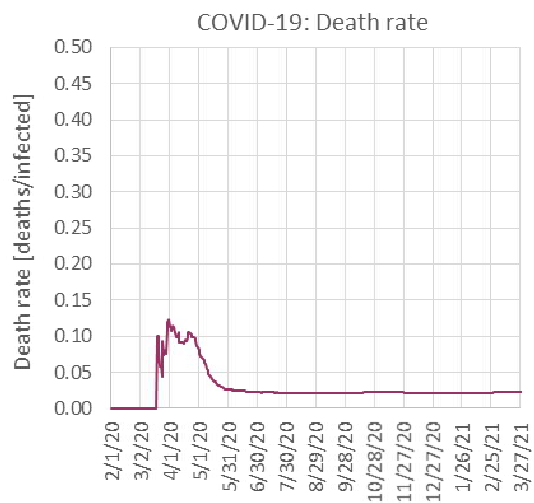

# Spain

| General parameters | | |
|---|---|---|
| Population | N | 46,754,778 |
| Infection rate | r | 7.15E-09 |
| Removal Time | 1/a [days] | 21 |
| Removal rate | a [1/days] | 0.047619 |
| Basic Reprod. Rate | Ro | 7.020 |

| Permissiveness Pattern | | | |
|---|---|---|---|
| Stage Number | Stage Start Date | Permissiveness | Transition time [days] |
| 0 | | 1 | |
| 1 | 1-Feb-20 | 0.7 | 3 |
| 2 | 22-Mar-20 | 0.01 | 7 |
| 3 | 13-Jun-20 | 0.32 | 5 |
| 4 | 4-Aug-20 | 0.2 | 6 |
| 5 | 25-Oct-20 | 0.05 | 15 |
| 6 | 15-Nov-20 | 0.23 | 30 |
| 7 | 13-Jan-21 | 0.03 | 14 |
| 8 | 10-Mar-21 | 0.25 | 14 |
| 9 | 12-Apr-21 | 0.06 | 14 |

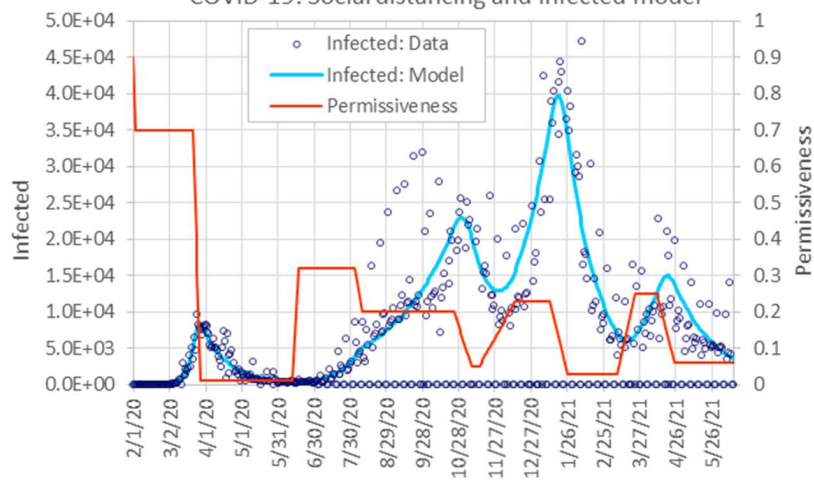

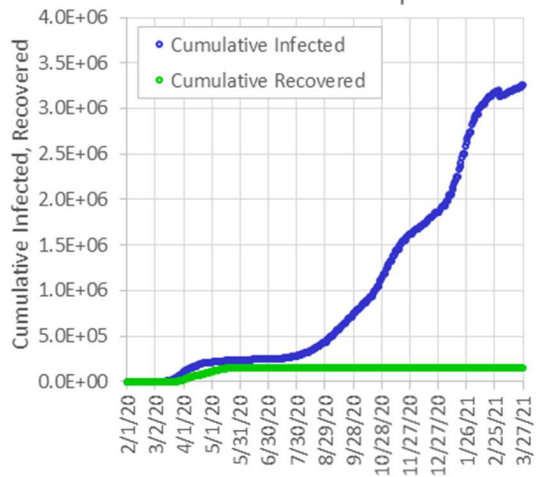

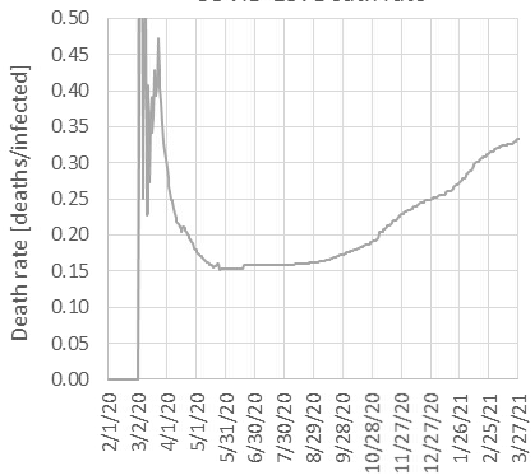

# Sweden

| General parameters | | |
|---|---|---|
| Population | N | 10,099,265 |
| Infection rate | r | 0.00000003 |
| Removal Time | 1/a [days] | 21 |
| Removal rate | a [1/days] | 0.047619 |
| Basic Reprod. Rate | Ro | 6.363 |

| Permissiveness Pattern | | | |
|---|---|---|---|
| Stage Number | Stage Start Date | Permissiveness | Transition time [days] |
| 0 | | 1 | |
| 1 | 20-Feb-20 | 0.7 | 6 |
| 2 | 31-Mar-20 | 0.17 | 2 |
| 3 | 25-May-20 | 0.23 | 5 |
| 4 | 5-Jun-20 | 0.04 | 5 |
| 5 | 1-Aug-20 | 0.3 | 20 |
| 6 | 30-Oct-20 | 0.23 | 20 |
| 7 | 14-Dec-20 | 0.1 | 20 |
| 8 | 28-Jan-21 | 0.19 | 14 |
| 9 | 15-Apr-21 | 0.1 | 14 |

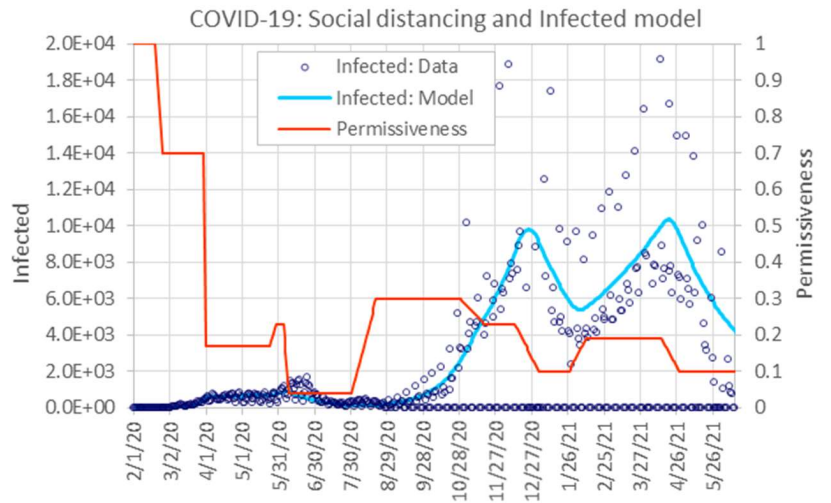

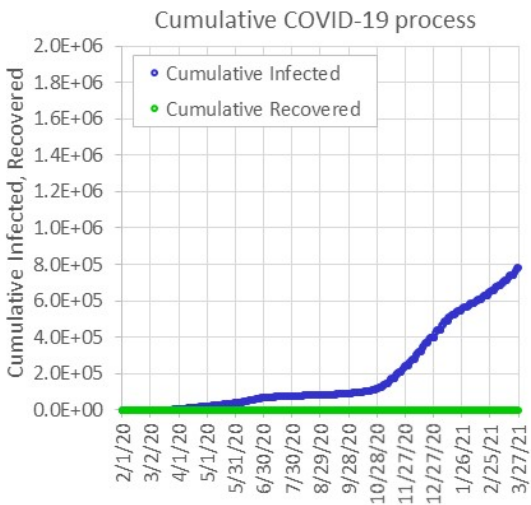

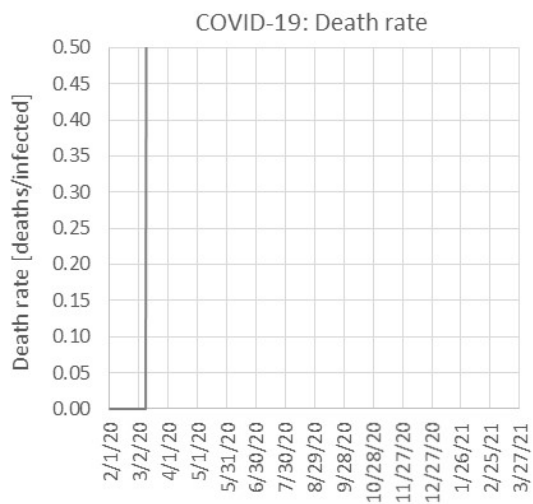

# Switzerland

## General parameters

| | | |
|---|---|---|
| Population | N | 8,654,622 |
| Infection rate | r | 3.67E-08 |
| Removal Time | 1/a [days] | 21 |
| Removal rate | a [1/days] | 0.047619 |
| Basic Reprod. Rate | Ro | 6.670 |

## Permissiveness Pattern

| Stage Number | Stage Start Date | Permissiveness | Transition time [days] |
|---|---|---|---|
| 0 | | 1 | |
| 1 | 5-Feb-20 | 0.7 | 7 |
| 2 | 19-Mar-20 | 0.02 | 4 |
| 3 | 30-Jun-20 | 0.33 | 40 |
| 4 | 15-Nov-20 | 0.1 | 21 |
| 5 | 30-Jan-21 | 0.2 | 14 |
| 6 | 15-Apr-21 | 0.07 | 14 |
| 7 | 15-Apr-22 | 0.07 | 14 |
| 8 | 18-Nov-22 | 0.04 | 20 |
| 9 | 21-Dec-22 | 0.16 | 7 |

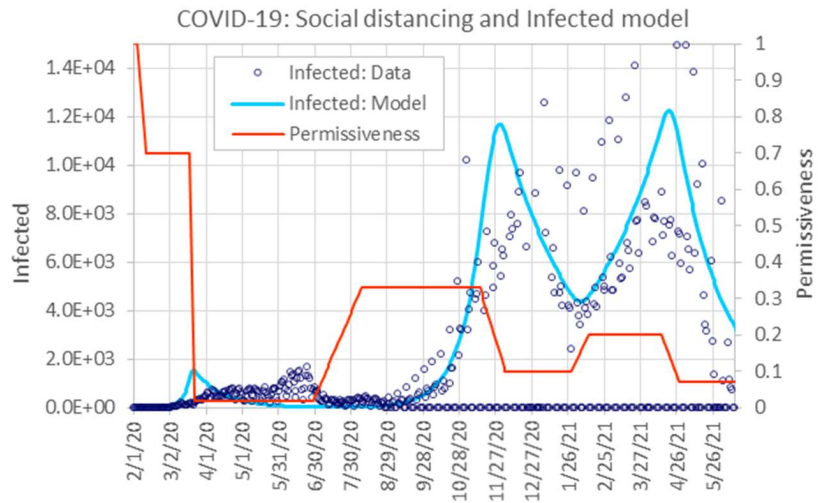

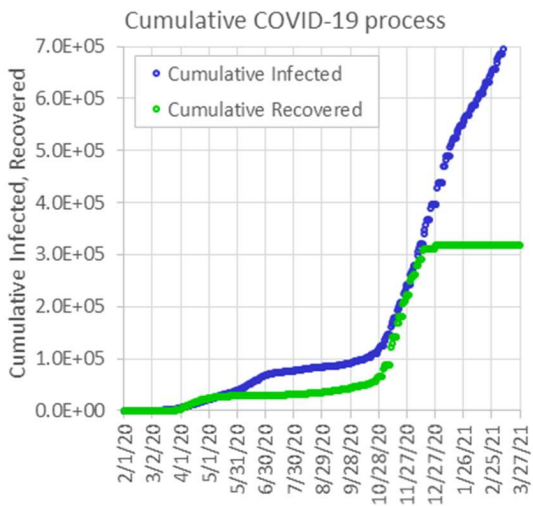

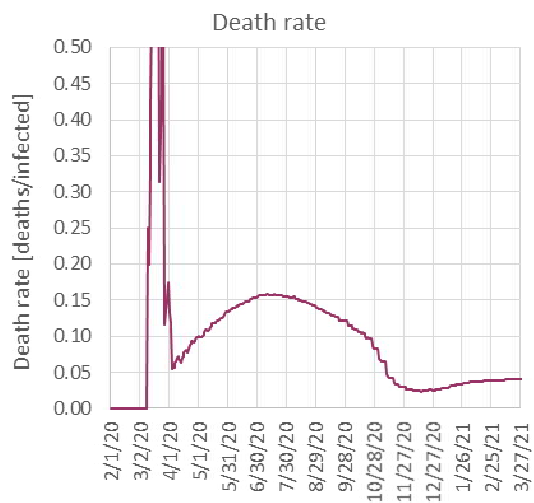

# UK

| General parameters | | |
|---|---|---|
| Population | N | 66,650,000 |
| Infection rate | r | 5.2E-09 |
| Removal Time | 1/a [days] | 21 |
| Removal rate | a [1/days] | 0.047619 |
| Basic Reprod. Rate | Ro | 7.278 |

| Permissiveness Pattern | | | |
|---|---|---|---|
| Stage Number | Stage Start Date | Permissiveness | Transition time [days] |
| 0 | | 1 | |
| 1 | 12-Feb-20 | 0.7 | 15 |
| 2 | 21-Mar-20 | 0.14 | 15 |
| 3 | 16-Apr-20 | 0.05 | 6 |
| 4 | 3-Jul-20 | 0.18 | 5 |
| 5 | 15-Aug-20 | 0.28 | 6 |
| 6 | 19-Oct-20 | 0.09 | 15 |
| 7 | 27-Nov-20 | 0.24 | 15 |
| 8 | 31-Dec-20 | 0.03 | 21 |
| 9 | 21-Apr-21 | 0.3 | 21 |

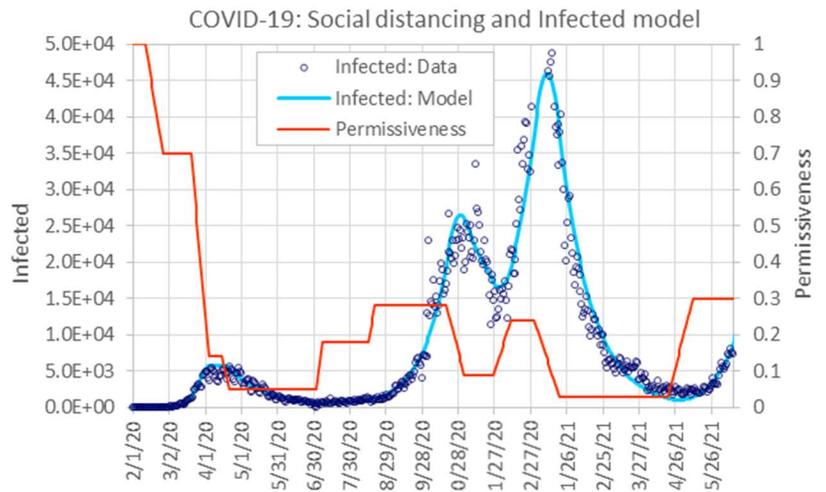

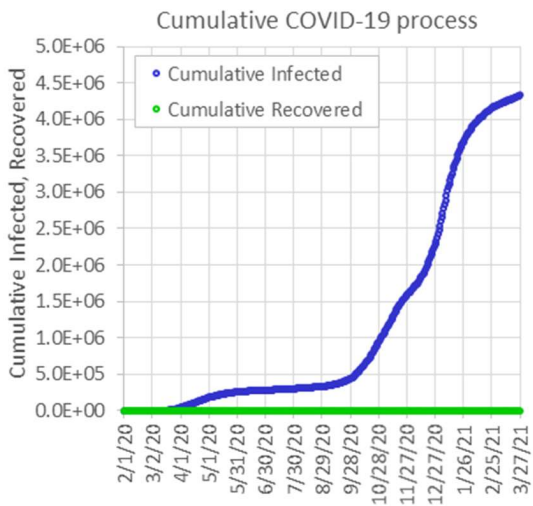

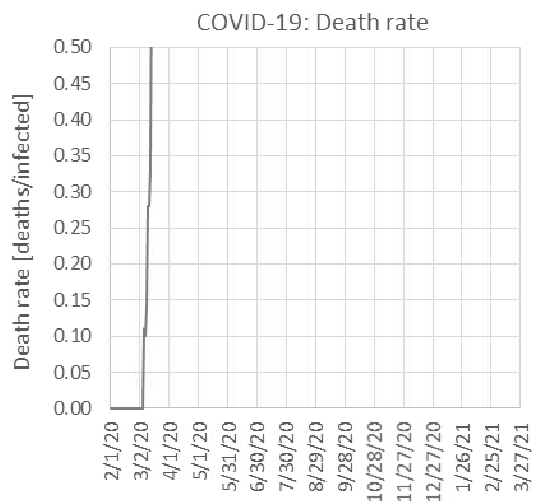

# Argentina

| General parameters | | |
|---|---|---|
| Population | N | 45,195,774 |
| Infection rate | r | 5.63E-09 |
| Removal Time | 1/a [days] | 14 |
| Removal rate | a [1/days] | 0.071429 |
| Basic Reprod. Rate | Ro | 3.562 |

| Permissiveness Pattern | | | |
|---|---|---|---|
| Stage Number | Stage Start Date | Permissiveness | Transition time [days] |
| 0 | | 1 | |
| 1 | 1-Apr-20 | 0.7 | 14 |
| 2 | 25-May-20 | 0.34 | 40 |
| 3 | 25-Sep-20 | 0.2 | 30 |
| 4 | 5-Nov-20 | 0.18 | 14 |
| 5 | 1-Dec-20 | 0.4 | 14 |
| 6 | 5-Jan-21 | 0.18 | 14 |
| 7 | 7-Feb-21 | 0.35 | 14 |
| 8 | 10-May-21 | 0.25 | 21 |
| 9 | 21-Dec-22 | 0.16 | 7 |

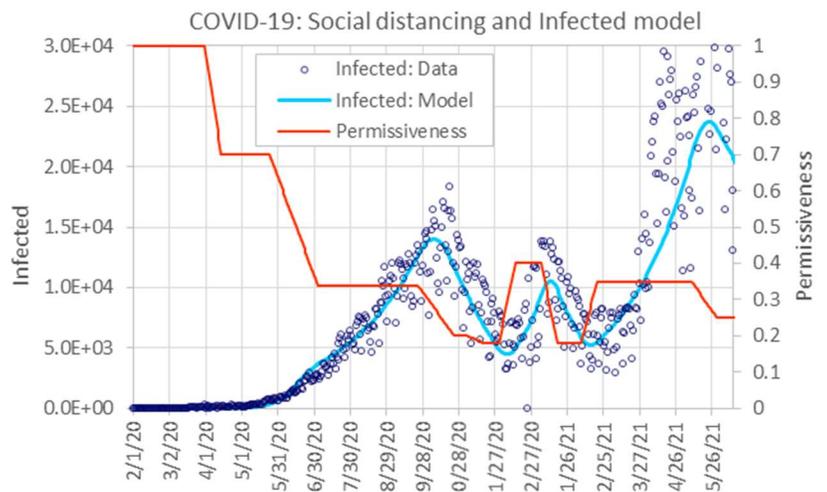

COVID-19: Social distancing and Infected model

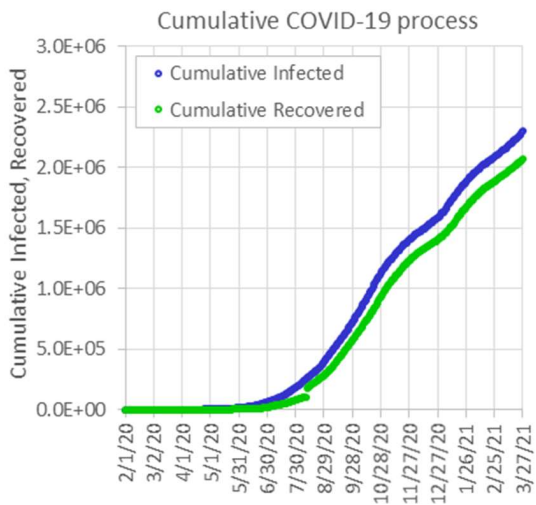

Cumulative COVID-19 process

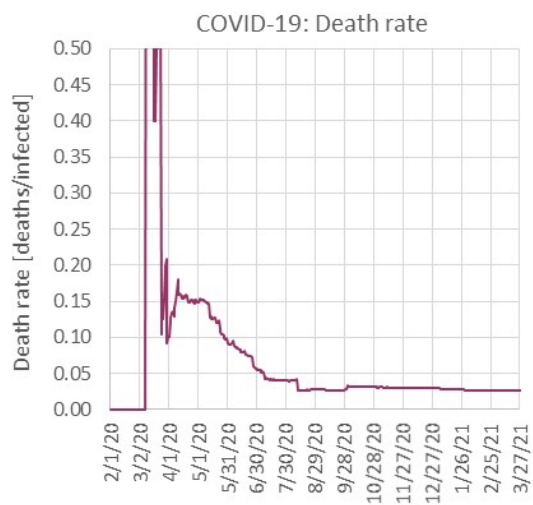

COVID-19: Death rate

# Brazil

## General parameters

| | | |
|---|---|---|
| Population | N | 212,559,417 |
| Infection rate | r | 1.71E-09 |
| Removal Time | 1/a [days] | 21 |
| Removal rate | a [1/days] | 0.047619 |
| Basic Reprod. Rate | Ro | 7.633 |

## Permissiveness Pattern

| Stage Number | Stage Start Date | Permissiveness | Transition time [days] |
|---|---|---|---|
| 0 | | 1 | |
| 1 | 7-Mar-20 | 0.7 | 5 |
| 2 | 14-Apr-20 | 0.24 | 10 |
| 3 | 30-May-20 | 0.1 | 60 |
| 4 | 15-Oct-20 | 0.25 | 60 |
| 5 | 15-Nov-20 | 0.16 | 30 |
| 6 | 5-Jan-21 | 0.1 | 21 |
| 7 | 31-Jan-21 | 0.17 | 21 |
| 8 | 15-Mar-21 | 0.1 | 21 |
| 9 | 20-Apr-21 | 0.15 | 21 |

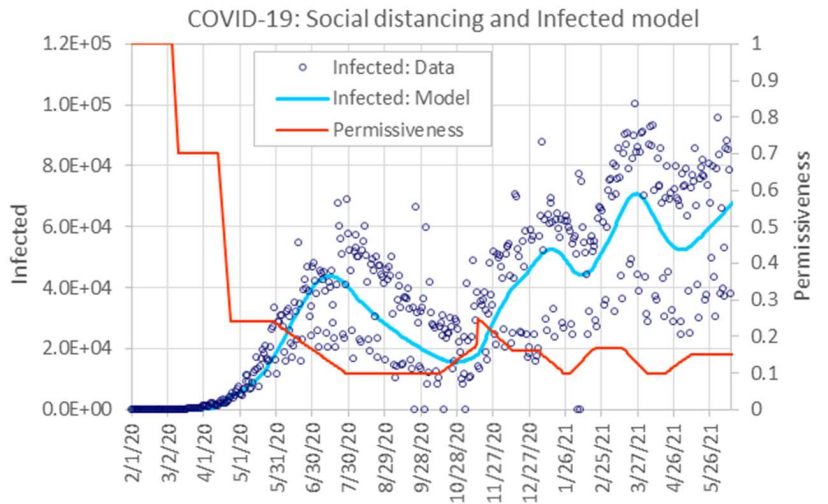

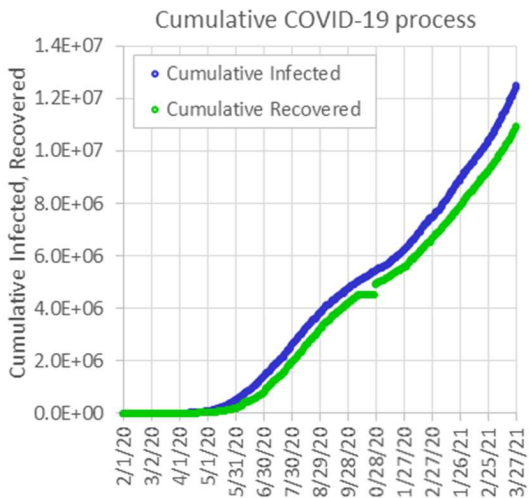

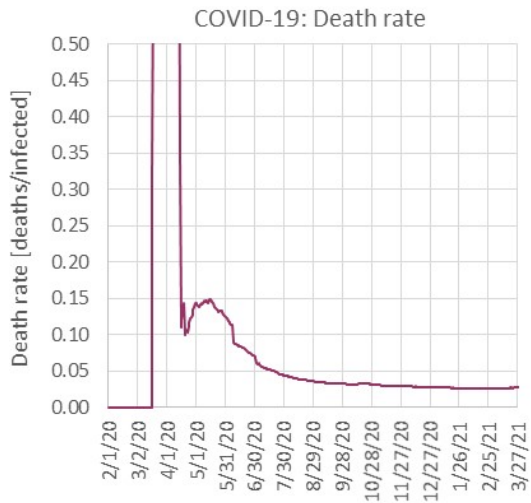

# Canada

| General parameters | | |
|---|---|---|
| Population | N | 37,742,154 |
| Infection rate | r | 1.2E-08 |
| Removal Time | 1/a [days] | 14 |
| Removal rate | a [1/days] | 0.071429 |
| Basic Reprod. Rate | Ro | 6.341 |

| Permissiveness Pattern | | | |
|---|---|---|---|
| Stage Number | Stage Start Date | Permissiveness | Transition time [days] |
| 0 | | 1 | |
| 1 | 28-Feb-20 | 0.7 | 6 |
| 2 | 31-Mar-20 | 0.15 | 2 |
| 3 | 3-May-20 | 0.06 | 5 |
| 4 | 23-Jun-20 | 0.21 | 5 |
| 5 | 6-Dec-20 | 0.1 | 45 |
| 6 | 15-Feb-21 | 0.22 | 21 |
| 7 | 13-Apr-21 | 0.09 | 14 |
| 8 | 18-Nov-22 | 0.04 | 20 |
| 9 | 21-Dec-22 | 0.16 | 7 |

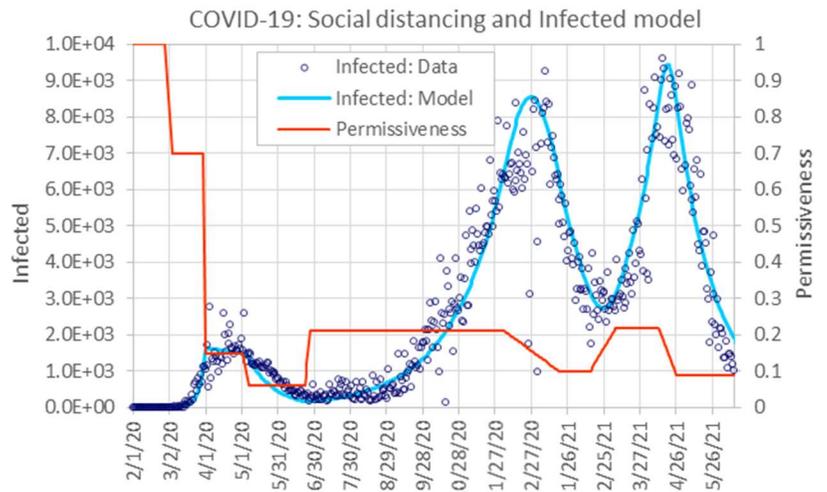

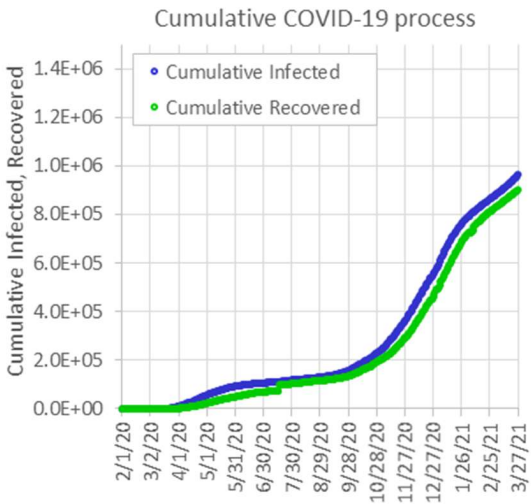

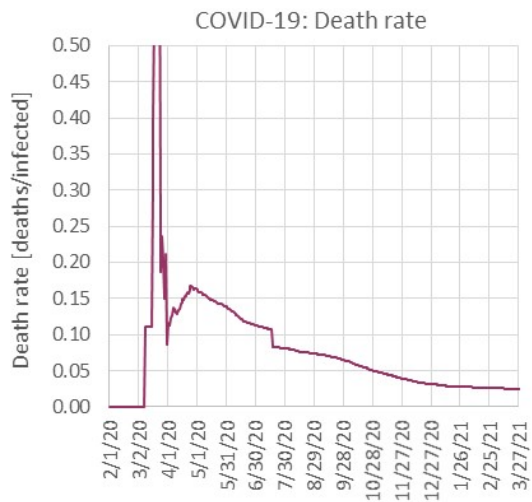

# Chile

## General parameters

| | | |
|---|---|---|
| Population | N | 19,116,201 |
| Infection rate | r | 1.24E-08 |
| Removal Time | 1/a [days] | 14 |
| Removal rate | a [1/days] | 0.071429 |
| Basic Reprod. Rate | Ro | 3.319 |

## Permissiveness Pattern

| Stage Number | Stage Start Date | Permissiveness | Transition time [days] |
|---|---|---|---|
| 0 | | 1 | |
| 1 | 10-Feb-20 | 0.7 | 5 |
| 2 | 30-Apr-20 | 0.45 | 5 |
| 3 | 5-Jun-20 | 0.2 | 7 |
| 4 | 30-Jul-20 | 0.31 | 25 |
| 5 | 22-Nov-20 | 0.39 | 25 |
| 6 | 10-Jan-21 | 0.2 | 21 |
| 7 | 8-Feb-21 | 0.39 | 7 |
| 8 | 30-Mar-21 | 0.25 | 14 |
| 9 | 10-May-21 | 0.35 | 7 |

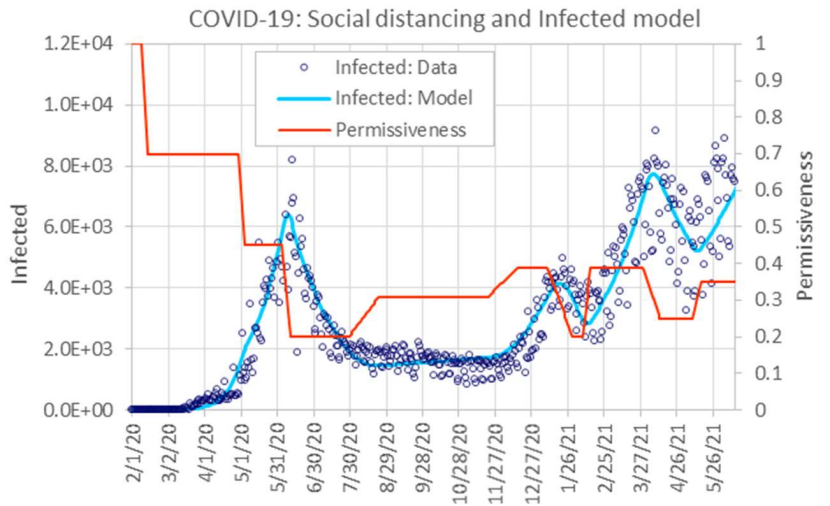

COVID-19: Social distancing and Infected model

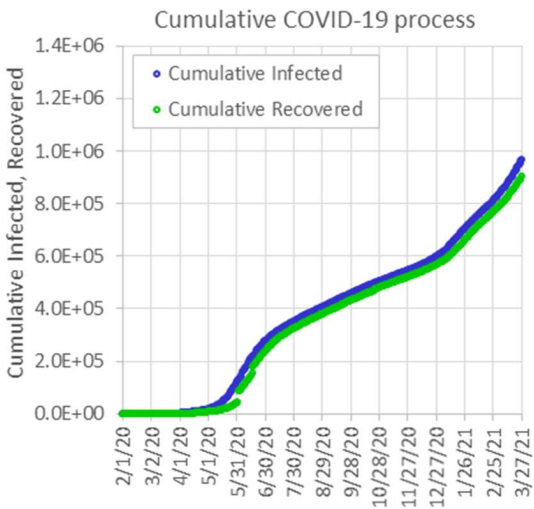

Cumulative COVID-19 process

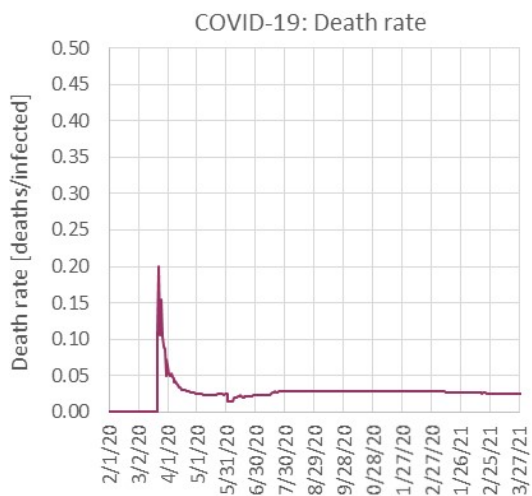

COVID-19: Death rate

# Colombia

## General parameters

| | | |
|---|---|---|
| Population | N | 50,882,891 |
| Infection rate | r | 4.77E-09 |
| Removal Time | 1/a [days] | 14 |
| Removal rate | a [1/days] | 0.071429 |
| Basic Reprod. Rate | Ro | 3.398 |

## Permissiveness Pattern

| Stage Number | Stage Start Date | Permissiveness | Transition time [days] |
|---|---|---|---|
| 0 | | 1 | |
| 1 | 15-Feb-20 | 0.7 | 5 |
| 2 | 25-Apr-20 | 0.43 | 5 |
| 3 | 5-Aug-20 | 0.19 | 16 |
| 4 | 1-Sep-20 | 0.31 | 25 |
| 5 | 1-Dec-20 | 0.4 | 15 |
| 6 | 31-Dec-20 | 0.15 | 21 |
| 7 | 25-Feb-21 | 0.4 | 7 |
| 8 | 18-Nov-22 | 0.04 | 20 |
| 9 | 21-Dec-22 | 0.16 | 7 |

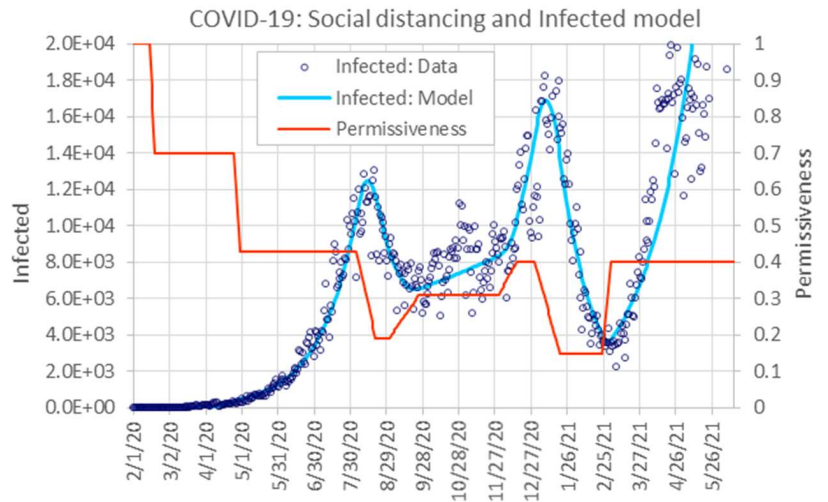

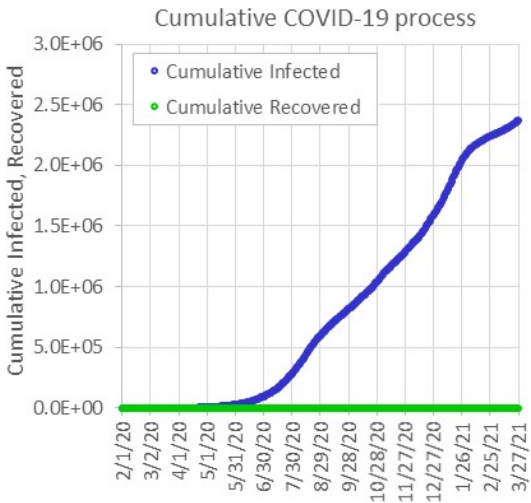

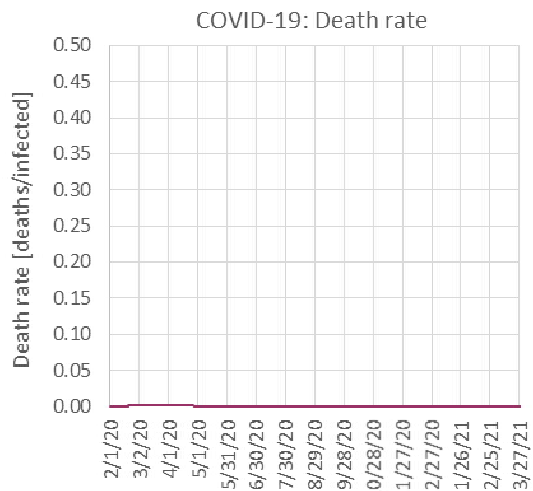

# Ecuador

| General parameters | | |
|---|---|---|
| Population | N | 17,643,054 |
| Infection rate | r | 1.28E-08 |
| Removal Time | 1/a [days] | 21 |
| Removal rate | a [1/days] | 0.047619 |
| Basic Reprod. Rate | Ro | 4.742 |

## Permissiveness Pattern

| Stage Number | Stage Start Date | Permissiveness | Transition time [days] |
|---|---|---|---|
| 0 | | 1 | |
| 1 | 13-Feb-20 | 0.7 | 7 |
| 2 | 7-Apr-20 | 0.26 | 7 |
| 3 | 10-Jun-20 | 0.35 | 14 |
| 4 | 17-Jul-20 | 0.18 | 14 |
| 5 | 1-Dec-20 | 0.25 | 15 |
| 6 | 10-Apr-21 | 0.12 | 21 |
| 7 | 30-May-21 | 0.2 | 21 |
| 8 | 10-Jun-22 | 0.24 | 20 |
| 9 | 21-Dec-22 | 0.16 | 7 |

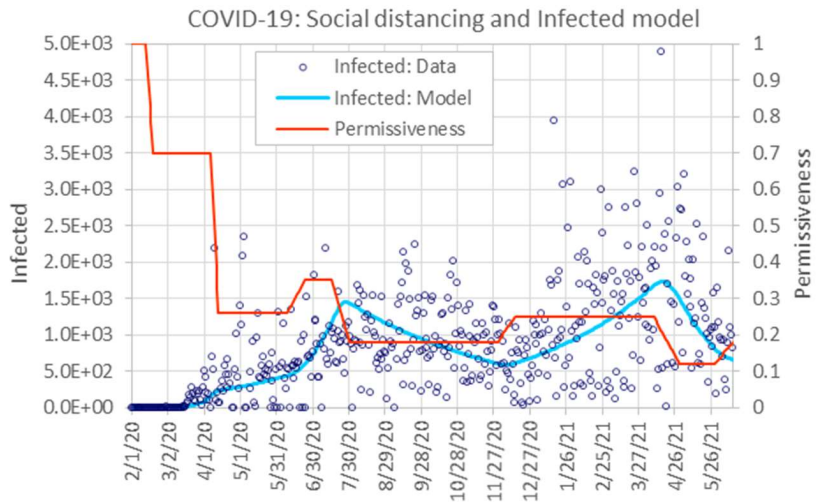

COVID-19: Social distancing and Infected model

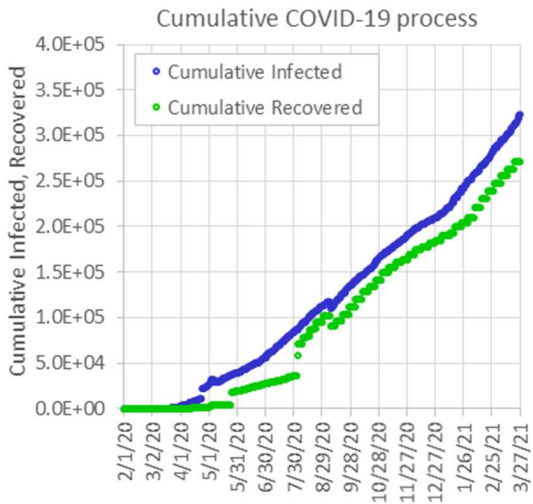

Cumulative COVID-19 process

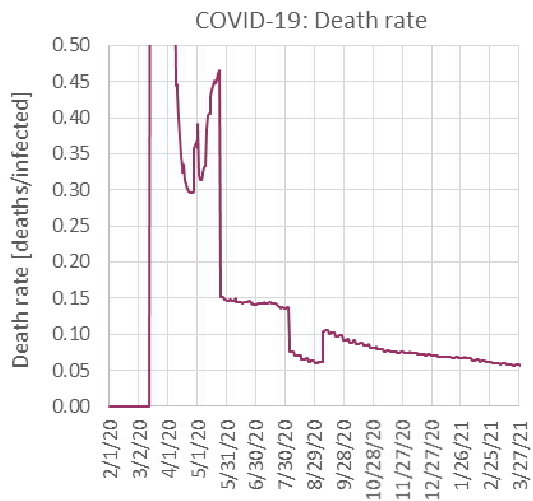

COVID-19: Death rate

# Guatemala

## General parameters

| | | |
|---|---|---|
| Population | N | 17,915,568 |
| Infection rate | r | 1.62E-08 |
| Removal Time | 1/a [days] | 14 |
| Removal rate | a [1/days] | 0.071429 |
| Basic Reprod. Rate | Ro | 4.063 |

## Permissiveness Pattern

| Stage Number | Stage Start Date | Permissiveness | Transition time [days] |
|---|---|---|---|
| 0 | | 1 | |
| 1 | 10-Apr-20 | 0.7 | 7 |
| 2 | 24-May-20 | 0.32 | 7 |
| 3 | 1-Jul-20 | 0.22 | 14 |
| 4 | 20-Sep-20 | 0.25 | 14 |
| 5 | 1-Dec-20 | 0.3 | 14 |
| 6 | 10-Jan-21 | 0.1 | 14 |
| 7 | 5-Feb-21 | 0.28 | 7 |
| 8 | 18-Nov-22 | 0.04 | 20 |
| 9 | 21-Dec-22 | 0.16 | 7 |

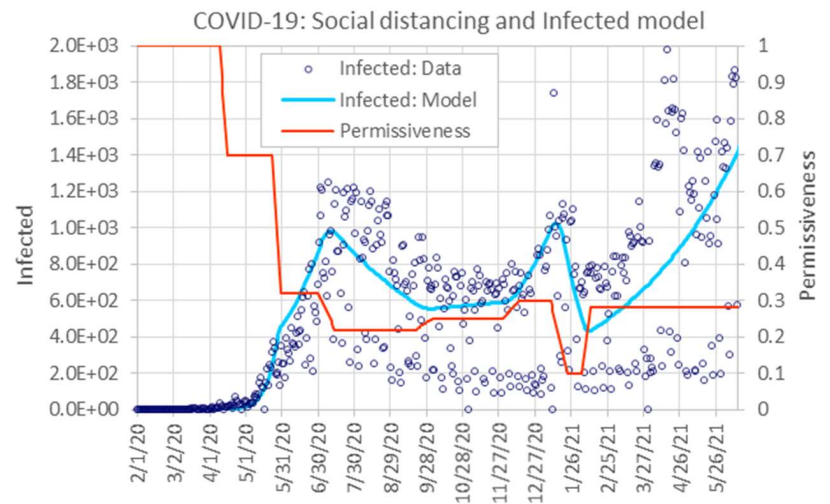

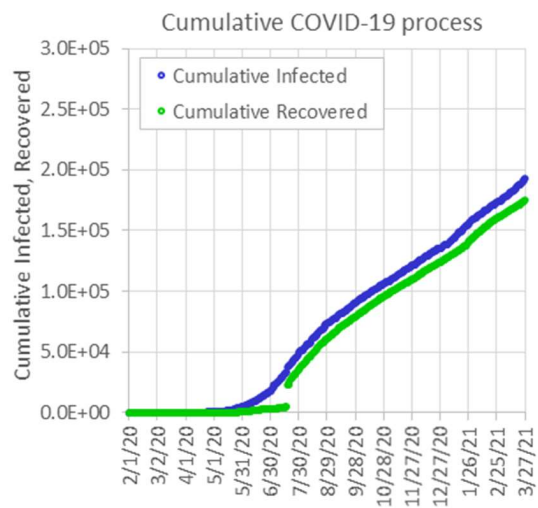

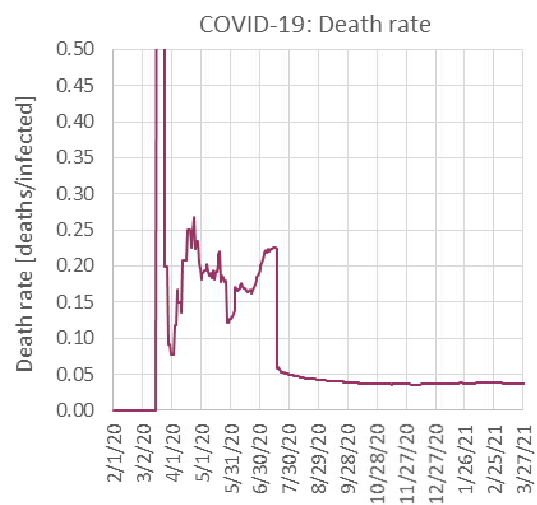

# Mexico

| General parameters | | |
|---|---|---|
| Population | N | 128,932,753 |
| Infection rate | r | 2.43E-09 |
| Removal Time | 1/a [days] | 21 |
| Removal rate | a [1/days] | 0.047619 |
| Basic Reprod. Rate | Ro | 6.579 |

| Permissiveness Pattern | | | |
|---|---|---|---|
| Stage Number | Stage Start Date | Permissiveness | Transition time [days] |
| 0 | | 1 | |
| 1 | 1-Mar-20 | 0.7 | 4 |
| 2 | 7-Apr-20 | 0.26 | 4 |
| 3 | 5-Jun-20 | 0.12 | 60 |
| 4 | 19-Sep-20 | 0.19 | 21 |
| 5 | 7-Jan-21 | 0.1 | 14 |
| 6 | 1-May-21 | 0.20 | 30 |
| 7 | 1-Sep-23 | 0.30 | 15 |
| 8 | | | |
| 9 | | | |

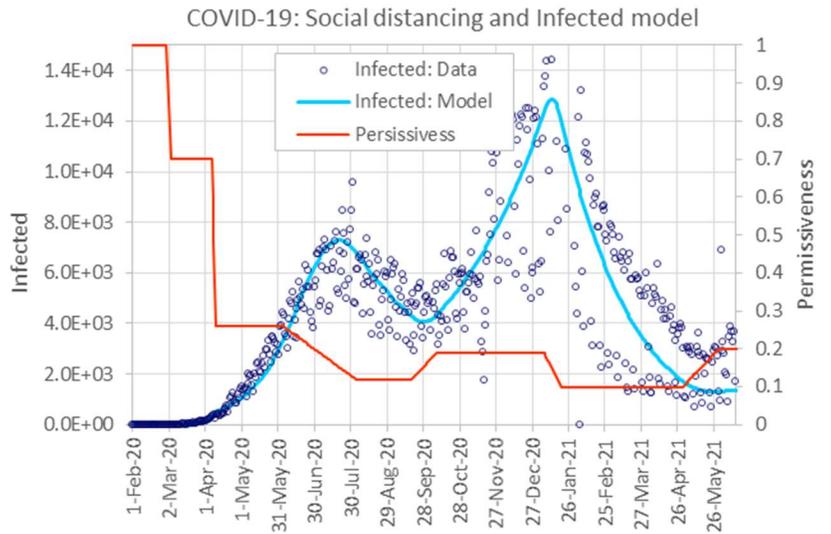

COVID-19: Social distancing and Infected model

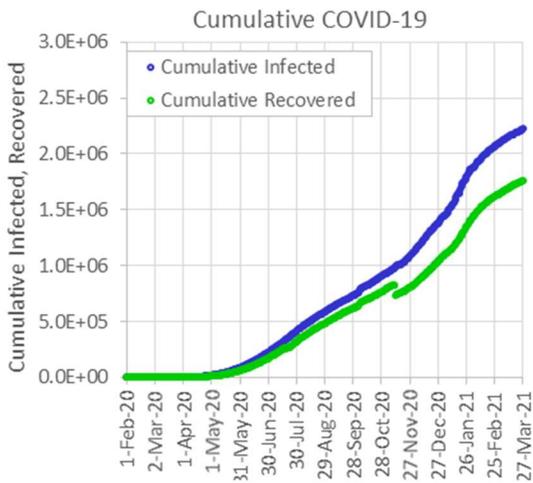

Cumulative COVID-19

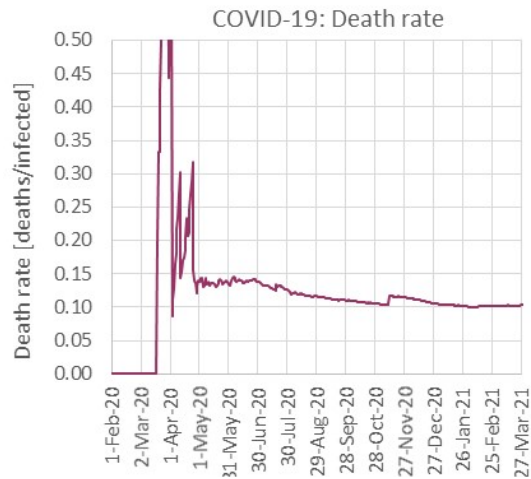

COVID-19: Death rate

# Panama

| General parameters | | |
|---|---|---|
| Population | N | 4,314,767 |
| Infection rate | r | 5.85E-08 |
| Removal Time | 1/a [days] | 21 |
| Removal rate | a [1/days] | 0.047619 |
| Basic Reprod. Rate | Ro | 5.301 |

| Permissiveness Pattern | | | |
|---|---|---|---|
| Stage Number | Stage Start Date | Permissiveness | Transition time [days] |
| 0 | | 1 | |
| 1 | 17-Feb-20 | 0.7 | 5 |
| 2 | 7-Apr-20 | 0.3 | 5 |
| 3 | 12-May-20 | 0.35 | 2 |
| 4 | 2-Jul-20 | 0.15 | 15 |
| 5 | 1-Oct-20 | 0.3 | 30 |
| 6 | 18-Dec-20 | 0.07 | 20 |
| 7 | 1-Apr-21 | 0.27 | 7 |
| 8 | 18-Nov-22 | 0.04 | 20 |
| 9 | 21-Dec-22 | 0.16 | 7 |

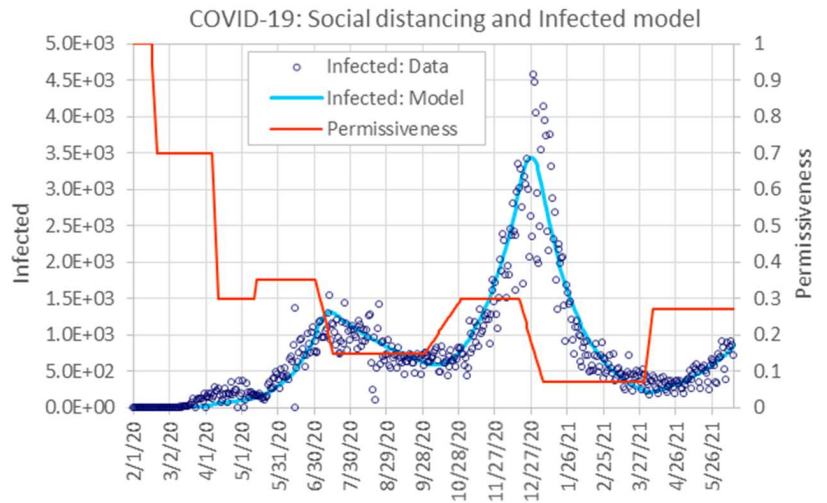

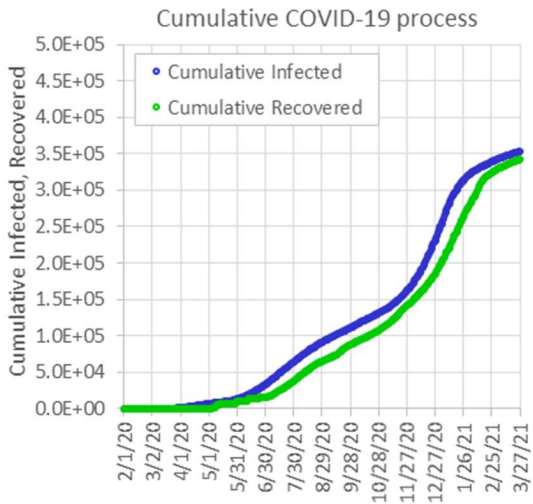

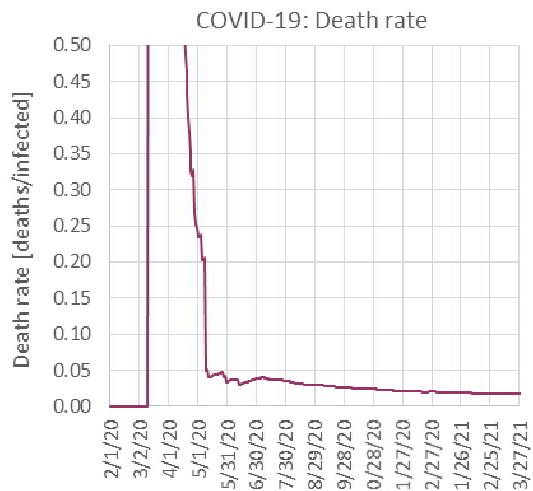

# Paraguay

| General parameters | | |
|---|---|---|
| Population | N | 7,132,538 |
| Infection rate | r | 2.75E-08 |
| Removal Time | 1/a [days] | 21 |
| Removal rate | a [1/days] | 0.047619 |
| Basic Reprod. Rate | Ro | 4.119 |

| Permissiveness Pattern | | | |
|---|---|---|---|
| Stage Number | Stage Start Date | Permissiveness | Transition time [days] |
| 0 | | 1 | |
| 1 | 10-Apr-20 | 0.7 | 5 |
| 2 | 10-May-20 | 0.41 | 5 |
| 3 | 10-Sep-20 | 0.19 | 7 |
| 4 | 5-Nov-20 | 0.33 | 7 |
| 5 | 25-Dec-20 | 0.17 | 7 |
| 6 | 20-Jan-21 | 0.32 | 21 |
| 7 | 7-Apr-21 | 0.2 | 60 |
| 8 | 18-Nov-22 | 0.04 | 20 |
| 9 | 21-Dec-22 | 0.16 | 7 |

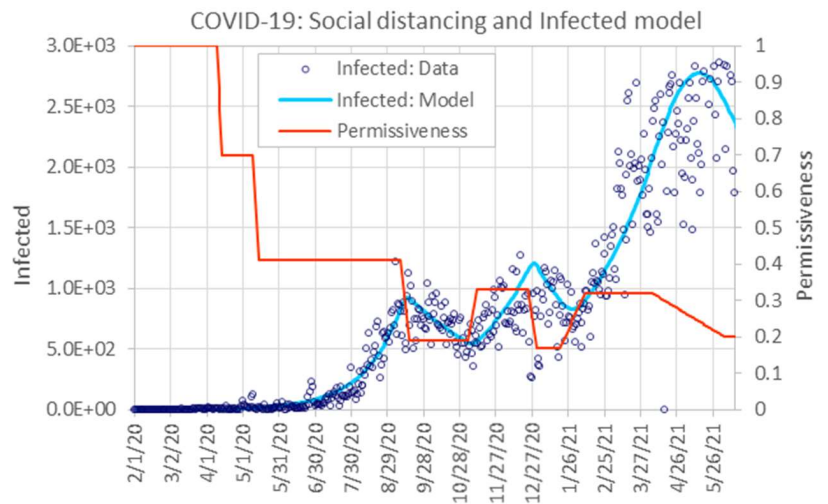

COVID-19: Social distancing and Infected model

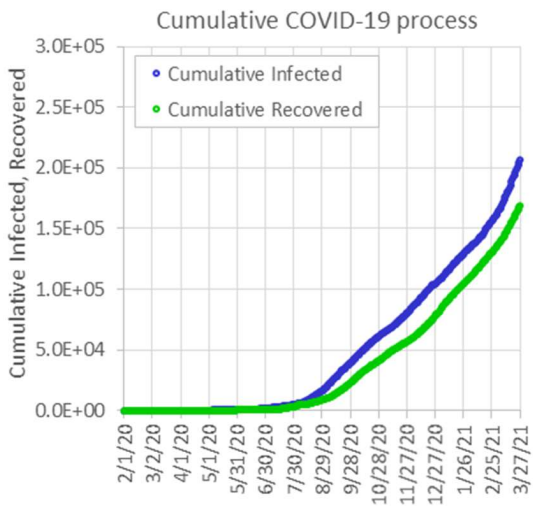

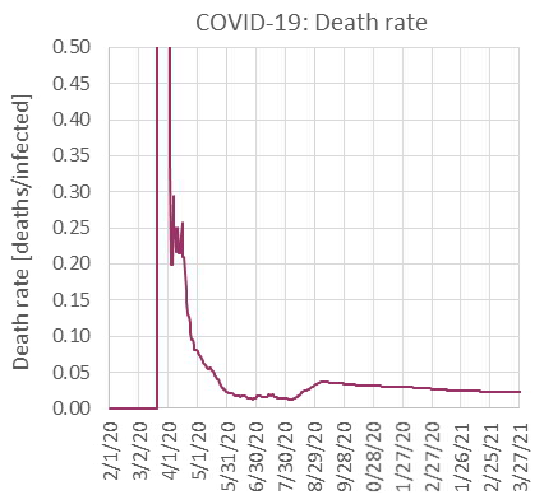

# Peru

| General parameters | | |
|---|---|---|
| Population | N | 32,971,854 |
| Infection rate | r | 7.71E-09 |
| Removal Time | 1/a [days] | 21 |
| Removal rate | a [1/days] | 0.047619 |
| Basic Reprod. Rate | Ro | 5.338 |

| Permissiveness Pattern | | | |
|---|---|---|---|
| Stage Number | Stage Start Date | Permissiveness | Transition time [days] |
| 0 | | 1 | |
| 1 | 17-Feb-20 | 0.7 | 5 |
| 2 | 24-Apr-20 | 0.4 | 5 |
| 3 | 10-May-20 | 0.16 | 16 |
| 4 | 2-Jul-20 | 0.33 | 25 |
| 5 | 6-Aug-20 | 0.13 | 10 |
| 6 | 1-Dec-20 | 0.27 | 14 |
| 7 | 15-Mar-21 | 0.15 | 14 |
| 8 | 18-Nov-22 | 0.04 | 20 |
| 9 | 21-Dec-22 | 0.16 | 7 |

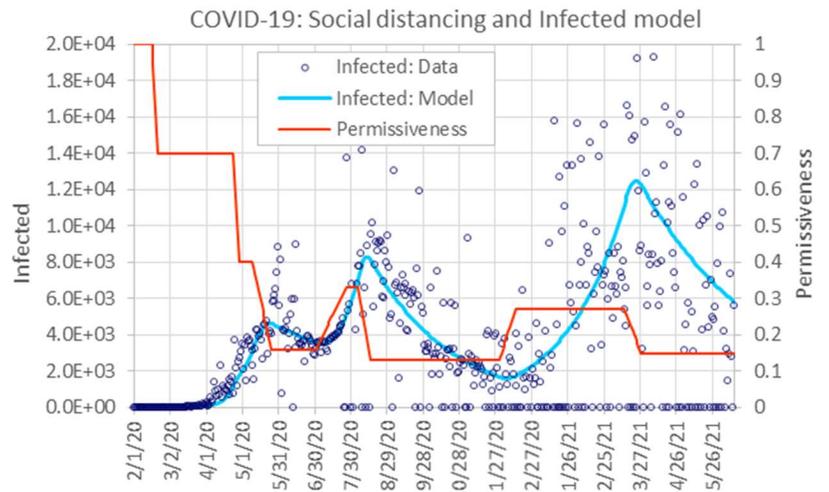

COVID-19: Social distancing and Infected model

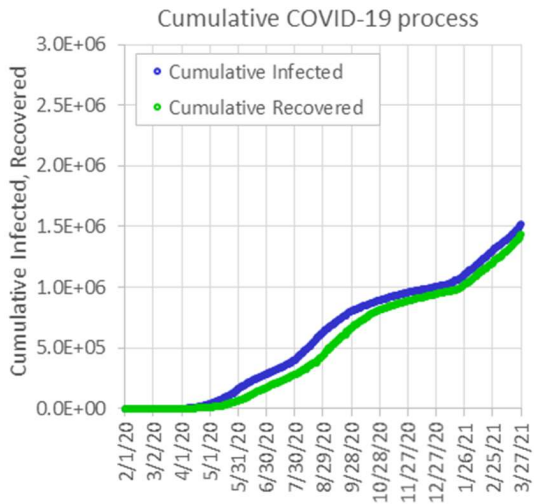

Cumulative COVID-19 process

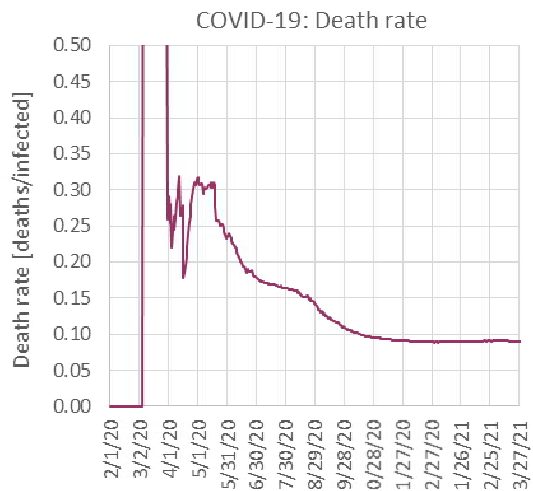

COVID-19: Death rate

# USA

| General parameters | | |
|---|---|---|
| Population | N | 331,002,651 |
| Infection rate | r | 1.26E-09 |
| Removal Time | 1/a [days] | 21 |
| Removal rate | a [1/days] | 0.047619 |
| Basic Reprod. Rate | Ro | 8.758 |

| Permissiveness Pattern | | | |
|---|---|---|---|
| Stage Number | Stage Start Date | Permissiveness | Transition time [days] |
| 0 | | 1 | |
| 1 | 11-Feb-20 | 0.7 | 9 |
| 2 | 20-Mar-20 | 0.09 | 17 |
| 3 | 6-Jun-20 | 0.22 | 7 |
| 4 | 8-Jul-20 | 0.09 | 7 |
| 5 | 4-Sep-20 | 0.19 | 40 |
| 6 | 25-Oct-20 | 0.14 | 30 |
| 7 | 18-Dec-20 | 0.06 | 30 |
| 8 | 10-Mar-21 | 0.2 | 21 |
| 9 | 1-Apr-21 | 0.05 | 21 |

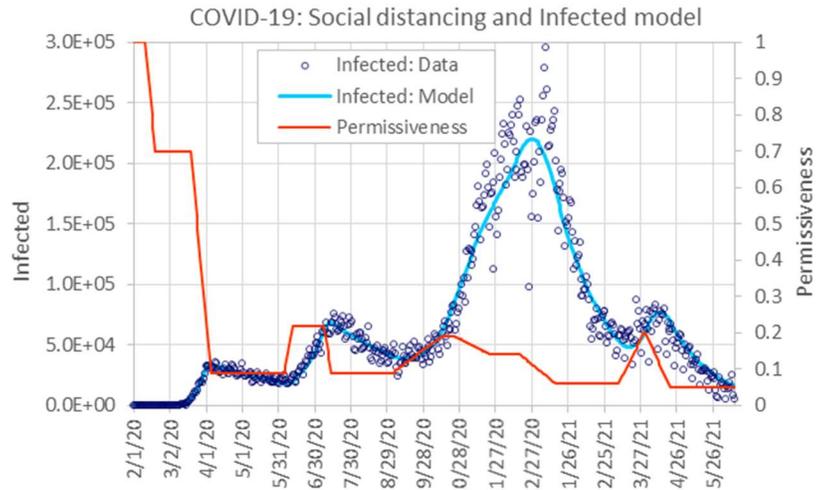

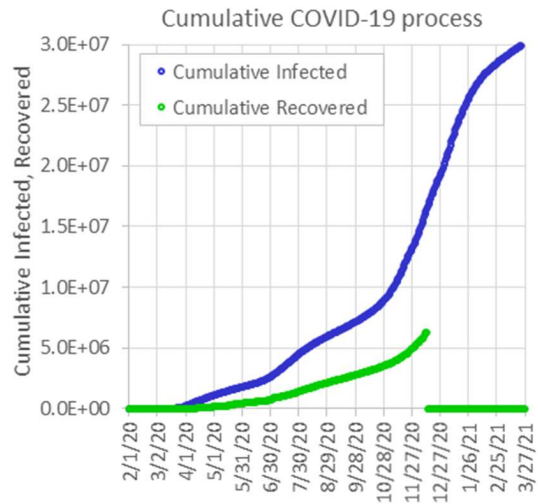

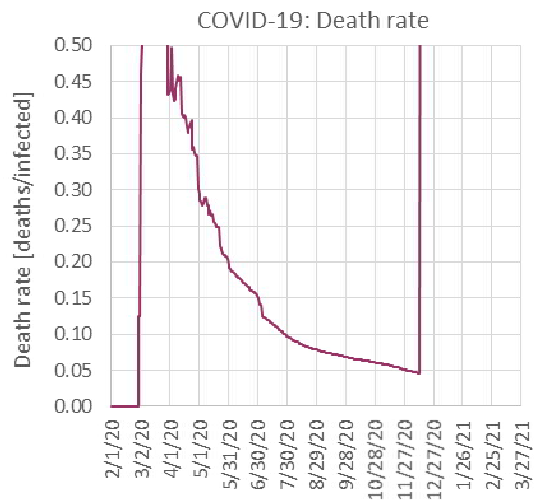

# India

| General parameters | | |
|---|---|---|
| Population | N | 1,380,004,385 |
| Infection rate | r | 3.46E-10 |
| Removal Time | 1/a [days] | 14 |
| Removal rate | a [1/days] | 0.071429 |
| Basic Reprod. Rate | Ro | 6.685 |

| Permissiveness Pattern | | | |
|---|---|---|---|
| Stage Number | Stage Start Date | Permissiveness | Transition time [days] |
| 0 | | 1 | |
| 1 | 16-Feb-20 | 0.7 | 5 |
| 2 | 10-Mar-20 | 0.23 | 3 |
| 3 | 13-Jul-20 | 0.17 | 25 |
| 4 | 13-Sep-20 | 0.12 | 4 |
| 5 | 10-Dec-20 | 0.1 | 14 |
| 6 | 12-Feb-21 | 0.31 | 21 |
| 7 | 16-Apr-21 | 0.07 | 21 |
| 8 | 18-Nov-22 | 0.04 | 20 |
| 9 | 21-Dec-22 | 0.16 | 7 |

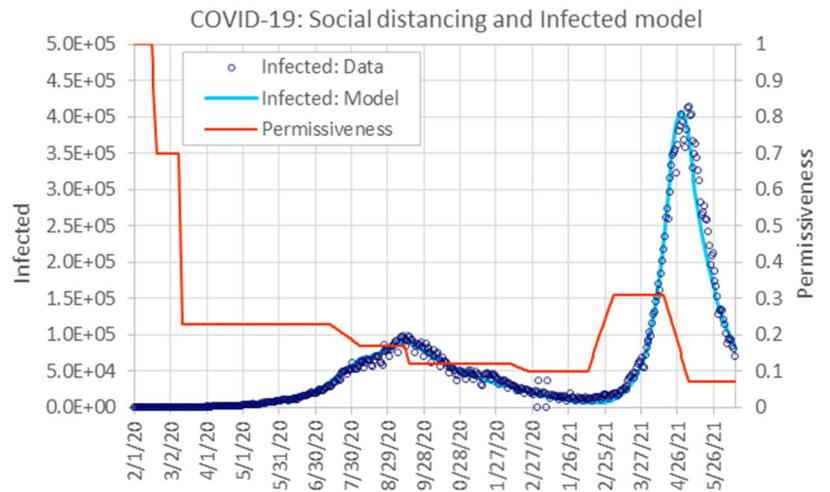

COVID-19: Social distancing and Infected model

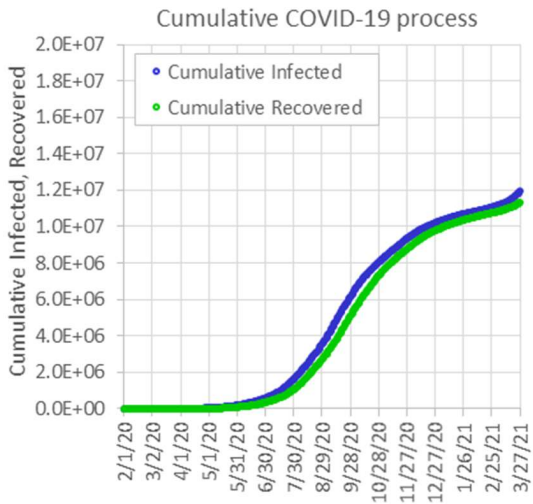

Cumulative COVID-19 process

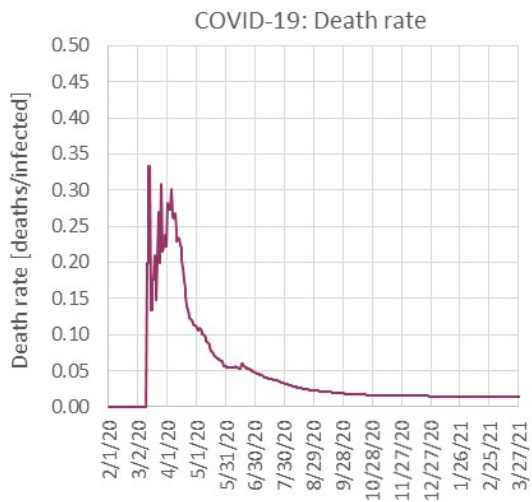

COVID-19: Death rate

# Iran

| General parameters | | |
|---|---|---|
| Population | N | 83,992,949 |
| Infection rate | r | 3.61E-09 |
| Removal Time | 1/a [days] | 21 |
| Removal rate | a [1/days] | 0.047619 |
| Basic Reprod. Rate | Ro | 6.368 |

| Permissiveness Pattern | | | |
|---|---|---|---|
| Stage Number | Stage Start Date | Permissiveness | Transition time [days] |
| 0 | - | 1 | - |
| 1 | 28-Jan-20 | 0.7 | 3 |
| 2 | 19-Mar-20 | 0.02 | 7 |
| 3 | 25-Apr-20 | 0.3 | 5 |
| 4 | 21-May-20 | 0.15 | 6 |
| 5 | 7-Sep-20 | 0.24 | 7 |
| 6 | 18-Nov-20 | 0.04 | 21 |
| 7 | 18-Dec-20 | 0.18 | 21 |
| 8 | 27-Feb-21 | 0.25 | 21 |
| 9 | 8-Apr-21 | 0.07 | 21 |

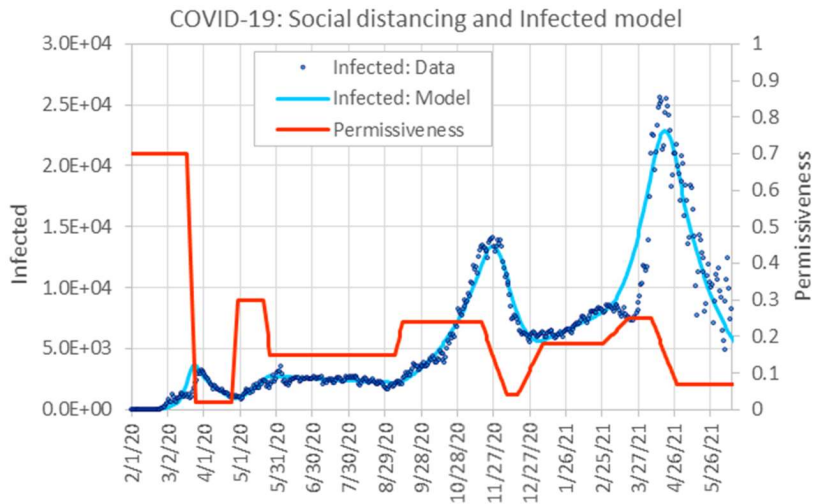

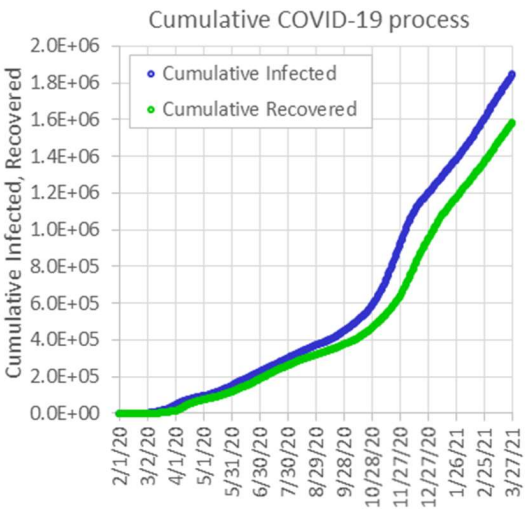

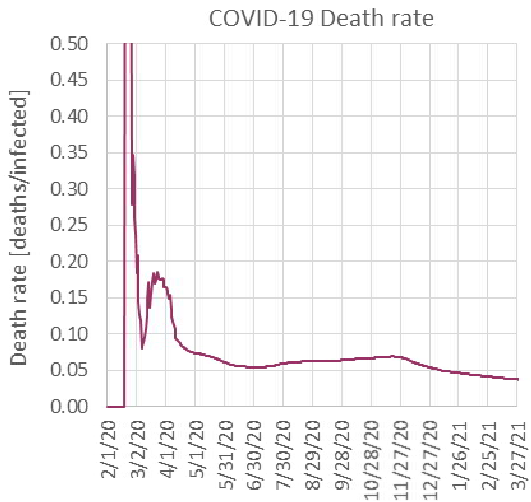

Japan

# Japan

### General parameters

| | | |
|---|---|---|
| Population | N | 126,476,461 |
| Infection rate | r | 4E-09 |
| Removal Time | 1/a [days] | 14 |
| Removal rate | a [1/days] | 0.071429 |
| Basic Reprod. Rate | Ro | 7.083 |

### Permissiveness Pattern

| Stage Number | Stage Start Date | Permissiveness | Transition time [days] |
|---|---|---|---|
| 0 | | 1 | |
| 1 | 11-Mar-20 | 0.7 | 5 |
| 2 | 5-Apr-20 | 0.001 | 3 |
| 3 | 12-Jun-20 | 0.4 | 3 |
| 4 | 23-Jul-20 | 0.07 | 14 |
| 5 | 16-Sep-20 | 0.22 | 21 |
| 6 | 10-Nov-20 | 0.19 | 21 |
| 7 | 7-Jan-21 | 0.06 | 14 |
| 8 | 18-Feb-21 | 0.2 | 28 |
| 9 | 5-May-21 | 0.08 | 7 |

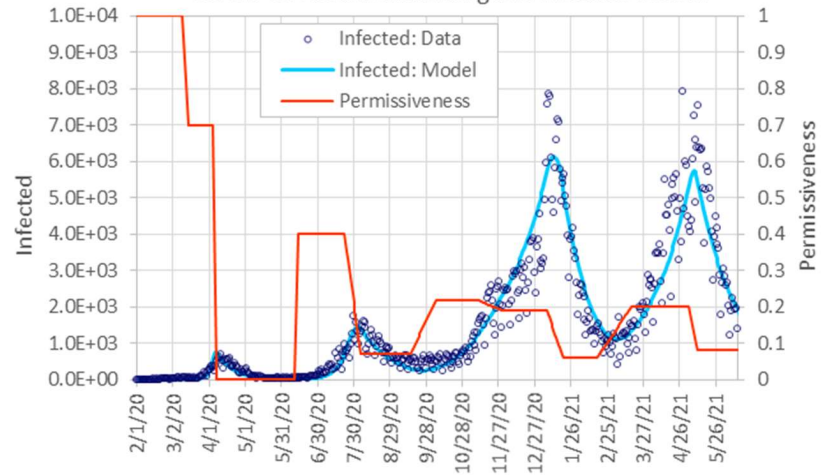

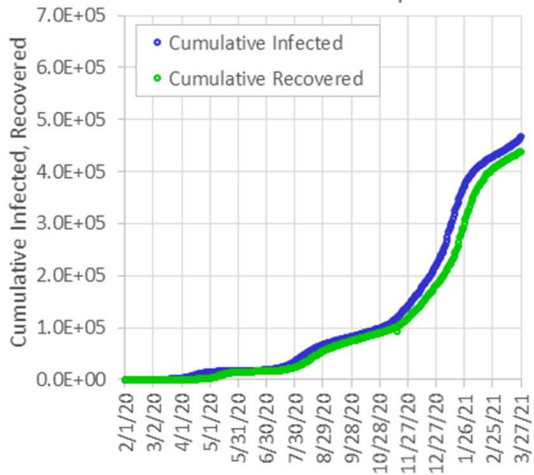

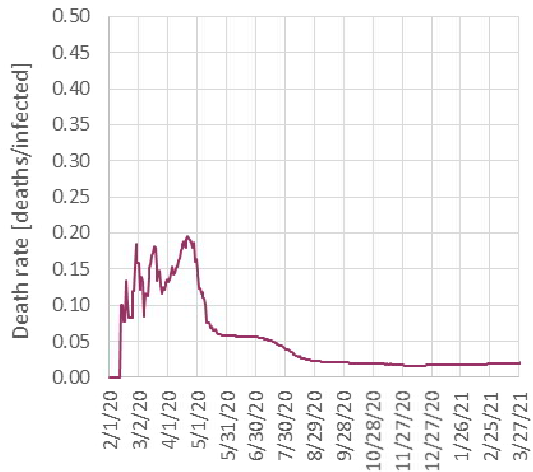

# Malaysia

## General parameters

| | | |
|---|---|---|
| Population | N | 32,365,999 |
| Infection rate | r | 1.14E-08 |
| Removal Time | 1/a [days] | 14 |
| Removal rate | a [1/days] | 0.071429 |
| Basic Reprod. Rate | Ro | 5.166 |

## Permissiveness Pattern

| Stage Number | Stage Start Date | Permissiveness | Transition time [days] |
|---|---|---|---|
| 0 | | 1 | |
| 1 | 20-Feb-20 | 0.7 | 7 |
| 2 | 10-Mar-20 | 0.05 | 30 |
| 3 | 25-Jul-20 | 0.45 | 7 |
| 4 | 9-Oct-20 | 0.25 | 7 |
| 5 | 25-Jan-21 | 0.12 | 15 |
| 6 | 16-Mar-21 | 0.28 | 21 |
| 7 | 27-May-21 | 0.1 | 7 |
| 8 | 18-Nov-22 | 0.04 | 20 |
| 9 | 21-Dec-22 | 0.16 | 7 |

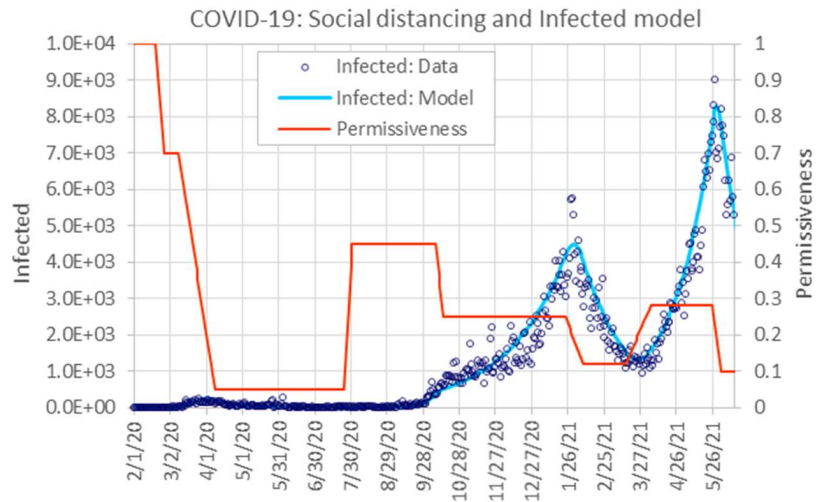

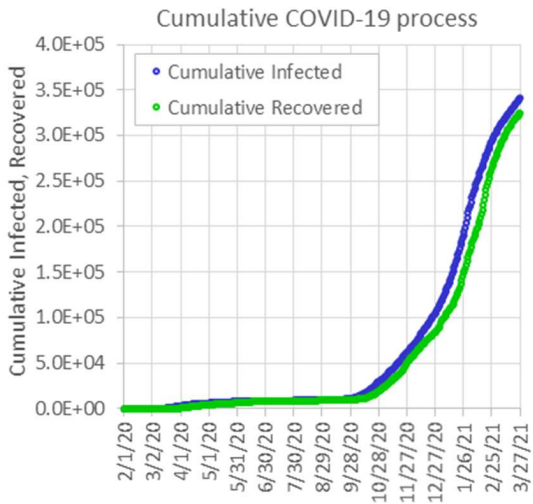

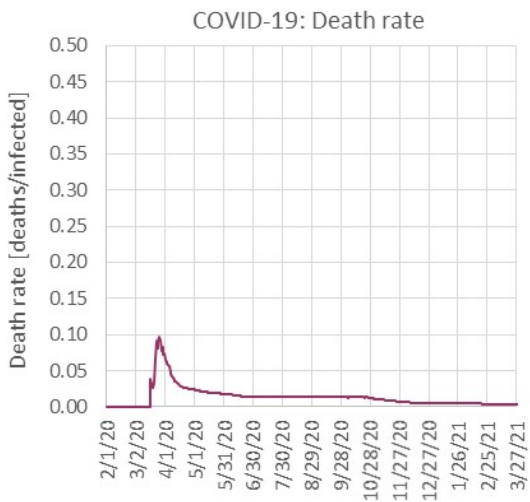

# South Korea

| General parameters | | |
|---|---|---|
| Population | N | 51,269,185 |
| Infection rate | r | 6.84E-09 |
| Removal Time | 1/a [days] | 21 |
| Removal rate | a [1/days] | 0.047619 |
| Basic Reprod. Rate | Ro | 7.364 |

| Permissiveness Pattern | | | |
|---|---|---|---|
| Stage Number | Stage Start Date | Permissiveness | Transition time [days] |
| 0 | | 1 | |
| 1 | 28-Jan-20 | 0.7 | 2 |
| 2 | 28-Feb-20 | 0.01 | 3 |
| 3 | 5-Jun-20 | 0.3 | 4 |
| 4 | 21-Aug-20 | 0.02 | 5 |
| 5 | 8-Oct-20 | 0.27 | 5 |
| 6 | 17-Dec-20 | 0.06 | 7 |
| 7 | 31-Jan-21 | 0.16 | 7 |
| 8 | 28-Apr-21 | 0.12 | 7 |
| 9 | 21-Dec-22 | 0.16 | 7 |

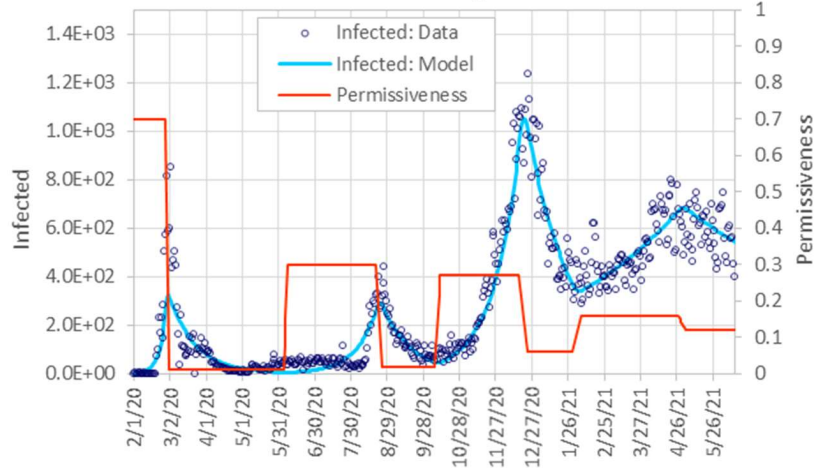

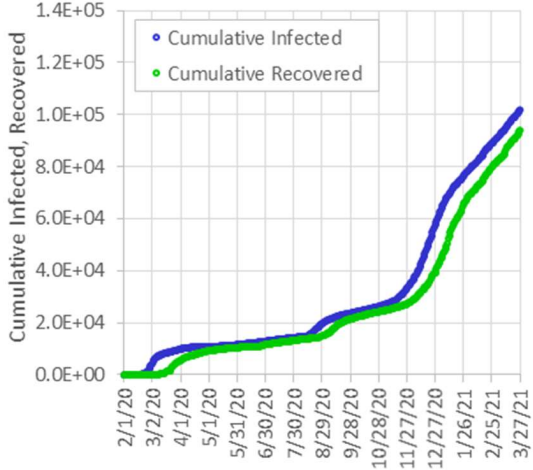

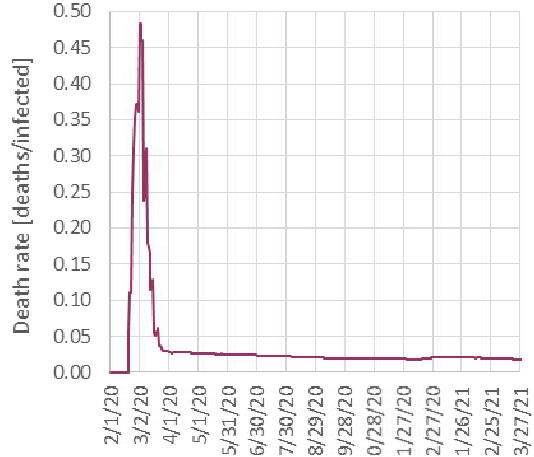

# Taiwan

| General parameters | | |
|---|---|---|
| Population | N | 32,365,999 |
| Infection rate | r | 8.4E-09 |
| Removal Time | 1/a [days] | 28 |
| Removal rate | a [1/days] | 0.035714 |
| Basic Reprod. Rate | Ro | 7.612 |

| Permissiveness Pattern | | | |
|---|---|---|---|
| Stage Number | Stage Date Start | Permissiveness | Transition time [days] |
| 0 | | 1 | |
| 1 | 5-Mar-20 | 0.7 | 14 |
| 2 | 20-Mar-20 | 0.001 | 3 |
| 3 | 20-May-20 | 0.14 | 14 |
| 4 | 1-Sep-20 | 0.16 | 7 |
| 5 | 10-Dec-20 | 0.08 | 15 |
| 6 | 15-Feb-22 | 0.17 | 21 |
| 7 | 7-Sep-22 | 0.24 | 7 |
| 8 | 18-Nov-22 | 0.04 | 20 |
| 9 | 21-Dec-22 | 0.16 | 7 |

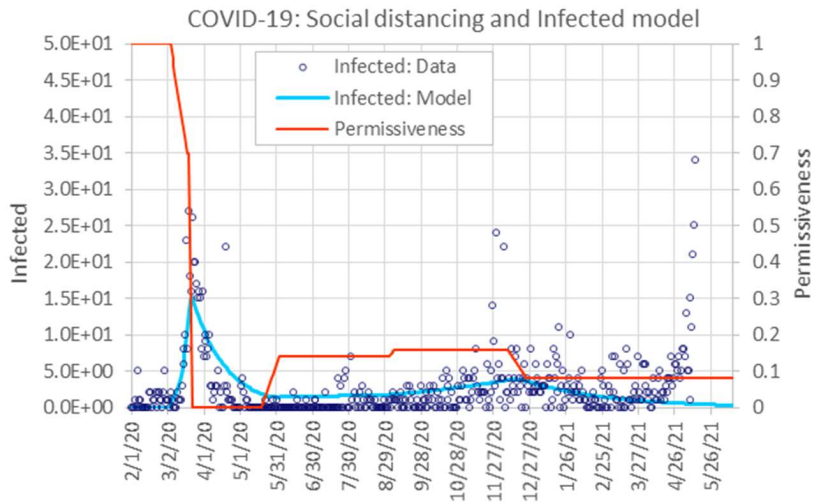

COVID-19: Social distancing and Infected model

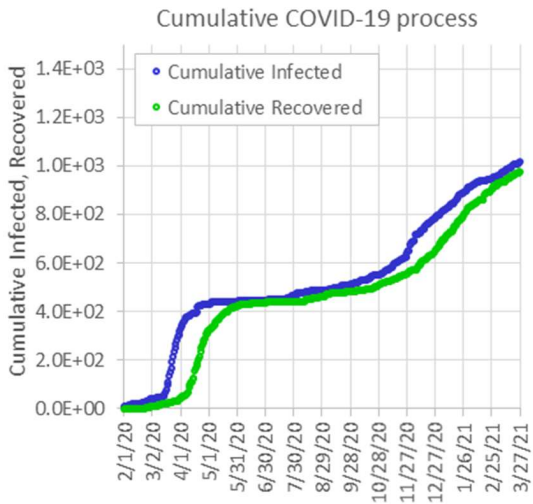

Cumulative COVID-19 process

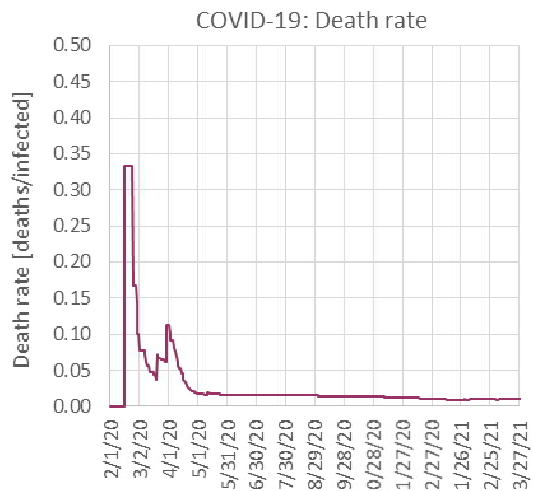

COVID-19: Death rate

# Thailand

### General parameters

| | | |
|---|---|---|
| Population | N | 69,799,978 |
| Infection rate | r | 4.01E-09 |
| Removal Time | 1/a [days] | 21 |
| Removal rate | a [1/days] | 0.047619 |
| Basic Reprod. Rate | Ro | 5.878 |

### Permissiveness Pattern

| Stage Number | Stage Start Date | Permissiveness | Transition time [days] |
|---|---|---|---|
| 0 | | 1 | |
| 1 | 17-Feb-20 | 0.7 | 2 |
| 2 | 22-Mar-20 | 0.001 | 5 |
| 3 | 1-Jul-20 | 0.25 | 30 |
| 4 | 10-Dec-20 | 0.7 | 3 |
| 5 | 3-Jan-21 | 0.001 | 3 |
| 6 | 10-Feb-21 | 0.3 | 7 |
| 7 | 7-Sep-22 | 0.24 | 7 |
| 8 | 18-Nov-22 | 0.04 | 20 |
| 9 | 21-Dec-22 | 0.16 | 7 |

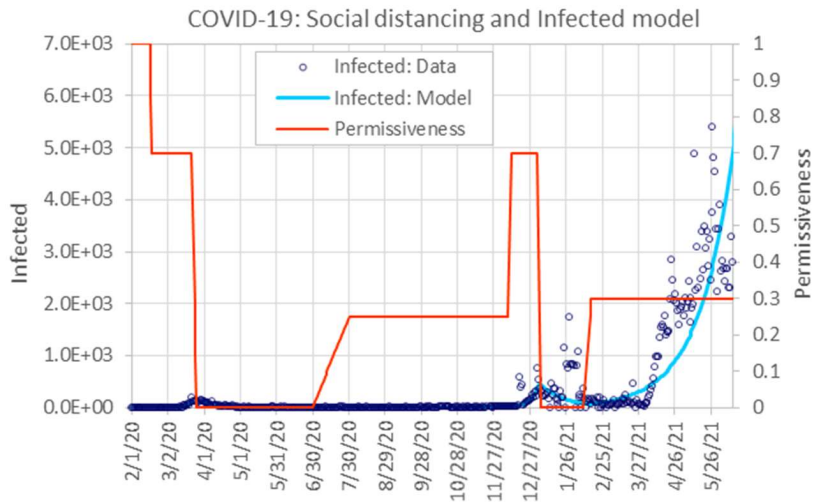

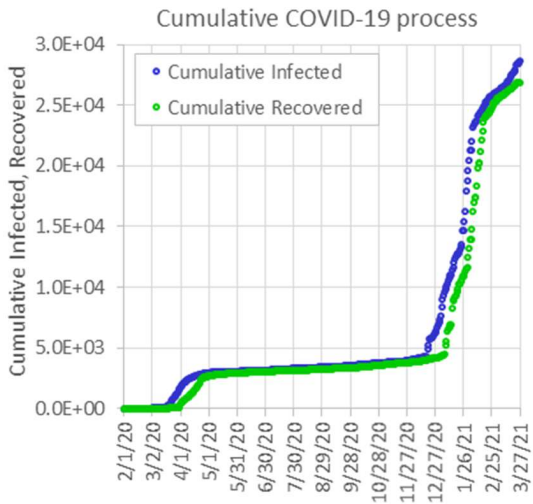

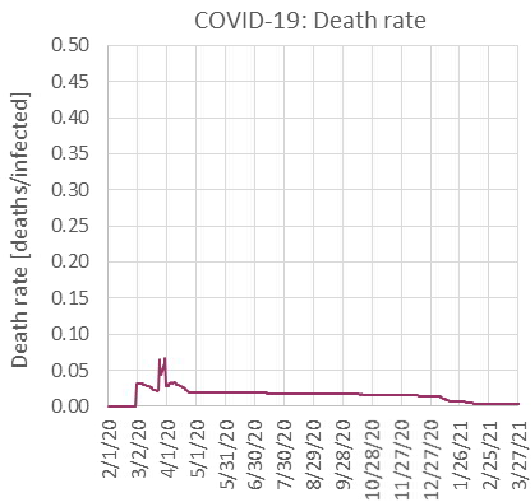

# Australia

## General parameters

| | | |
|---|---|---|
| Population | N | 25,499,884 |
| Infection rate | r | 1.21E-08 |
| Removal Time | 1/a [days] | 21 |
| Removal rate | a [1/days] | 0.047619 |
| Basic Reprod. Rate | Ro | 6.480 |

## Permissiveness Pattern

| Stage Number | Stage Start Date | Permissiveness | Transition time [days] |
|---|---|---|---|
| 0 | | 1 | |
| 1 | 21-Feb-20 | 0.7 | 7 |
| 2 | 22-Mar-20 | 0.01 | 7 |
| 3 | 27-May-20 | 0.39 | 5 |
| 4 | 25-Jul-20 | 0.01 | 5 |
| 5 | 10-Dec-20 | 0.1 | 15 |
| 6 | 15-Feb-21 | 0.17 | 21 |
| 7 | 7-Sep-22 | 0.24 | 7 |
| 8 | 18-Nov-22 | 0.04 | 20 |
| 9 | 21-Dec-22 | 0.16 | 7 |

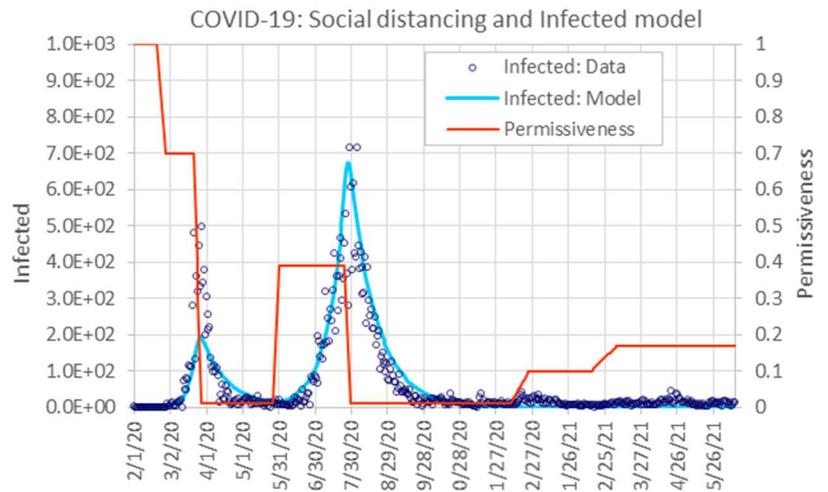

COVID-19: Social distancing and Infected model

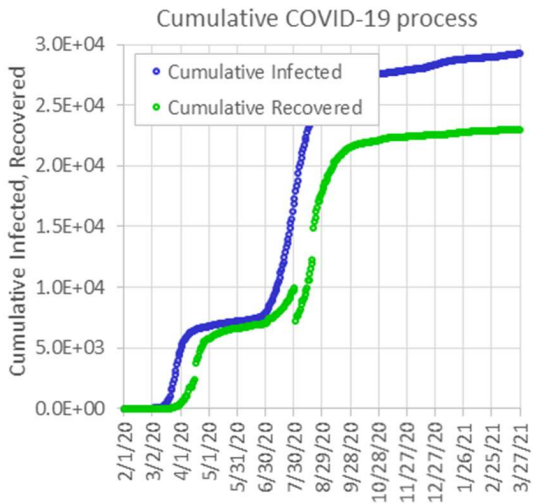

Cumulative COVID-19 process

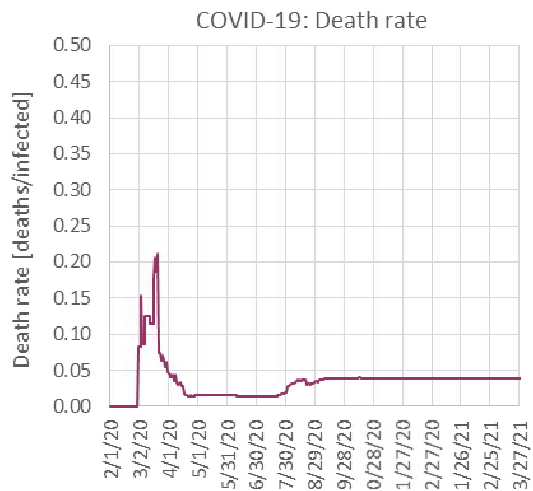

COVID-19: Death rate

# New Zealand

| General parameters | | |
|---|---|---|
| Population | N | 4,822,233 |
| Infection rate | r | 8.97E-08 |
| Removal Time | 1/a [days] | 14 |
| Removal rate | a [1/days] | 0.071429 |
| Basic Reprod. Rate | Ro | 6.056 |

| Permissiveness Pattern | | | |
|---|---|---|---|
| Stage Number | Stage Start Date | Permissiveness | Transition time [days] |
| 0 | | 1 | |
| 1 | 7-Mar-20 | 0.7 | 7 |
| 2 | 23-Mar-20 | 0.001 | 10 |
| 3 | 15-Jun-20 | 0.3 | 7 |
| 4 | 15-Aug-20 | 0.1 | 7 |
| 5 | 15-Oct-20 | 0.18 | 7 |
| 6 | 15-Feb-21 | 0.17 | 21 |
| 7 | 7-Sep-22 | 0.24 | 7 |
| 8 | 18-Nov-22 | 0.04 | 20 |
| 9 | 21-Dec-22 | 0.16 | 7 |

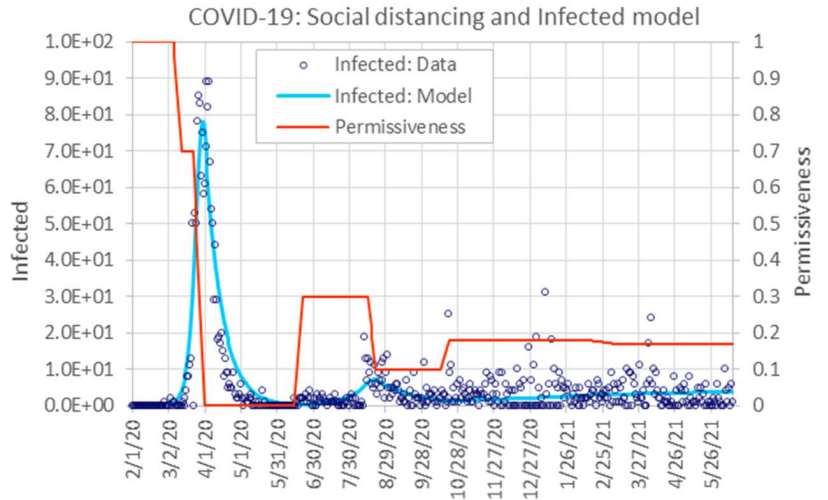

COVID-19: Social distancing and Infected model

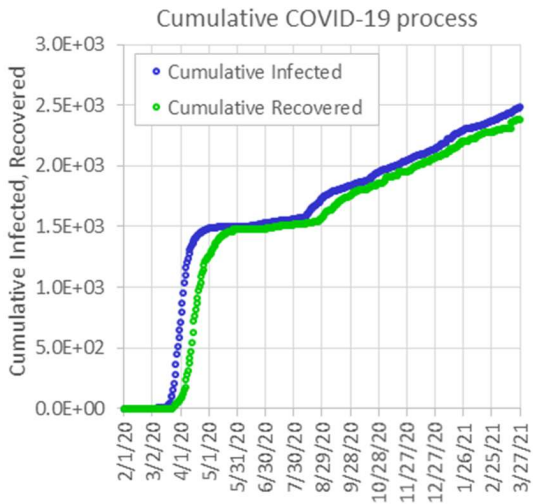

Cumulative COVID-19 process

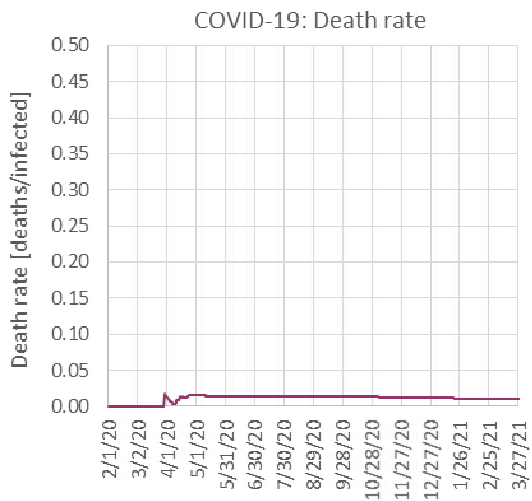

COVID-19: Death rate

# Papua New Guinea

| General parameters | | |
|---|---|---|
| Population | N | 9,904,607 |
| Infection rate | r | 2.97E-08 |
| Removal Time | 1/a [days] | 14 |
| Removal rate | a [1/days] | 0.071429 |
| Basic Reprod. Rate | Ro | 4.118 |

| Permissiveness Pattern | | | |
|---|---|---|---|
| Stage Number | Stage Start Date | Permissiveness | Transition time [days] |
| 0 | | 1 | |
| 1 | 13-Jul-20 | 0.7 | 7 |
| 2 | 10-Aug-20 | 0.01 | 7 |
| 3 | 30-Sep-20 | 0.37 | 7 |
| 4 | 5-Dec-20 | 0.1 | 7 |
| 5 | 10-Dec-20 | 0.25 | 15 |
| 6 | 31-Jan-21 | 0.4 | 21 |
| 7 | 5-Apr-21 | 0.2 | 21 |
| 8 | 18-Nov-22 | 0.04 | 20 |
| 9 | 21-Dec-22 | 0.16 | 7 |

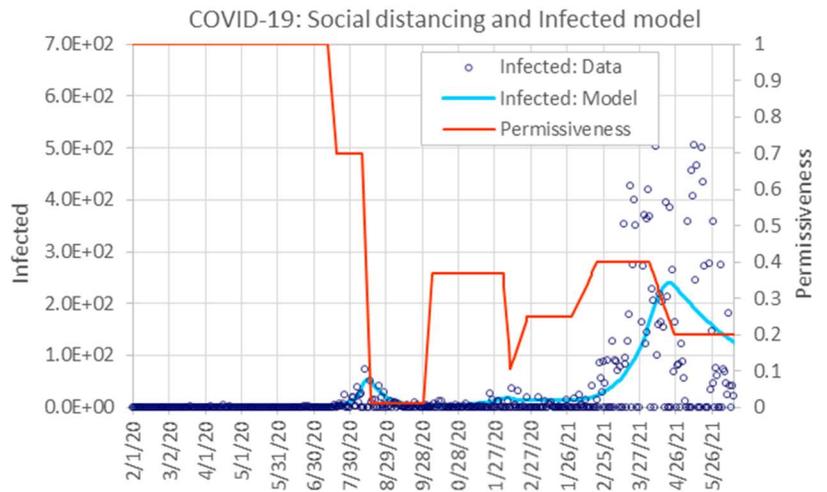

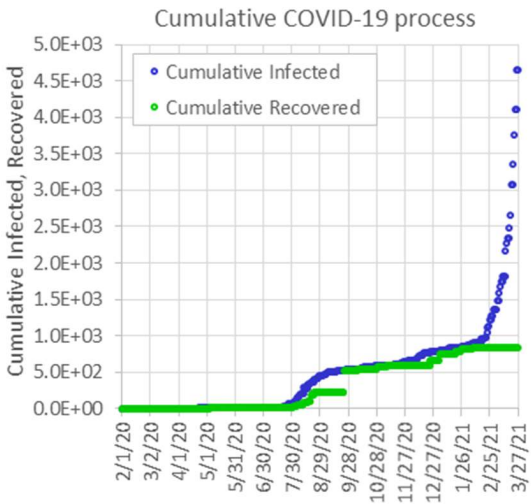

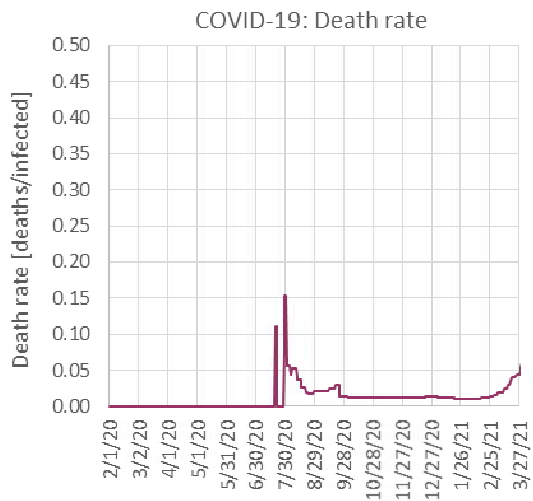

# Algeria

### General parameters

| | | |
|---|---|---|
| Population | N | 53,771,296 |
| Infection rate | r | 4.9E-09 |
| Removal Time | 1/a [days] | 28 |
| Removal rate | a [1/days] | 0.035714 |
| Basic Reprod. Rate | Ro | 7.377 |

### Permissiveness Pattern

| Stage Number | Stage Start Date | Permissiveness | Transition time [days] |
|---|---|---|---|
| 0 | | 1 | |
| 1 | 25-Feb-20 | 0.7 | 7 |
| 2 | 10-Mar-20 | 0.3 | 7 |
| 3 | 7-May-20 | 0.06 | 14 |
| 4 | 9-Jun-20 | 0.34 | 7 |
| 5 | 9-Jul-20 | 0.07 | 20 |
| 6 | 11-Oct-20 | 0.38 | 7 |
| 7 | 11-Nov-20 | 0.05 | 14 |
| 8 | 10-Mar-21 | 0.2 | 21 |
| 9 | 21-Dec-22 | 0.16 | 7 |

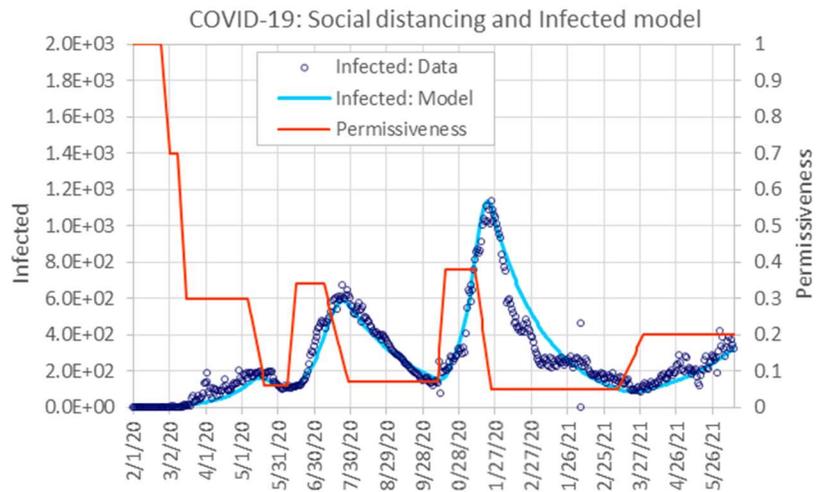

COVID-19: Social distancing and Infected model

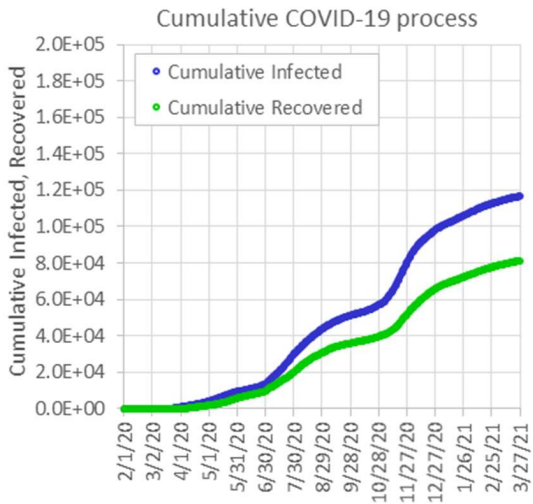

Cumulative COVID-19 process

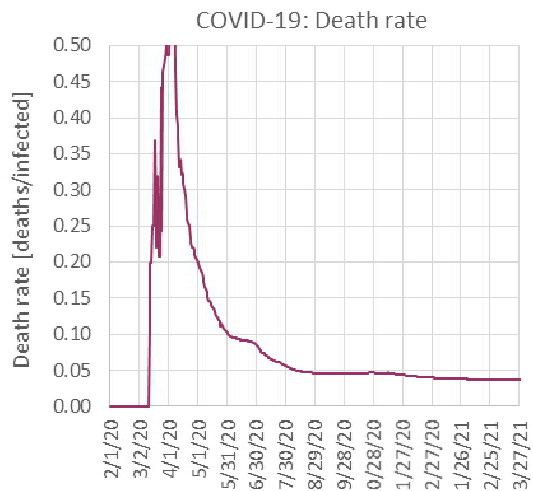

COVID-19: Death rate

# Egypt

## General parameters

| | | |
|---|---|---|
| Population | N | 102,334,404 |
| Infection rate | r | 2.15E-09 |
| Removal Time | 1/a [days] | 28 |
| Removal rate | a [1/days] | 0.035714 |
| Basic Reprod. Rate | Ro | 6.161 |

## Permissiveness Pattern

| Stage Number | Stage Start Date | Permissiveness | Transition time [days] |
|---|---|---|---|
| 0 | | 1 | |
| 1 | 1-Mar-20 | 0.7 | 6 |
| 2 | 6-Apr-20 | 0.39 | 6 |
| 3 | 29-May-20 | 0.02 | 30 |
| 4 | 18-Sep-20 | 0.25 | 7 |
| 5 | 15-Dec-20 | 0.48 | 2 |
| 6 | 31-Dec-20 | 0.05 | 4 |
| 7 | 5-Feb-21 | 0.2 | 7 |
| 8 | 17-May-21 | 0.08 | 7 |
| 9 | 21-Dec-22 | 0.16 | 7 |

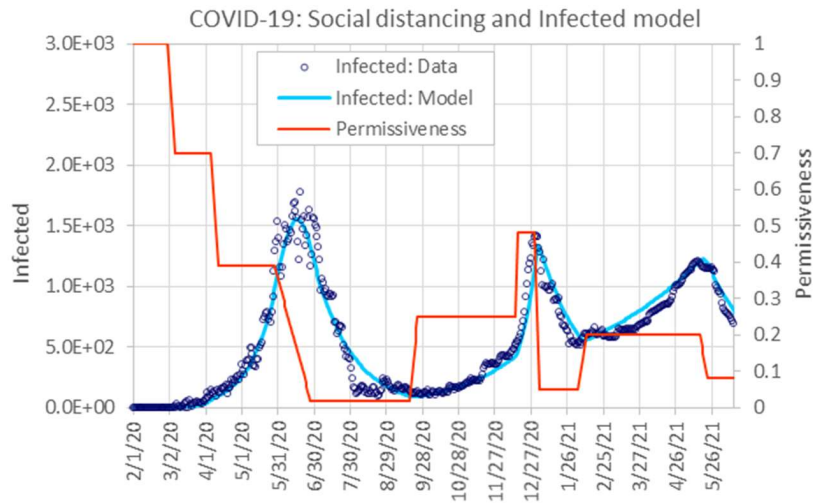

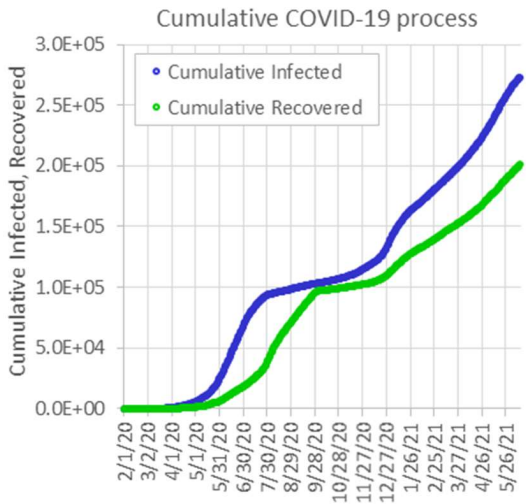

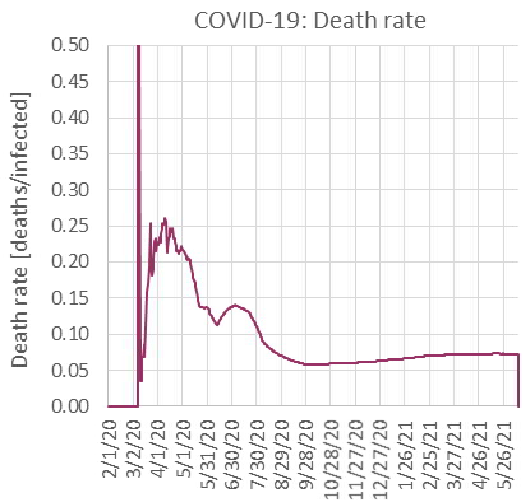

# Kenya

| General parameters | | |
|---|---|---|
| Population | N | 53,771,296 |
| Infection rate | r | 3.3E-09 |
| Removal Time | 1/a [days] | 28 |
| Removal rate | a [1/days] | 0.035714 |
| Basic Reprod. Rate | Ro | 4.968 |

| Permissiveness Pattern | | | |
|---|---|---|---|
| Stage Number | Stage Start Date | Permissiveness | Transition time [days] |
| 0 | | 1 | |
| 1 | 3-Apr-20 | 0.7 | 14 |
| 2 | 7-May-20 | 0.41 | 14 |
| 3 | 25-Jul-20 | 0.06 | 7 |
| 4 | 20-Sep-20 | 0.5 | 7 |
| 5 | 2-Nov-20 | 0.05 | 7 |
| 6 | 26-Jan-21 | 0.47 | 21 |
| 7 | 20-Mar-21 | 0.05 | 21 |
| 8 | 18-May-21 | 0.3 | 21 |
| 9 | 21-Dec-22 | 0.16 | 7 |

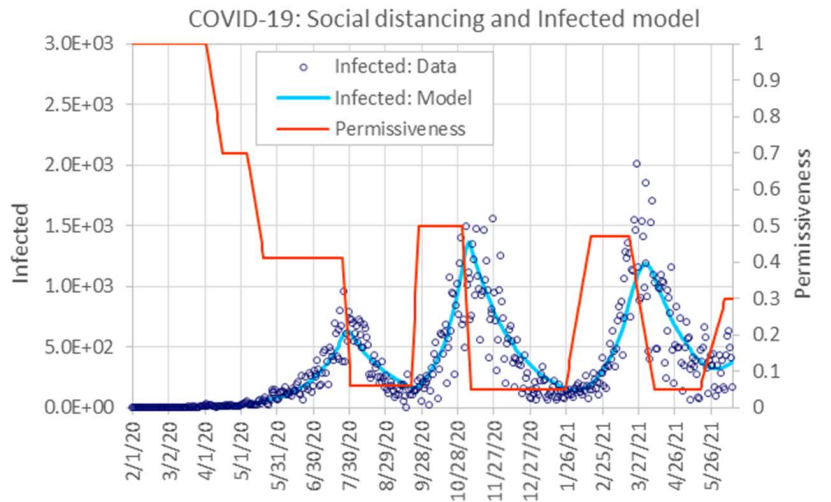

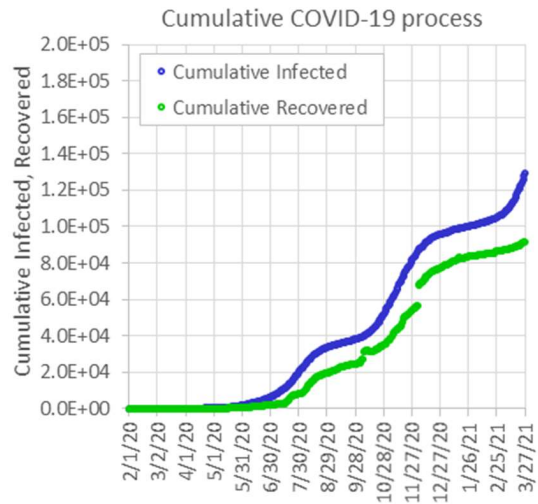

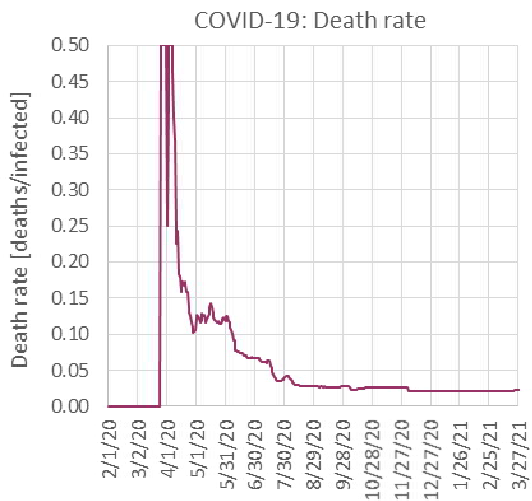

# Nigeria

| General parameters | | |
|---|---|---|
| Population | N | 206,139,589 |
| Infection rate | r | 1.35E-09 |
| Removal Time | 1/a [days] | 21 |
| Removal rate | a [1/days] | 0.047619 |
| Basic Reprod. Rate | Ro | 5.844 |

| Permissiveness Pattern | | | |
|---|---|---|---|
| Stage Number | Stage Start Date | Permissiveness | Transition time [days] |
| 0 | | 1 | |
| 1 | 25-Mar-20 | 0.7 | 6 |
| 2 | 20-Apr-20 | 0.3 | 14 |
| 3 | 15-Jun-20 | 0.11 | 20 |
| 4 | 30-Oct-20 | 0.33 | 7 |
| 5 | 13-Jan-21 | 0.03 | 7 |
| 6 | 15-May-21 | 0.17 | 21 |
| 7 | 7-Sep-22 | 0.24 | 7 |
| 8 | 18-Nov-22 | 0.04 | 20 |
| 9 | 21-Dec-22 | 0.16 | 7 |

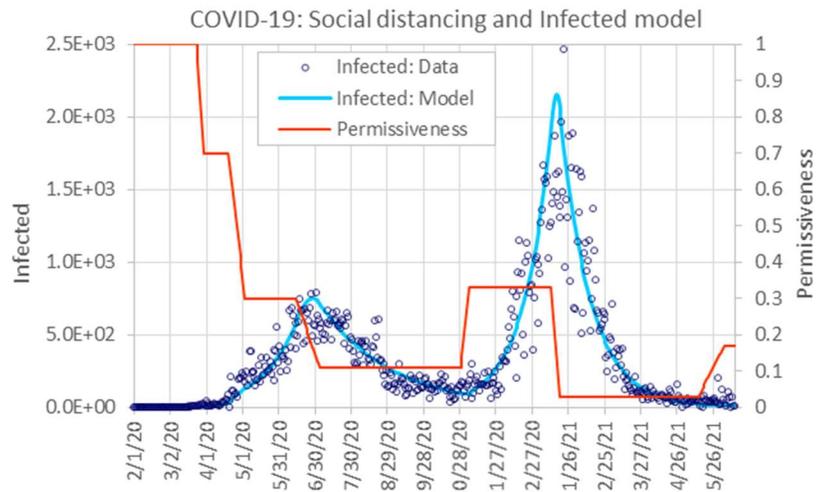

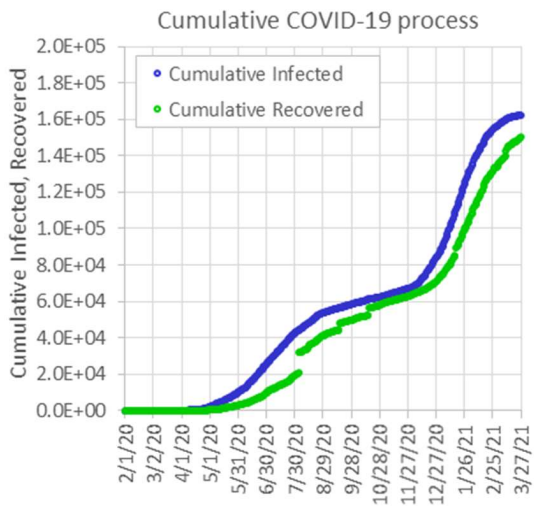

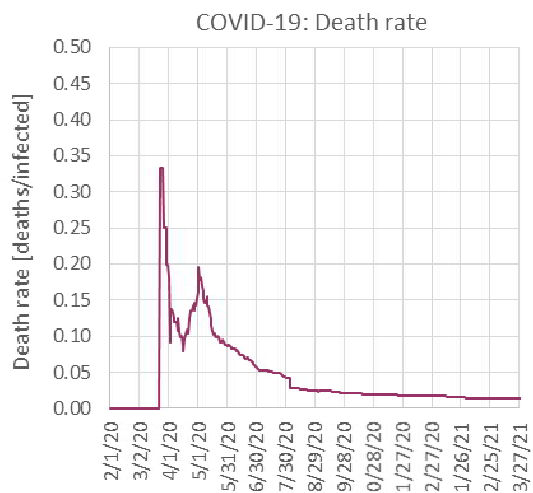

# South Africa

## General parameters

| General parameters | | |
|---|---|---|
| Population | N | 59,308,690 |
| Infection rate | r | 4.45E-09 |
| Removal Time | 1/a [days] | 21 |
| Removal rate | a [1/days] | 0.047619 |
| Basic Reprod. Rate | Ro | 5.542 |

## Permissivenes Pattern

| Stage Number | Stage Start Date | Permissiveness | Transition time [days] |
|---|---|---|---|
| 0 | | 1 | |
| 1 | 7-Mar-20 | 0.7 | 6 |
| 2 | 10-Apr-20 | 0.38 | 6 |
| 3 | 13-Jul-20 | 0.03 | 5 |
| 4 | 4-Oct-20 | 0.34 | 7 |
| 5 | 27-Dec-20 | 0.01 | 15 |
| 6 | 15-Feb-21 | 0.17 | 21 |
| 7 | 7-Apr-21 | 0.25 | 7 |
| 8 | 18-Nov-22 | 0.04 | 20 |
| 9 | 21-Dec-22 | 0.16 | 7 |

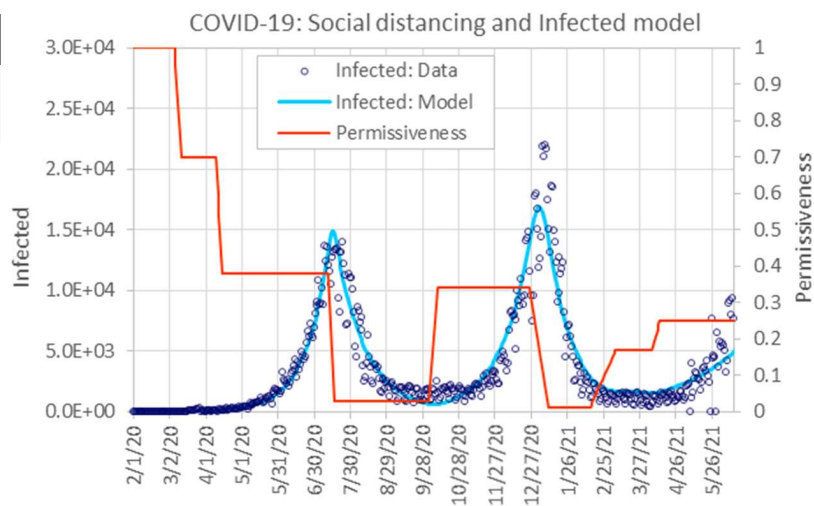

COVID-19: Social distancing and Infected model

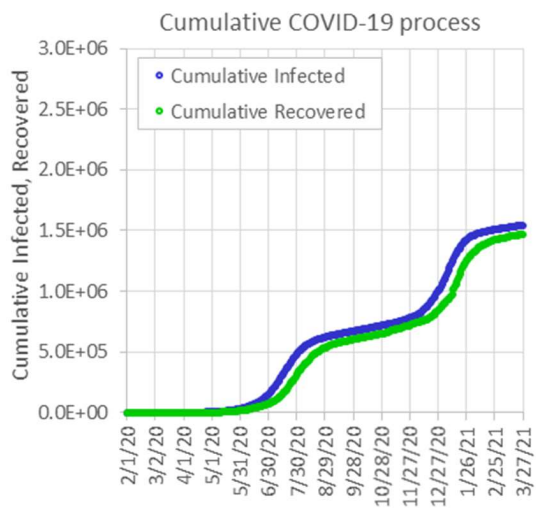

Cumulative COVID-19 process

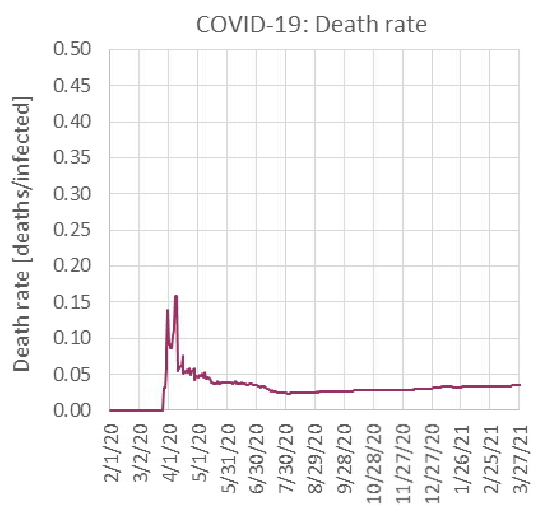

COVID-19: Death rate